\documentclass{article}

\usepackage{arxiv}

\usepackage[utf8]{inputenc} 
\usepackage[T1]{fontenc}    
\usepackage{hyperref}       
\usepackage{url}            
\usepackage{booktabs}       
\usepackage{amsfonts}       
\usepackage{nicefrac}       
\usepackage{microtype}      
\usepackage{lipsum}		
\usepackage{graphicx}
\usepackage{natbib}
\usepackage{doi}
\usepackage{siunitx}
\usepackage{amsmath}
\usepackage{amsthm}

\usepackage{mathabx}
\usepackage{subcaption}
\usepackage{csquotes}
\usepackage{booktabs}
\usepackage{xcolor}
\usepackage{algorithm}
\usepackage{algpseudocode}
\algblock{Input}{EndInput}
\algnotext{EndInput}
\algblock{Output}{EndOutput}
\algnotext{EndOutput}
\newtheorem*{remark}{Remark}
\newtheorem{theorem}{Theorem}
\providecommand{\keywords}[1]
{
  \small	
  \textbf{\textit{Keywords---}} #1
}

\usepackage{cleveref}

\title{A Score-based Generative Solver for PDE-constrained Inverse Problems with Complex Priors}

\author{ \href{https://orcid.org/0000-0003-2284-1030}{\includegraphics[scale=0.06]{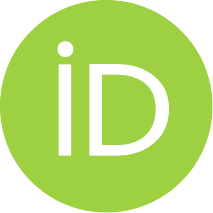}\hspace{1mm}Yankun Hong}\\
	Centre for Analysis, Scientific Computing \\ and Applications\\
	Eindhoven University of Technology\\
	Eindhoven, 5600 MB \\
	\texttt{y.hong@tue.nl} \\
	\And
	\href{https://orcid.org/0000-0003-3606-499X}{\includegraphics[scale=0.06]{orcid.pdf}\hspace{1mm}Harshit Bansal} \\Centre for Analysis, Scientific Computing \\ and Applications\\
	Eindhoven University of Technology\\
	Eindhoven, 5600 MB \\
	\texttt{h.bansal@tue.nl} \\
        \And
	\href{https://orcid.org/0000-0003-3947-5882}{\includegraphics[scale=0.06]{orcid.pdf}\hspace{1mm}Karen Veroy} \\Centre for Analysis, Scientific Computing \\ and Applications\\
	Eindhoven University of Technology\\
	Eindhoven, 5600 MB \\
	\texttt{k.p.veroy@tue.nl} \\
}

\renewcommand{\shorttitle}{A score-based diffusion model for PDE-constrained inverse problems}

\hypersetup{
pdftitle={A Score-based Generative Solver for PDE-constrained Inverse Problems with Complex Priors},
pdfauthor={Y.~Hong, H.~Bansal, K.~P.~Veroy},
pdfkeywords={Inverse Estimation, Score-based Generative Diffusion Model, Bayesian Inference, Solid Mechanics, Physics-informed Deep Learning},
}

\begin{document}
\maketitle

\begin{abstract}
In the field of inverse estimation for system modeled by partial differential equation (PDE), challenges arise when estimating high- (or even infinite-) parameters. Typically, the ill-posed nature of such problems necessitates leveraging prior information to achieve well-posedness. In most existing inverse solvers, the prior distribution is assumed to be of either Gaussian or Laplace form which, in many practical scenarios, is an oversimplification. In case the prior is complex and the likelihood model is computationally expensive (e.g., due to an expensive forward model), drawing the sample from such posteriors can be computationally intractable, especially when the unknown parameter is high-dimensional. 
In this work, we address the issue of complex and unknown prior distributions by proposing an efficient sampling method, namely, a score-based diffusion model. This model combines a score-based generative sampling tool with a noising and denoising process driven by stochastic differential equations (SDEs). This tool is used for iterative sample generation in accordance with the posterior distribution, while simultaneously learning and leveraging the underlying information and constraints inherent in the given complex prior. A time-varying time schedule is proposed to adapt the method for posterior sampling. Secondly, to expedite the simulation of non-parameterized PDEs and enhance the generalization capacity, we introduce a physics-informed convolutional neural network surrogate for the forward model. Finally, numerical experiments, including a hyper-elastic problem and a multi-scale mechanics problem, demonstrate the efficacy of the proposed approach. In particular, the score-based generative diffusion model, coupled with the physics-informed convolutional neural network surrogate, effectively learns geometrical features from provided prior samples, yielding better inverse estimation results compared to the state-of-the-art techniques.
\end{abstract}


\section{Introduction}\label{sec:Intro}
The field of inverse problems is of great interest in numerous areas of research and industry, since it plays a crucial role in inferring parameters that cannot be directly observed \citep{benning_modern_2018, engl_regularization_2000}. With the growing relevance of digital twins, for example, inverse problems constrained by partial differential equations are of particular importance. In such problems, given noised partial observations of the unknown, (typically) high-dimensional states, our aim is to estimate parameters which may also be high-dimensional. Due to the high dimensionality of the state or parameter to be inferred, such inverse problems are generally strongly ill-posed. Furthermore, additional challenges also arise in such scenarios due to the high computational expense associated with solving the forward problem \citep{dasgupta_conditional_2023}. We now take a closer look at these two challenges -- \emph{(1)} ill-posedness and \emph{(2)} computational expense of forward problems. 

\noindent \textbf{Ill-posed nature of inverse estimation}: The methods for solving inverse problems can be broadly categorized into classical \textit{knowledge-driven} approaches and more recently proposed \textit{data-driven} methods \citep{arridge_solving_2019, runkel_learning_2023}. On the one hand, knowledge-driven methods leverage forward models to minimize the misfit between simulated results and observations, and seek solutions akin to the maximum likelihood point, which may not be unique. To mitigate ill-posedness, knowledge-driven methods are often used in conjunction with regularization techniques. (For a comprehensive understanding of knowledge-driven approaches and regularization, we refer the reader to review works such as \citep{benning_modern_2018, engl_regularization_2000}.) However, in many practical scenarios, most regularization techniques that applied in konwledge-driven inverse methods (e.g., $L_1$ and $L2$ regularization) are often an oversimplification of the available auxiliary data (e.g., some complex but informative prior knowledge likes the geometrical structure of the unknown parameter) that may be pivotal for inverse estimation and needs to be considered in the inverse framework. On the other hand, data-driven methods offer an alternative perspective to extend variational regularization techniques for ill-posed problems with complex (prior) data \citep{arridge_solving_2019, dasgupta_conditional_2023}. In the Bayesian framework, the (noisy) forward model functions as the likelihood of the observation, and the variational regularization term serves as the prior \citep{benning_modern_2018, dasgupta_conditional_2023}. This perspective allows for greater flexibility in choosing priors to handle ill-posedness, ranging from classical Gaussian priors in analytical formulations, to more sophisticated priors in statistical formulations, making use of recent techniques such as machine learning \citep{arridge_solving_2019, scarlett_theoretical_2023, ulyanov_deep_2018}.
Data-driven inverse estimation methods with machine learning-based priors were initially proposed and predominantly applied for inverse problems arising in the field of computer vision, e.g., medical imaging \citep{jalal_robust_2021, song_solving_2022} and image processing \citep{bohra_bayesian_2023, chung_diffusion_2023, jaiswal_physics-driven_2023, runkel_learning_2023, ulyanov_deep_2018}. These aforementioned works demonstrate that, by leveraging complex priors, data-driven inverse methods perform exceptionally well for strongly ill-posed inverse problems, such as for the extremely high- or infinite-dimensional inverse problems of interest in this work. 

In Bayesian inversion, solving the inverse problem translates into seeking the posterior distribution. Practically, the posterior distribution can be complex and can not be expressed analytically. Sampling methods are thus pivotal, particularly in data-driven inverse methods or when dealing with complex priors or posteriors. Comprehensive reviews of this strategy can be found in works such as \citep{arridge_solving_2019, scarlett_theoretical_2023}. With the emergence of data science and machine learning, the computer vision field has witnessed the development of various sampling tools, including variational autoencoders \citep{kingma_adam_2017}, generative adversarial networks \citep{goodfellow_generative_2020}, and score-based\footnote{In statistics, the score, also called informant, is the gradient of the logarithm of a probability density ($p$) with respect to the parameter ($\boldsymbol{\mu}$), i.e., $\nabla_{\boldsymbol{\mu}} \ln{p}$. It indicates the steepness of the log-likelihood function and thereby the sensitivity to infinitesimal changes to the parameter values.} methods like Langevin Monte Carlo \citep{grenander_representations_1994} and score-based diffusion models \citep{ho_denoising_2020, sohl-dickstein_deep_2015, song_generative_2020}. These sampling tools are designed to draw, or generate, new samples by learning the distribution for the given dataset of samples.

Among these sampling methods, diffusion models have been recognized for their superior performance and ability to serve as an efficient tool for solving inverse problems and to facilitate posterior sampling \citep{song_solving_2022, song_score-based_2021}. Our study thus focuses on the score-based diffusion models to solve PDE-constrained inverse problems.
Furthermore, the advantage of a score-based method over a likelihood-based technique is its independence from the normalization function, which is often challenging and potentially intractable. The score-based diffusion model, in conjunction with the Langevin Monte Carlo method, introduces a noising-denoising process based on a stochastic differential equation (SDE) to enhance the accuracy of the learned surrogate for the score. Historically, the diffusion model was independently proposed and referred to as Score Matching with Langevin Dynamics (SMLD) \citep{song_generative_2020} and Denoising Diffusion Probabilistic Modeling (DDPM) \citep{ho_denoising_2020, sohl-dickstein_deep_2015}. Both SMLD and DDPM are based on the score matching algorithm \citep{efron_tweedies_2011, hyvarinen_estimation_2005, vincent_connection_2011}, and categorized as more general \textit{score-based generative diffusion models}; they differ only in the way SDEs are used to apply noise \citep{song_score-based_2021}.

The application of the score-based diffusion model in inverse problems is predominantly active in the field of computer vision, as seen in \citep{bohra_bayesian_2023, chung_diffusion_2023, song_solving_2022}. In such applications, the prior is not presented in an analytical formulation but as a sample, e.g., a set of images, following the underlying prior distribution, which is not necessarily in Gaussian or Laplace form, nor available in the closed form expression. 
The diffusion model has been demonstrated to excel at learning the score and, consequently, the underlying distribution of the sample, as the prior; see \citep{song_generative_2020, song_score-based_2021}. By leveraging the score of the likelihood of the observations, with the prior learned via score-based generative diffusion model, a sampling method for the posterior is obtained, thereby rendering the inverse estimation and uncertainty quantification to be feasible \citep{chung_diffusion_2023, song_solving_2022}. 

In recent years, the theory and methodology of diffusion models, both for solving inverse problems and for generative modeling, have developed rapidly and is continually being improved; see, e.g., \citep{de_bortoli_riemannian_2022, ma_preconditioned_2023}. Most of the works (see, e.g., \citep{jalal_robust_2021, song_solving_2022}) deal with solving linear inverse problems. However, since most practical problems are non-linear, \citet{chung_diffusion_2023} have proposed an improved way to sample posteriors for inverse problems under noise, which can efficiently handle the nonlinearity of the likelihood given by the forward model. This improvement is a big step forward in employing the theory of diffusion models for solving nonlinear inverse problems. 
Despite the rapid development and success of diffusion models in the community of machine learning and computer vision (see, e.g., the framework proposed in \citep{chung_diffusion_2023}), its application in the field of computational science and physics-constrained inverse problems is still under-explored. Recently, the \textit{original} diffusion model has been applied to estimate the initial condition of a dynamical system \citep{dasgupta_conditional_2023, legin_posterior_2023}, extended to an audio inverse problem \citep{moliner_solving_2023}, and applied in the context of micro-structure optimal tuning \citep{vlassis_denoising_2023}. \citet{rozet_score-based_2023} have leveraged the original diffusion model to develop a score-based data assimilation method. Also, the strategy of a noising-denoising process to mimic the development of a diffusion problem and the subsequent use of the reverse process to infer the initial condition has been applied in \citep{holzschuh_solving_2023}.
However, research in the field of applying diffusion models, even with recent improvements, e.g., in \citep{chung_diffusion_2023}, in PDE-constrained inverse problems, is still limited. This is true especially in the field of computational mechanics, where often only boundary observations are available for inverse estimation. Furthermore, there still exist potential issues and difficulties that are not well understood, e.g., the instabilities that arise in the conditional sampling see, for instance, \citep{baldassari_conditional_2023}.

In this work, building upon standard methods for Bayesian inference, we propose two main modifications to the diffusion model for solving PDE-constrained inverse problems. First, we adapt and extend the score-based diffusion model for conditional distributions driven by the PDE-constrained inverse problems, specifically, the posterior in physics-constrained Bayesian inversion, by building on the recently-developed theory for complex, non-linear forward model \citep{chung_diffusion_2023}. Second, to address the instability and divergence of the diffusion model in solving PDE-constrained inverse problems, we propose a new noising stochastic process, which works as a time-varying step size in the sampling process.  This overcomes potential instabilities during denoising and further refines the theory behind score-based diffusion models for posterior sampling.

\noindent \textbf{High computational expense of forward problems}: The computational cost to solve the forward problem plays a significant role in the performance of inverse problem methods. The diffusion model is a Markov Chain Monte Carlo method that converges slowly \citep{song_score-based_2021} and relies on the derivative of the forward model. Hence, it is generally computationally intensive, especially for extremely high-dimensional PDE-constrained inverse problems. The associated large computational expense to solve the forward models necessitates the use of surrogate or reduced order models to improve efficiency.  

In this work, we consider problems involving an infinite-dimensional input parameter. An example of the later is a spatially dependent physical property, such as the conductivity in thermal problems or Young's modulus in solid mechanics. To expedite the forward simulation, an inexpensive surrogate for the forward problem is imperative. In this scenario, however, parameterization of the input can be challenging, rendering the class of parametric model order reduction techniques, such as the reduced basis method \citep{hesthaven_certified_2016, quarteroni_reduced_2016}, and (several) ML-based model order reduction techniques, such as \citep{guo_reduced_2018, hesthaven_non-intrusive_2018}, inapplicable. Another class of (ML-based) surrogates utilizes neural networks as a regression tool to approximate the parameter-to-state map. The widely-known approaches in this class include, e.g., physics-informed neural networks (PINNs) \citep{lagaris_artificial_1998, mao_physics-informed_2020, raissi_physics-informed_2019}, and DeepONets \citep{lu_learning_2021}. However, as shown in Table 1 of \citep{kovachki_neural_2023}, some ML-based reduced surrogate approaches learn functions for evaluating the state at a given spatial point. This means that one single network evaluation yields information only at a small batch of input points. This approach thus tends to be inefficient when we need to predict the whole field of the unknown state for a given parameter field. The physics-informed PointNet \citep{kashefi_physics-informed_2023} is an improved variant of PINN to address this issue. It allows the learned network to take the input at any point. Alternative methods, e.g., Neural Operators \citep{kovachki_neural_2023} and convolutional neural networks (CNNs) \citep{gao_phygeonet_2021, kharazmi_variational_2019, sharma_weakly-supervised_2018} are also suitable for mapping the infinite-dimensional input quantity to the infinite-dimensional unknown quantity of the PDE. However, as a method based on ML, it is possible that the learned neural network has a poor generalization accuracy due to the high dimensionality of the input parameter and limited training data \citep{sharma_weakly-supervised_2018}. To enhance the generalization accuracy, a physics-informed strategy has been introduced into the training process by incorporating the governing PDE as a part of the loss function, see, e.g., \citep{gao_phygeonet_2021, kharazmi_variational_2019, sharma_weakly-supervised_2018}, where a finite difference method is wrapped into a convolutional kernel to determine the gradient. However, such methods to obtain derivatives face challenges near the boundaries of the domain of the underlying problem \citep{kashefi_physics-informed_2023}. 

In this work, the forward model maps a high-dimensional \textit{input} parameter (e.g., a distributed or spatially-dependent material property) to a high-dimensional \textit{output} (e.g., the state). Assuming that the input (or \textit{features}) and the output (or \textit{labels}) are represented using images of the same resolution, then the forward model essentially maps between two high-dimensional images. Given their documented success in learning maps between images, we propose to exploit CNNs \citep{lecun_deep_2015, schmidhuber_deep_2015} as surrogates for the forward model. Furthermore, we also leverage the physics-informed strategy \citep{karniadakis_physics-informed_2021} to improve the generalization ability of the network. Contrary to the finite difference method used in \citep{gao_phygeonet_2021, kharazmi_variational_2019, sharma_weakly-supervised_2018}, we apply the finite element method to incorporate all relevant operators into the linear layers and activation functions of the convolutional neural network, thus alleviating the challenges at boundaries for gradient operators associated with finite difference-based methods. This approach is easily implemented using existing finite element packages and is compatible with finite element-based supervised learning data. Furthermore, since the gradient via the finite element method does not rely on a regular mesh, it can also be readily deployed for methods, such as graph CNNs, that fare well in irregular mesh domains \citep{gruber_comparison_2022, pichi_graph_2023}.

\noindent \textbf{Summary of contributions:} The main contributions of this work are highlighted as follows.

\begin{itemize}
    \item To address the instabilities encountered by the score-based generative diffusion model in solving PDE-constrained inverse problems with high-dimensionality and complex statistical priors, we propose a time-varying step size strategy which leads to a new denoising stochastic process for the denoising algorithm and improves the theoretical understanding of the score-based generative diffusion model.
    \item To enhance generalization accuracy, we leverage the CNN surrogate of the forward model with a physics-informed strategy by incorporating the governing PDE in the training stage.
    \item To demonstrate the efficiency of the proposed inverse solver compared to state-of-the-art methods, we perform numerical experiments on a hyper-elastic mechanics model and a more complex multi-scale mechanics model.
\end{itemize}

\subsection{Outline}\label{subsec:outline}
The remaining sections of this paper deal with the score-based generative diffusion model and the physics-informed CNN. Beginning with \cref{sec:Prob_set}, we present a concise problem description of the forward and inverse problems, and a brief discussion of Bayesian inversion as a blueprint for subsequent numerical experiments. Next, \cref{sec:GDM} discusses the score-based generative diffusion model. In \cref{subsec:GDMprior}, we present the theory of the diffusion model and how to draw samples from it. In \cref{subsec:GDMposterior}, built upon the theory in \cref{subsec:GDMprior}, the adaptations of the diffusion model to the inverse estimation are discussed. \Cref{sec:PICNN} then presents the details of the numerical experiments and the problem-tailored CNNs as the inexpensive surrogate for the forward models. In \cref{subsec:CNN_PS}, we focus on the problem description of our numerical experiments. In Sections \ref{subsec:CNN} and \ref{subsec:PICNN}, we then discuss the CNN surrogate and its physics-informed counterpart for the specific test problems. The results of the numerical experiments are presented in \cref{sec:Res}. In \cref{subsec:hemp} we discuss the performance of the diffusion model in a hyper-elastics mechanics problem, while in \cref{subsec:msmp} we delve into a multi-scale problem. In \cref{subsec:compcost}, we present briefly the computational cost for the numerical implementation. Finally, we conclude in \cref{sec:Concl} with a summary of the work, along with a perspective on potential directions for future study.

\section{Problem description and Bayesian inversion}\label{sec:Prob_set}

In this section, we show a concise problem description of the forward model and the associated inverse problem. Next, we discuss the Bayesian formulation of the inverse problem and present the necessary terminology. 

We consider an unknown field of physical quantity, $u(x)$, within a computational domain $x \in \Omega$, governed by PDEs or, more generally, by a corresponding action functional $A(u(x); {\boldsymbol{\mu}}(x))$. This action, such as the Lagrangian in a kinetics problem or the virtual work in mechanics problems, is influenced by the physical property or {\it parameter field}, ${\boldsymbol{\mu}}(x)$,  that controls the system. The determination of the unknown physical quantity $u$ is given by
\begin{equation}\label{min_act}
    u(x) = \arg \min_{\tilde{u}(x)} A(\tilde{u}(x); {\boldsymbol{\mu}}(x)) =: \mathcal{A}({\boldsymbol{\mu}}(x)).
\end{equation}
Eqn.~\eqref{min_act} defines a map $\mathcal{A}$ from the underlying physical property $\boldsymbol{\mu}$ (e.g., a parameter) to the (unknown) quantity $u$ via $u = \mathcal{A}({\boldsymbol{\mu}})$, i.e., the parameter to unknown state map. Assume a partially noisy observation of $u$ is available: 
\begin{equation}\label{eq:2}
    y = \mathcal{H}(u) + \varepsilon,
\end{equation}
\noindent where $\mathcal{H}$ is the observation operator, and $\varepsilon \sim \mathcal{N}(0, \Sigma_\varepsilon)$ represents Gaussian noise in the observations. 
Using $u = \mathcal{A}({\boldsymbol{\mu}})$, \eqref{eq:2} can be expressed as
\begin{equation}\label{for_mod}
    y = \mathcal{H} \circ \mathcal{A} ({\boldsymbol{\mu}}) + \varepsilon.
\end{equation}
Under such a setting, the likelihood of $y$ conditioned on ${\boldsymbol{\mu}}$ is given by
\begin{equation}\label{for_mod_dis}
    y|{\boldsymbol{\mu}} \sim \mathcal{N}(\mathcal{H} \circ \mathcal{A} ({\boldsymbol{\mu}}), \Sigma_\varepsilon). 
\end{equation}

In the inverse problem, given the observation data $y$, the aim is to estimate the physical property or parameter ${\boldsymbol{\mu}}(x)$ as a function on $\Omega$. To address the ill-posedness, a prior distribution $\Tilde{p}({\boldsymbol{\mu}})$ for the parameter to be estimated ${\boldsymbol{\mu}}$ is introduced. From standard Bayesian theory, given the observation $y$ and the prior $\Tilde{p}$, the posterior of ${\boldsymbol{\mu}}$ is given by
\begin{equation}\label{mu_post}
    p({\boldsymbol{\mu}}|y) = \frac{p(y|{\boldsymbol{\mu}})\Tilde{p}({\boldsymbol{\mu}})}{p(y)} \ \propto \ 
    \exp{\left(-\frac{1}{2\sigma_{\varepsilon}^2}\left(y-\mathcal{H} \circ \mathcal{A} ({\boldsymbol{\mu}})\right)^T \cdot \left(y-\mathcal{H} \circ \mathcal{A} ({\boldsymbol{\mu}})\right)\right)} \Tilde{p}({\boldsymbol{\mu}}),
\end{equation}

\noindent where i.i.d. noise for each observation, i.e., $\Sigma_\varepsilon = \sigma_\varepsilon^2 \textbf{I}$, has been imposed. 

If the prior $\Tilde{p}$ is of Gaussian or Laplace form, seeking the maximum a posteriori (MAP) point in \eqref{mu_post} is equivalent to the conventional variational method for the inverse problem in which we minimize the misfit between the simulated observation, $\mathcal{H} \circ \mathcal{A} ({\boldsymbol{\mu}})$, and the true observation $y$ with $L_2$ or $L_1$ regularization, respectively. However, challenges arise when a more informative yet complex prior is available, albeit not given in a closed-form expression. 

In this work, we consider a scenario where no analytical expression of the prior is provided, which is the case for many practical scientific applications. Instead, we are given a sample $S := \{\Tilde{{\boldsymbol{\mu}}}_i\}_{i=1}^{\Tilde{N}}$ drawn from the \enquote{hidden} prior which respects the underlying features, for instance, geometrical constraints. Such a prior is more informative than the often assumed Gaussian priors or similar simple priors. Furthermore, it is well-known that a more informative prior can better estimate the underlying features of the field of the parameter ${\boldsymbol{\mu}}(x)$. The inverse problem, investigated in this work, aims to estimate the unknown infinite-dimensional parameter $\boldsymbol{\mu}$ using the given observation $y$, forward model $\mathcal{H} \circ \mathcal{A}$ and the prior sample $S$.

\section{Score-based generative diffusion model}\label{sec:GDM}

We begin this section by motivating the need for the score-based diffusion model for solving the high-dimensional inverse problem, specifically for tackling the challenges associated with the complex posterior distribution in Eqn.~\eqref{mu_post}. In \cref{subsec:GDMprior}, we present the basic elements of diffusion models for sampling from a general distribution, as well as the method used for unconditional sampling. Next, in \cref{subsec:GDMposterior}, we discuss the diffusion model for conditional sampling from the posterior distribution in the scope of physics-constrained inverse problems, and propose several improvements. 

In our problem, Eqn.\,\eqref{mu_post} generally lacks a closed form due to unknown, complex, high-dimensional priors $\widetilde{p}$. In such settings, analytical analysis of the posterior becomes difficult. Consequently, a tractable sampling method is essential for solving the inverse problem. Using likelihood-based sampling methods is impractical as they often rely on accurate posterior probabilities. However, the later becomes computationally intractable when dealing with a high-dimensional ${\boldsymbol{\mu}}$ since the normalization function can not be readily computed. 

The conventional score-based Markov Chain Monte Carlo (MCMC) sampling method, often referred to as the Langevin Monte Carlo method, or Langevin dynamics sampling \citep{grenander_representations_1994}, provides the recursive sample step which describes the transition of the Markov chain, given by 
\begin{equation}\label{LD}
    {\boldsymbol{\mu}}_{k} \gets {\boldsymbol{\mu}}_{k-1} + \epsilon \nabla_{\boldsymbol{\mu}} \ln p({\boldsymbol{\mu}}_{k-1}) + \sqrt{2\epsilon} \, \textbf{z}_{k-1}. 
\end{equation}
This provides the state at step $k$, for $k \in \{1, ..., K\}$, where $K$ is the total number of MCMC steps, given the probability density $p$ for ${\boldsymbol{\mu}}$. Here, numerical discretization introduces a large number of degrees of freedom for the unknown field of the property, denoted as ${\boldsymbol{\mu}} \in \mathbb{R}^{N_{\text{d}}}$, where $N_{\text{d}}$ is the dimensionality of $\boldsymbol{\mu}$. Furthermore, $\mathbf{z}_k \sim \mathcal{N}(0, \mathbf{I})$, $\mathbf{z}_k \in \mathbb{R}^{N_{\rm d}}$, has the same data shape as ${\boldsymbol{\mu}}$, and the hyper-parameter $\epsilon$, which approaches zero as $K \to \infty$, controls the step size. The recursive formula \eqref{LD} is initialized with ${\boldsymbol{\mu}}_0$ drawn from an arbitrary prior distribution, such as a standard Gaussian, $\mathcal{N}(0, \mathbf{I})$. The term, $\nabla_{\boldsymbol{\mu}} \ln p({\boldsymbol{\mu}})$, referred to as the \textit{score}, is computationally intractable but can be approximated using a machine learning surrogate through the score-matching algorithm; see, e.g., \citep{hyvarinen_estimation_2005, vincent_connection_2011}. However, such surrogates typically perform poorly in the lower probability areas due to limited training data in that region, leading to poor generalization accuracy, making the sampling unstable. In the score-based method, the so-called manifold hypothesis \citep{song_generative_2020} states that the distribution of practical quantities is generally far from the uniform distribution and, hence, this poor performance due to the low likelihood area is likely to be more severe. The aforementioned issues have prompted the development of the \textit{score-based generative diffusion model} \citep{ho_denoising_2020, sohl-dickstein_deep_2015, song_generative_2020, song_score-based_2021}. 

\subsection{Diffusion model for unconditional distribution}\label{subsec:GDMprior} 

We first delve into \textit{unconditional sampling}, i.e., the standard score-based generative diffusion model which, based on the given prior samples, can be used to generate new samples that respect the underlying prior distribution. The method presented in this subsection is primarily based on \citep{song_score-based_2021, vincent_connection_2011}. The main idea is discussed next.

The score-matching algorithm is a method that builds a robust surrogate to predict the score at given parameter points. The score-based generative diffusion model then incorporates a noising-denoising process into the score-matching algorithm to enhance the quality of score prediction. The \textit{noising process}, defined by a stochastic differential equation (SDE), transforms the target distribution $p$, given by a sample $S_0$, into a distribution, $p_T$, referred to as the distribution at the terminal time $T$ of noising process and we call it the terminated distribution. This terminated distribution is designed such that it is easy to draw samples from. Furthermore, it should be suited for the score-matching algorithm, which requires that the terminated distribution is almost Gaussian. Subsequently, a sample is drawn from the terminated distribution $p_T$ using both the Langevin Monte Carlo method in \eqref{LD} and the reverse SDE \citep{anderson_reverse-time_1982}, facilitating the transition of the sample back to the initial distribution $p_0 := p$. Governed by the reverse SDE, this transition process is commonly known as the \textit{denoising process}.

\subsubsection{Noising-denoising process}\label{subsubsec:N-DN_p} 

The noising process involves perturbing the given sample $S_0$ via an the governing SDE in the following manner:
\begin{equation}\label{SDE_kernel}
    {\rm d} {\boldsymbol{\mu}} = \textbf{f}({\boldsymbol{\mu}}, t) {\rm d}t + g(t) {\rm d} \textbf{w}, \qquad t \in (0, T].
\end{equation}
This yields a time-dependent noised sample, $S(t)$, in which different time steps correspond to varying noise levels imposed on the given prior sample. Here, $\mathbf{w}$ is an $N_{\rm d}$-dimensional Wiener process, $\textbf{f}: \mathbb{R}^{N_{\rm d}} \times \mathbb{R} \mapsto \mathbb{R}^{N_{\rm d}}$ represents the drift of the SDE, and $g: \mathbb{R} \mapsto \mathbb{R}$ is the diffusion of the SDE. The noising process requires the SDE to be such that, for any given initial condition $S_0$, the obtained terminated distribution $p_T := p(t = T)$ approximately follows a tractable distribution, such as a Gaussian distribution.

Intuitively, $S$ at $t \to 0$ corresponds to an unaltered sample, while at $t = T$, it corresponds to a sample with the largest noise scale and the one that approximately follows a Gaussian distribution. Solving \eqref{SDE_kernel} with $p(t \to 0) = p_0$ (or alternatively $S(t \to 0) = S_0$) as the initial condition results in $p(t)$ (or $S(t)$). 

\begin{figure}[t!]
\centering
    \captionsetup{width=0.95\linewidth}
    \includegraphics[width=0.98\linewidth]{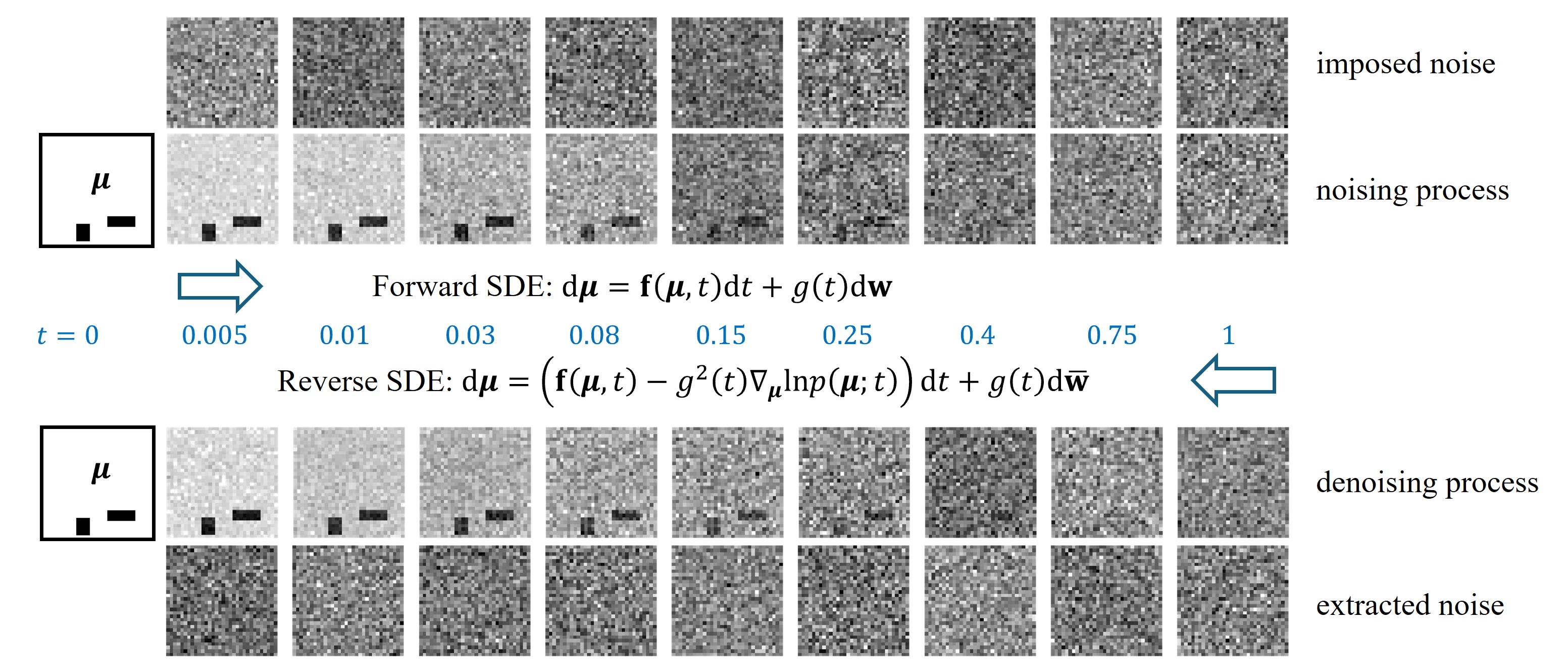}
    \caption{The schematic diagram for the noising and denoising processes.}
    \label{fig:SDERSDE}
\end{figure}

The reverse SDE \citep{anderson_reverse-time_1982} corresponding to \eqref{SDE_kernel}, is given by
\begin{equation}\label{RSDE_kernel}
    {\rm d} {\boldsymbol{\mu}} = \left(\textbf{f}({\boldsymbol{\mu}}, t) - g^2(t) \nabla_{\boldsymbol{\mu}} \ln{p(\boldsymbol{\mu}; t)} \right){\rm d}t + g(t) {\rm d} \bar{\textbf{w}}, \qquad t \in (0, T],
\end{equation}
for the case where the diffusion $g$ is independent of the state $\boldsymbol{\mu}$. Here, $\bar{\mathbf{w}}$ is a reverse $N_{\rm d}$-dimensional Wiener process. 

Using the Euler-Maruyama discretization technique, at the $n$-th time step, with $n \in (1, \ldots, N_T)$ and $t_0 = 0$, $t_{N_T} = T$, we can rewrite \eqref{SDE_kernel} and \eqref{RSDE_kernel} as:
\begin{align}
    \boldsymbol{\mu}_{n} & \gets \boldsymbol{\mu}_{n-1}  + \textbf{f}(\boldsymbol{\mu}_{n-1}, t_{n-1}) \Delta t + g(t_{n-1}) \sqrt{\Delta t} \, \textbf{z}_{n-1}, \label{EM_SDE} \\
    \boldsymbol{\mu}_{n-1} & \gets \boldsymbol{\mu}_{n}  - \left(\textbf{f}(\boldsymbol{\mu}_{n}, t_{n}) - g^2(t_{n}) \nabla_{\boldsymbol{\mu}} \ln{p(\boldsymbol{\mu}_{n}; t_{n})}\right) \Delta t + g(t_{n}) \sqrt{\Delta t} \, \bar{\textbf{z}}_{n}, \label{EM_RSDE}
\end{align}
respectively, with $\mathbf{z}_n$ and $\bar{\mathbf{z}}_n$, following $\mathcal{N}(0, \mathbf{I})$, and being independent and identically distributed (i.i.d.) for all $n \in (1, \ldots, N_T)$. The schematic diagram for the noising process, using forward SDE \eqref{SDE_kernel}, and the denoising process, using reverse SDE \eqref{RSDE_kernel}, is illustrated in \cref{fig:SDERSDE}. 

We now formulate the sampling algorithm in \cref{algo:GDM_prior} based on the reverse SDE described by \eqref{EM_RSDE}, the Langevin Monte Carlo \eqref{LD}, and the approximated score $s_\theta (\boldsymbol{\mu}, t) \approx \nabla_{\boldsymbol{\mu}} \ln{p_t(\boldsymbol{\mu})}$ from the score-matching algorithm; note that here, $\theta$ denotes the learnable parameter for $s_\theta$ and $p_t(\boldsymbol{\mu}) := p(\boldsymbol{\mu}; t)$.

\begin{algorithm}[b!]
    \begin{algorithmic}[1] 
        \Input
        \State $\textbf{f}$ and $g$  \Comment{the drift and diffusion of the SDE in \eqref{SDE_kernel}}
        \State $s_\theta$  \Comment{the approximated score function for $\nabla_{\boldsymbol{\mu}} \ln{p_t(\boldsymbol{\mu})}$}
        \State $p_T$  \Comment{the tractable terminated distribution} 
        \EndInput
        
        \State draw a sample $\{\boldsymbol{\mu}^i_{N_T}\}_{i=1}^{N_S} \sim p_T$
        \For {$n = N_T$ \texttt{:} $1$}
            \For {$i = 1$ \texttt{:} $N_S$}
                \State $\hat{\boldsymbol{\mu}}_{n-1}^i \gets \boldsymbol{\mu}_{n}^i  - \left(\textbf{f}(\boldsymbol{\mu}_{n}^i, t_{n}) - g^2(t_{n}) s_\theta(\boldsymbol{\mu}_{n}^i, t_{n})\right) \Delta t$
                \State $\textbf{z} \sim \mathcal{N}(0, \textbf{I})$
                \State $\boldsymbol{\mu}_{n-1}^i \gets \hat{\boldsymbol{\mu}}_{n-1}^i + g(t_{n}) \sqrt{\Delta t} \, \textbf{z}$ \Comment{reverse SDE in \eqref{EM_RSDE}}
                \For {$j = 1$ \texttt{:} $K$}  \Comment{Langevin Monte Carlo in \eqref{LD}}
                    \State $\textbf{z} \sim \mathcal{N}(0, \textbf{I})$
                    \State $\epsilon \gets 2 (r \|\textbf{z}\|/\|s_\theta(\boldsymbol{\mu}_{n}^i, t_{n})\|)^2$  \Comment{Langevin step size}
                    \State $\boldsymbol{\mu}_{n-1}^i \gets \boldsymbol{\mu}_{n-1}^i + \epsilon \, s_\theta(\boldsymbol{\mu}_{n}^i, t_{n}) + \sqrt{2\epsilon} \, \textbf{z}$
                \EndFor
            \EndFor
        \EndFor
        \Output
        \State $\{\boldsymbol{\mu}^i_0\}_{i=1}^{N_S}$  \Comment{the samples following $p_0$}
        \EndOutput
    \end{algorithmic}
    \caption{Score-based generative diffusion model for unconditional distribution}
    \label{algo:GDM_prior}
\end{algorithm} 

In \cref{algo:GDM_prior}, we commence with Line 5, which involves drawing a sample following the terminated distribution $p_T$. For each sample point $\boldsymbol{\mu}^i$, we next apply the reverse SDE, i.e., \eqref{EM_RSDE}, successively to perturb the sample point. This transforms the distribution $p_{t_{n}}$ to the preceding one $p_{t_{n-1}}$. This process is outlined in Lines 8 -- 10. We then employ a corrector process to enhance the sampling quality using Langevin Monte Carlo, i.e., \eqref{LD}, as an MCMC method, as detailed in Lines 11 -- 15. In this algorithm, the Langevin step size, $\epsilon$, which, as suggested in \citep{song_score-based_2021}, plays a pivotal role in balancing the weight of information from noise $\textbf{z}$ and the signal, i.e., the score, is determined by the norm of the Gaussian noise, the norm of the score, and a tunable hyper-parameter $r$ that stands for the signal-to-noise ratio. 

To comprehend why the score-based generative diffusion model enhances sampling quality compared to using only Langevin Monte Carlo in this context, we underscore that the bottleneck for Langevin Monte Carlo in \eqref{LD} arises from the poor approximation of $s_\theta$ to $\nabla_{\boldsymbol{\mu}} \ln p({\boldsymbol{\mu}})$ in areas with a smaller likelihood. This deficiency hampers convergence from the initial state, drawn following another tractable distribution (typically a Gaussian distribution), which is likely to be situated in those less likely areas. In MCMC terminology, this is referred to as \textit{slow mixing}.

In the diffusion model, the score surrogate $s_\theta$ is learned not only for the true distribution $p_0$ but also for all time steps $t \in (0, T]$. Specifically, at $t=T$, the distribution is close to a Gaussian distribution, from which we draw the initial state for Langevin Monte Carlo. Since the approximation for $s_\theta(\cdot; t=T)$ to $\nabla_{\boldsymbol{\mu}} \ln p_T({\boldsymbol{\mu}})$ is of better quality compared to the approximation obtained by Langevin Monte Carlo alone, and since $p_T$ is nearly the same as the Gaussian distribution for the initial state, mixing at time step $t=T$ is ensured to be fast. Subsequently, at each time step $t = t_n$, the sample from the previous step $t = t_{n+1}$ often serves as an excellent initial state for MCMC. This alleviates concerns about the low likelihood area since the distributions $p_{t_n}$ and $p_{t_{n-1}}$ often resemble each other. Hence, a sample point from $p_{t_{n+1}}$ frequently has a high likelihood in $p_{t_{n}}$.

We now introduce two most representative variants of diffusion models based on the SDEs employed in the process. These two variants, historically proposed independently, have been demonstrated to be efficient in computer vision \citep{ho_denoising_2020, sohl-dickstein_deep_2015, song_generative_2020}. The first, known as Score Matching with Langevin Dynamics (SMLD) \citep{song_generative_2020, song_score-based_2021}, employs an SDE with exploding variance as $t \to \infty$ and is thus referred to as a Variance Exploding (VE) SDE. The second, named Denoising Diffusion Probabilistic Model (DDPM) \citep{ho_denoising_2020, sohl-dickstein_deep_2015, song_score-based_2021}, utilizes an SDE with bounded variance for any $t$ and is hence referred to as a Variance Preserving (VP) SDE. On the one hand, SMLD utilizes an SDE-based noising process, the score matching algorithm, and Langevin Monte Carlo at each noise level to approach the true sample. On the other hand, DDPM trains the probabilistic model at each noise level to ensure tractable sampling via also score-matching algorithm, where the latter provides a way to learn the score of the underlying distribution from a given set of samples. Essentially, they are both categorized as score-based generative diffusion models \citep{song_score-based_2021}. To better understand the performance of diffusion models, we will implement these two variants in this work. Further details regarding the execution of the noising/denoising processes in both variants, (e.g., the SDEs and the stochastic processes), are briefly summarized next. 

\noindent \textbf{SMLD with VE SDE:} The general VE SDE used in the SMLD method, along with its forward and reverse Euler-Maruyama discretizations, are expressed as follows:
\begin{align}
    {\rm d} \boldsymbol{\mu} & = \sqrt{\frac{{\rm d} \sigma^2(t)}{{\rm d} t}}\, {\rm d} \textbf{w}, & & t \in (0,T]; \label{SMLD_g} \\
    \boldsymbol{\mu}_{n} & \gets \boldsymbol{\mu}_{n-1} + \sqrt{\sigma^2(t_{n}) - \sigma^2(t_{n-1})}\, \textbf{z}_{n-1}, & & n \in (1, ..., N_T); \label{SMLD_EM_g} \\
    \boldsymbol{\mu}_{n-1} & \gets \boldsymbol{\mu}_{n} + (\sigma^2(t_{n}) - \sigma^2(t_{n-1})) \nabla_{\boldsymbol{\mu}} \ln{p_{t_{n}}(\boldsymbol{\mu}_{n})} & & \nonumber\\
    & \qquad \qquad \qquad + \sqrt{\sigma^2(t_{n}) - \sigma^2(t_{n-1})}\, \bar{\textbf{z}}_{n}, & & n \in (1, ..., N_T). \label{SMLDR_EM_g}
\end{align}
Here, $\sigma: [0, T] \mapsto \mathbb{R}$ is a given increasing function. The solution in a conditional formulation of the above SDE \eqref{SMLD_g}, denoted by $p_{0t}$, represents the distribution at $t$ conditioned on $\boldsymbol{\mu}(t=0)$ and is given by 
\begin{equation}\label{SMLD_c_g}
    p_{0t}(\boldsymbol{\mu}(t) | \boldsymbol{\mu}(0)) = \mathcal{N}\left( \boldsymbol{\mu}(t); \boldsymbol{\mu}(0), (\sigma^2(t) - \sigma^2(0)) \textbf{I} \right). 
\end{equation}
From \eqref{SMLD_c_g}, we can see that, since the variance, $\sigma(t)$, is an increasing with $t$ function with respect to $t$, the variance of the stochastic process, $(\sigma^2(t) - \sigma^2(0))$, also increases and, hence, the SDE \eqref{SMLD_g} is called a VE SDE. 

In this work, we use $\sigma(t) = \frac{\hat{\sigma}^t}{\sqrt{2\ln{\hat{\sigma}}}}$ \citep{song_score-based_2021}, where $\hat{\sigma}$ is a given constant. Given the form of the SDE and by using the Euler-Maruyama discretization, following \citep{song_score-based_2021}, we arrive at the following expressions:
\begin{align}
    {\rm d} \boldsymbol{\mu} & = \hat{\sigma}^t {\rm d} \textbf{w}, & & t \in (0,T]; \label{SMLD} \\
    \boldsymbol{\mu}_{n} & \gets \boldsymbol{\mu}_{n-1} + \hat{\sigma}^{t_{n-1}} \sqrt{\Delta t}\, \textbf{z}_{n-1}, & & n \in (1, ..., N_T); \label{SMLD_EM} \\
    \boldsymbol{\mu}_{n-1} & \gets \boldsymbol{\mu}_{n} + \hat{\sigma}^{2 t_{n}} \nabla_{\boldsymbol{\mu}} \ln{p_{t_{n}}(\boldsymbol{\mu}_{n})} \Delta t + \hat{\sigma}^{t_{n}} \sqrt{\Delta t}\, \bar{\textbf{z}}_{n}, & & n \in (1, ..., N_T). \label{SMLDR_EM}
\end{align}
The conditional solution is also provided by:
\begin{equation}\label{SMLD_c}
    p_{0t}(\boldsymbol{\mu}(t) | \boldsymbol{\mu}(0)) = \mathcal{N}\left( \boldsymbol{\mu}(t); \boldsymbol{\mu}(0), \frac{\hat{\sigma}^{2t} - 1}{2\ln{\hat{\sigma}}} \textbf{I} \right). 
\end{equation}

\noindent \textbf{DDPM with VP SDE:} In the scope of DDPM, the VP SDE, the corresponding Euler-Maruyama discretization of the forward process, and the Euler-Maruyama discretization for the reverse SDE are, respectively, expressed as follows:
\begin{align}
    {\rm d} \boldsymbol{\mu} & = -\frac{1}{2} \beta(t) \boldsymbol{\mu} {\rm d}t + \sqrt{\beta(t)}\, {\rm d} \textbf{w}, & & t \in (0,T]; \label{DDPM} \\
    \boldsymbol{\mu}_{n} & \gets \sqrt{1 - \beta_{n-1}} \boldsymbol{\mu}_{n-1} + \sqrt{\beta_{n-1}}\, \textbf{z}_{n-1}, & & n \in (1, ..., N_T); \label{DDPM_EM} \\
    \boldsymbol{\mu}_{n-1} & \gets \frac{\boldsymbol{\mu}_{n}}{\sqrt{1 - \beta_n}} + \beta_n \nabla_{\boldsymbol{\mu}} \ln{p_{t_{n}}(\boldsymbol{\mu}_{n})} + \sqrt{\beta_n}\, \bar{\textbf{z}}_{n}, & & n \in (1, ..., N_T). \label{DDPMR_EM}
\end{align}
Here, $\beta: [0,T] \mapsto \mathbb{R}^+$ is a given function and is generally increasing \citep{song_score-based_2021}, and $\beta_n := \beta(t_{n}) \Delta t$. In this work, as suggested in \citep{song_score-based_2021}, $\beta(t) = 32 t$. In this way, we impose noise with increasing strengths during the noising process. Furthermore, the aforementioned form of $\beta(t)$ guarantees that the terminated distribution is almost Gaussian. The conditional solution for \eqref{DDPM} is given by:
\begin{equation}\label{DDPM_c}
    p_{0t}(\boldsymbol{\mu}(t) | \boldsymbol{\mu}(0)) = \mathcal{N}\left( \boldsymbol{\mu}(t); \boldsymbol{\mu}(0) e^{-\frac{1}{2} \int_0^t \beta(s) \, {\rm d}s}, \left(1 - e^{- \int_0^t \beta(s) \, {\rm d}s}\right) \textbf{I} \right). 
\end{equation}
In \eqref{SMLD_c_g}, we can see that the variance $\left(1 - e^{- \int_0^t \beta(s) \, {\rm d}s}\right)$ is bounded by $1$ as $t$ increases. Hence, the SDE \eqref{DDPM} is called a VP SDE. 

\subsubsection{Denoising score matching}\label{subsubsec:DSM}

To obtain an estimated score $s_\theta(\boldsymbol{\mu}, t)$ that approximates $\nabla_{\boldsymbol{\mu}} \ln{p_t(\boldsymbol{\mu})}$ for all $\boldsymbol{\mu}$ and all $t \in [0, T]$, we use a loss function as proposed in \citep{song_score-based_2021}. This loss function, which serves as the target for minimization in the training stage, is given by
\begin{equation}\label{loss_score}
\begin{split}
    \mathcal{L}_s(\theta) := & \ \mathbb{E}(\left\|s_\theta(\boldsymbol{\mu}(t), t) - \nabla_{\boldsymbol{\mu}} \ln{p_{0t}(\boldsymbol{\mu}(t)|\boldsymbol{\mu}(0))} \right\|^2) \\
    = & \ \mathbb{E}_{t \sim \mathcal{U}(0,T)} \left( \lambda(t) \mathbb{E}_{\boldsymbol{\mu}(0) \sim p_0} \mathbb{E}_{\boldsymbol{\mu}(t) \sim p_{0t}} (\left\|s_\theta(\boldsymbol{\mu}(t), t) - \nabla_{\boldsymbol{\mu}} \ln{p_{0t}(\boldsymbol{\mu}(t)|\boldsymbol{\mu}(0))} \right\|^2) \right),
\end{split}
\end{equation}
where $\mathcal{U}(0,T)$ denotes the uniform distribution over $[0, T]$. $\mathcal{L}_s(\theta)$ differs slightly from the loss functions used in the standard denoising score-matching algorithms developed in \citep{hyvarinen_estimation_2005, vincent_connection_2011} since it introduces a temporal weighting function $\lambda: (0, T] \mapsto \mathbb{R}$ that, as in \citep{song_score-based_2021}, possesses the following property:
\begin{equation*}
    \frac{1}{\lambda(t)} \ \propto \ \mathbb{E} \left(\left\|\nabla_{\boldsymbol{\mu}} \ln{p_{0t}(\boldsymbol{\mu}(t)|\boldsymbol{\mu}(0))} \right\|^2\right).
\end{equation*}
This property can be readily determined from \eqref{SMLD_c_g}, \eqref{SMLD_c}, or \eqref{DDPM_c} since we have the analytical expression of the distribution. This weight aims to balance the magnitude of different score-matching losses across time. This can be understood from the fact that the time steps that have a large score expectation naturally have a large absolute approximated misfit \citep{song_score-based_2021}.

Furthermore, as demonstrated in \citep{efron_tweedies_2011, hyvarinen_estimation_2005, vincent_connection_2011}, for any noising SDEs corresponding to Gauss–Markov processes, e.g., the processes \eqref{SMLD_c_g} and \eqref{DDPM_c} associated to VE and VP SDEs, respectively, the loss function $\mathcal{L}_s$ in \eqref{loss_score} is equivalent to 
\begin{equation}\label{loss_s_eq1}
    \mathcal{L}_s(\theta) = \mathbb{E}_{t \sim \mathcal{U}(0,T)} \left( \lambda(t) \mathbb{E} \left(\left\|s_\theta(\boldsymbol{\mu}(t), t) + \frac{\textbf{z}_{0t}}{\sigma_{0t}} \right\|^2\right) \right),
\end{equation}
where $\sigma_{0t}$, the standard deviation of the conditional distribution $p_{0t}$, can easily be determined using \eqref{SMLD_c_g}, \eqref{SMLD_c}, or \eqref{DDPM_c}. Additionally, $\textbf{z}_{0t}$ represents the standard Gaussian fluctuation used to generate the data point $\boldsymbol{\mu}(t)$ conditional on $\boldsymbol{\mu}(0)$ according to \eqref{SMLD_c_g}, \eqref{SMLD_c}, or \eqref{DDPM_c}. 

To train the estimated score $s_\theta$ using the given $S_0 =  \{\tilde{\boldsymbol{\mu}}_{0}^{i}\}_{i=1}^{\tilde{N}}$, we employ the conditional distribution \eqref{SMLD_c_g}, \eqref{SMLD_c}, or \eqref{DDPM_c}, depending on the noising SDE to be used. This allows us to generate $S_n = \{\tilde{\boldsymbol{\mu}}_{n}^{i}\}_{i=1}^{\tilde{N}}$ such that each $\tilde{\boldsymbol{\mu}}_{n}^{i}$ follows the conditional distribution $p_{0t}(\cdot|\tilde{\boldsymbol{\mu}}_{0}^{i})$. Specifically, since $p_{0t}$ is a Gaussian distribution according to \eqref{SMLD_c_g}, \eqref{SMLD_c}, or \eqref{DDPM_c}, we determine the conditional mean denoted by $\bar{\boldsymbol{\mu}}_{n}^{i}|\tilde{\boldsymbol{\mu}}_0^i$ and the conditional standard deviation $\sigma_{0t}$, and then generate $\textbf{z}_{0t} \sim \mathcal{N}(0, \textbf{I})$. Finally, we generate the training data $\tilde{\boldsymbol{\mu}}_{n}^{i} = (\bar{\boldsymbol{\mu}}_{n}^{i}|\tilde{\boldsymbol{\mu}}_0^i) + \sigma_{0t} \textbf{z}_{0t}$. Additionally, this procedure provides the values for $\sigma_{0t}$ and $\textbf{z}_{0t}$ that appear in the loss \eqref{loss_s_eq1}.

The prevailing choice for the architecture of the surrogate $s_\theta$ backbone is the U-net architecture \citep{ronneberger_u-net_2015}, a neural network design that has proven effective in several works on diffusion models, e.g., \citep{chung_diffusion_2023, song_generative_2020, song_solving_2022, song_score-based_2021}. The U-net architecture is widely acknowledged for its strong performance, particularly in tasks related to machine learning and computer vision. For a comprehensive understanding of the training process of the U-net, which is a special architecture of the convolutional neural network, detailed information can be found in standard references on machine learning and computer vision, such as \citep{goodfellow_deep_2016, he_deep_2015, ronneberger_u-net_2015, tancik_fourier_2020}.

\begin{remark}
In \citep{hyvarinen_estimation_2005}, an alternative loss function, which is equivalent to \eqref{loss_score}, is employed under the name of the Implicit Score Matching (ISM) method. This loss function is expressed in the form:
\begin{equation*}
    \mathcal{L}_s(\theta) = \mathbb{E} \left(\left\|s_\theta(\boldsymbol{\mu}(t), t)\right\|^2 + 2 {\rm div}_{\boldsymbol{\mu}}(s_\theta(\boldsymbol{\mu}(t), t))\right) + \mathbb{E} \left(\left\|\nabla_{\boldsymbol{\mu}} \ln{p_{0t}(\boldsymbol{\mu}(t)|\boldsymbol{\mu}(0))}\right\|^2\right).
\end{equation*}
Since the last term in the above equation does not depend on $\theta$, minimizing $\mathcal{L}_s(\theta)$ with respect to $\theta$ is equivalent to minimizing
\begin{equation*}
    \mathbb{E} \left(\left\|s_\theta(\boldsymbol{\mu}(t), t)\right\|^2 + 2 {\rm div}_{\boldsymbol{\mu}}(s_\theta(\boldsymbol{\mu}(t), t))\right).
\end{equation*}
Essentially, the term \enquote{implicit} in ISM signifies that the above-defined loss function does not explicitly contain any terms relevant to the score $\nabla_{\boldsymbol{\mu}} \ln{p_{0t}(\boldsymbol{\mu}(t)|\boldsymbol{\mu}(0))}$. Compared to \eqref{loss_s_eq1}, ISM is useful when the dataset of $\boldsymbol{\mu}(t)$ is available, without explicitly requiring $\frac{\textbf{z}_{0t}}{\sigma_{0t}}$. However, delving into ISM is outside the scope of this work.
\end{remark}

\subsection{Diffusion model for posterior distribution}\label{subsec:GDMposterior}

In \cref{subsec:GDMprior}, we delved into the generative diffusion model designed for sampling from an unconditional distribution $p$. In this subsection, we shift our focus to the posterior distribution given by \eqref{mu_post} which, unlike in \cref{subsec:GDMprior}, is a conditional distribution. We present how to adjust the sampling method described in the previous subsection to this conditional case. In addition, we 
show how the conditional sampling method can be adapted to nonlinear forward models in PDE-constrained inverse problems.

\subsubsection{Standard posterior sampling}\label{subsubsec:standardGDMp}

The approach outlined in this subsection builds upon prior works, drawing inspiration particularly from \citep{chung_diffusion_2023, efron_tweedies_2011}.

The diffusion model tailors the posterior distribution through its score function given by
\begin{equation}\label{post_score}
    \nabla_{\boldsymbol{\mu}} \ln{p(\boldsymbol{\mu}(t)|y)} = \nabla_{\boldsymbol{\mu}} \ln{p(y|\boldsymbol{\mu}(t))} + \nabla_{\boldsymbol{\mu}} \ln{\Tilde{p}_t(\boldsymbol{\mu}(t))}. 
\end{equation}
The second term in \eqref{post_score}, $\nabla_{\boldsymbol{\mu}} \ln{\Tilde{p}_t(\boldsymbol{\mu}(t))} := \nabla_{\boldsymbol{\mu}} \ln{\Tilde{p}(\boldsymbol{\mu}(t); t)}$, represents the score function of the prior distribution $\Tilde{p}$, denoted as $s_\theta$, and is approximated by the U-net architecture, as outlined in \cref{subsubsec:DSM}.

The first term in \eqref{post_score}, which corresponds to the likelihood, is contingent on the forward model $\mathcal{A}$ and is influenced by the noising process through \eqref{SMLD_EM_g}, \eqref{SMLD_EM}, or \eqref{DDPM_EM}. At the $n$-th time step, i.e., at $t = t_n$, the likelihood can be factorized as follows:
\begin{equation}
\begin{split}
    p(y|\boldsymbol{\mu}_n) & = \int p(y|\boldsymbol{\mu}_0, \boldsymbol{\mu}_n) p(\boldsymbol{\mu}_0|\boldsymbol{\mu}_n) \, {\rm d} \boldsymbol{\mu}_0 \\
    & = \int p(y|\boldsymbol{\mu}_0) p(\boldsymbol{\mu}_0|\boldsymbol{\mu}_n) \, {\rm d} \boldsymbol{\mu}_0.  \label{post_fact}
\end{split}
\end{equation}
This factorization is justified by the conditional independence of $y$ and $\boldsymbol{\mu}_n$ given $\boldsymbol{\mu}_0$, see \citep{chung_diffusion_2023}. It is important to note that the observation $y$ always corresponds to the true posterior, i.e., $t=0$, rather than the noised distribution at any intermediate time step. Consequently, upon observing $\boldsymbol{\mu}_0$, since $y$ is dependent solely on $\boldsymbol{\mu}_0$, it becomes independent of $\boldsymbol{\mu}_n$.

The computation of the integral in \eqref{post_fact} is known to be intractable \citep{chung_diffusion_2023}. To address this, in \citep{chung_diffusion_2023}, an approximation of the form
\begin{align}
    & p(y|\boldsymbol{\mu}_n) = \int p(y|\boldsymbol{\mu}_0) p(\boldsymbol{\mu}_0|\boldsymbol{\mu}_n) \, {\rm d} \boldsymbol{\mu}_0 \approx p(y|\bar{\boldsymbol{\mu}}_0(\boldsymbol{\mu}_n)) \label{llh_proxy} \\
    & \text{where } \bar{\boldsymbol{\mu}}_0(\boldsymbol{\mu}_n) := \mathbb{E} (\boldsymbol{\mu}_0|\boldsymbol{\mu}_n) = \int \boldsymbol{\mu}_0 p(\boldsymbol{\mu}_0|\boldsymbol{\mu}_n) \, {\rm d} \boldsymbol{\mu}_0, \nonumber
\end{align}
is suggested. Furthermore, in \citep{chung_diffusion_2023}, it is demonstrated that the difference between the proxy $p(y|\bar{\boldsymbol{\mu}}_0(\boldsymbol{\mu}_n))$ and the true value $p(y|\boldsymbol{\mu}_n)$ constitutes a Jensen Gap\footnote{Jensen Gap is the difference between the two sides of the Jensen inequality, and is given by $\mathbb{E}(\varphi(\boldsymbol{\mu})) - \varphi(\mathbb{E}(\boldsymbol{\mu}))$, which is positive for any convex function $\varphi$ according to Jensen inequality \citep{gao_bounds_2020}. Here, $\boldsymbol{\mu}$ is a random variable. }, controlled by the Lipschitz constant of the distribution \citep{chung_diffusion_2023, gao_bounds_2020}. It is noteworthy that \emph{(i)} when the forward model is linear (i.e., $y$ is linear with respect to $\boldsymbol{\mu}_0$) and the noising process is a Gauss–Markov process, the Jensen Gap is zero, and the proxy is accurate; \emph{(ii)} as the denoising process progresses, and $t_n$ approaches $0$, the conditional distribution $p(\boldsymbol{\mu}_0|\boldsymbol{\mu}_n)$ degrades to a deterministic value $\boldsymbol{\mu}_0$, i.e., $\bar{\boldsymbol{\mu}}_0(\boldsymbol{\mu}_n) = \boldsymbol{\mu}_0$, causing the Jensen Gap to tend to $0$. In other words, the error of the approximation of $p(y|\boldsymbol{\mu}_n)$ by $p(y|\bar{\boldsymbol{\mu}}_0(\boldsymbol{\mu}_n))$ decreases and tends to $0$ as the denoising process progresses.

The computation of the expectation $\bar{\boldsymbol{\mu}}_0(\boldsymbol{\mu}_n)$ is facilitated by providing the conditional distribution $p_{0t}(\boldsymbol{\mu}_{t_n}|\boldsymbol{\mu}_0)$ in \eqref{SMLD_c_g}, \eqref{SMLD_c}, or \eqref{DDPM_c} via Tweedie's formula \citep{chung_diffusion_2023, efron_tweedies_2011, kim_noise2score_2021}, which is presented next.

\begin{theorem}[Tweedie's formula \citep{efron_tweedies_2011}]
    Let the conditional distribution $p(\boldsymbol{\mu}_{t_n}|\boldsymbol{\mu}_0)$ belong to the exponential family that has the following formulation:
    \begin{equation*}
        p(\boldsymbol{\mu}_{t_n}|\boldsymbol{\mu}_0) = p_0(\boldsymbol{\mu}_{t_n}) \exp{(\boldsymbol{\mu}_0^T F(\boldsymbol{\mu}_{t_n}) - \phi(\boldsymbol{\mu}_0))},
    \end{equation*}
    where $F$ is a function of $\boldsymbol{\mu}_{t_n}$, $\phi(\boldsymbol{\mu}_0)$ is a function that normalizes the density, and $p_0$ is the density when $\boldsymbol{\mu}_0 = 0$. Under the aforementioned setting, the posterior mean $\bar{\boldsymbol{\mu}}_0 := \mathbb{E} (\boldsymbol{\mu}_0|\boldsymbol{\mu}_n)$ should satisfy:
    \begin{equation*}
        (\nabla_{\boldsymbol{\mu}_{t_n}} F(\boldsymbol{\mu}_{t_n}))^T \bar{\boldsymbol{\mu}}_0 = \nabla_{\boldsymbol{\mu}_{t_n}} \ln{p(\boldsymbol{\mu}_{t_n})} - \nabla_{\boldsymbol{\mu}_{t_n}} \ln{p_0(\boldsymbol{\mu}_{t_n})}.
    \end{equation*}
\end{theorem} 

Applying Tweedie's formula to the SMLD and DDPM methods described in Section \ref{subsubsec:N-DN_p}, following the method shown in \citep{chung_diffusion_2023}, we arrive at the following results
\begin{itemize}
\item For SMLD with the general VE SDE given in \eqref{SMLD_g}, let $F(\boldsymbol{\mu}) := \frac{\boldsymbol{\mu}}{\sigma^2({t_n}) - \sigma^2(0)}$, we have
    \begin{equation}\label{SMLD_pos_g}
        \bar{\boldsymbol{\mu}}_0(\boldsymbol{\mu}_n) = \boldsymbol{\mu}_n + (\sigma^2({t_n}) - \sigma^2(0)) \nabla_{\boldsymbol{\mu}} \ln{p_{t_n}(\boldsymbol{\mu}_n)}.
    \end{equation}
\item For SMLD with the VE SDE given in \eqref{SMLD}, let $F(\boldsymbol{\mu}) := \frac{2\ln{\hat{\sigma}}}{\hat{\sigma}^{2{t_n}} - 1} \boldsymbol{\mu}$, we have
    \begin{equation}\label{SMLD_pos}
        \bar{\boldsymbol{\mu}}_0(\boldsymbol{\mu}_n) = \boldsymbol{\mu}_n + \frac{\hat{\sigma}^{2{t_n}} - 1}{2\ln{\hat{\sigma}}} \nabla_{\boldsymbol{\mu}} \ln{p_{t_n}(\boldsymbol{\mu}_n)}.
    \end{equation}
\item For DDPM with the VP SDE given in \eqref{DDPM}, let $F(\boldsymbol{\mu}) := \frac{e^{- \frac{1}{2} \int_0^{t_n} \beta(s) \, {\rm d}s }}{1 - e^{- \int_0^{t_n} \beta(s) \, {\rm d}s}} \boldsymbol{\mu}$, we have
    \begin{equation}\label{DDPM_pos}
        \bar{\boldsymbol{\mu}}_0(\boldsymbol{\mu}_n) = \frac{1}{e^{- \frac{1}{2} \int_0^{t_n} \beta(s) \, {\rm d}s }} \left(\boldsymbol{\mu}_n + \left(1 - e^{- \int_0^{t_n} \beta(s) \, {\rm d}s}\right) \nabla_{\boldsymbol{\mu}} \ln{p_{t_n}(\boldsymbol{\mu}_n)}\right).
    \end{equation}
\end{itemize}

Note that the function $\nabla_{\boldsymbol{\mu}} \ln{p_t(\boldsymbol{\mu}_n)}$ that appears in \eqref{SMLD_pos_g}, \eqref{SMLD_pos}, and \eqref{DDPM_pos} corresponds to the unconditional score function for the temporal noised distribution based on the prior $\Tilde{p}$, and thus, $\Tilde{p}_t = p_t$. These are estimated using the method in Section \ref{subsubsec:DSM}. Hence, the term can be substituted by the score surrogate: $\nabla_{\boldsymbol{\mu}} \ln{\Tilde{p}_{t_n}(\boldsymbol{\mu}_n)} = \nabla_{\boldsymbol{\mu}} \ln{p_{t_n}(\boldsymbol{\mu}_n)} \approx s_\theta(\boldsymbol{\mu}_n, t)$. 

Similar to \eqref{mu_post}, we substitute \eqref{for_mod_dis} into \eqref{llh_proxy} and then into \eqref{post_score} to obtain the posterior score, given by
\begin{equation}\label{score_post_fin}
\begin{split}
    \nabla_{\boldsymbol{\mu}} \ln{p(\boldsymbol{\mu}_n|y)} & \approx \nabla_{\boldsymbol{\mu}} \ln{p(y|\bar{\boldsymbol{\mu}}_0(\boldsymbol{\mu}_n))} + \nabla_{\boldsymbol{\mu}} \ln{\Tilde{p}_{t_n}(\boldsymbol{\mu}_n)}\\
    & \approx \nabla_{\boldsymbol{\mu}} \frac{-\left(y-\mathcal{H} \circ \mathcal{A} (\bar{\boldsymbol{\mu}}_0(\boldsymbol{\mu}_n))\right)^T \cdot \left(y-\mathcal{H} \circ \mathcal{A} (\bar{\boldsymbol{\mu}}_0(\boldsymbol{\mu}_n))\right)}{2\sigma_{\varepsilon}^2} + s_\theta(\boldsymbol{\mu}_n, t_n) \\
    & = \rho \left(\nabla_{\boldsymbol{\mu}} \mathcal{H} \circ \mathcal{A} (\bar{\boldsymbol{\mu}}_0(\boldsymbol{\mu}_n))\right)^T \cdot \left(y-\mathcal{H} \circ \mathcal{A} (\bar{\boldsymbol{\mu}}_0(\boldsymbol{\mu}_n))\right) + s_\theta(\boldsymbol{\mu}_n, t_n)
\end{split}
\end{equation}
where $\rho = \frac{1}{\sigma_{\varepsilon}^2}$ can be regarded as the step size of the posterior denoising process \citep{chung_diffusion_2023}, and $\nabla_{\boldsymbol{\mu}}$ is the gradient with respect to the variable at $\boldsymbol{\mu}_n$. Eqn.~\eqref{score_post_fin} thus gives us the posterior score that allows us to sample from a conditional distribution using the sampling algorithm of score-based diffusion model described in \cref{subsec:GDMprior}. 

However, we have observed in numerical experiments that this standard setting $\rho = \frac{1}{\sigma_{\varepsilon}^2}$ can lead to divergence during the denoising process when replacing $s_\theta$ in \cref{algo:GDM_prior} with the conditional counterpart $\nabla_{\boldsymbol{\mu}} \ln{p(\boldsymbol{\mu}_n|y)}$ given by \eqref{score_post_fin}. In the next subsection, we consider an alternative to address this issue.  

\subsubsection{Time-varying step size}\label{subsubsec:TVSS}

In this subsection, we will propose two improvements for  SMLD and DDPM to address the issue of instability in the denoising process when applying score-based diffusion in PDE-constrained inverse problems. 

\noindent \textbf{Time-decreasing step size for $\rho$:} To understand the divergence risk linked to the denoising process, we investigate the actual noising-denoising process corresponding to the posterior. For the standard conditional sampling algorithm with constant step size $\rho = \frac{1}{\sigma_{\varepsilon}^2}$, the posterior stochastic process at time step $t_n$ is given by
\begin{equation}\label{nondecay_post}
    p_{\rm p}(\boldsymbol{\mu}_n; t_n|y) \ \propto \ p(y|\bar{\boldsymbol{\mu}}_0(\boldsymbol{\mu}_n)) \cdot \Tilde{p}_{t_n}(\boldsymbol{\mu}_n) \approx p(y|\boldsymbol{\mu}_n) \cdot \Tilde{p}_{t_n}(\boldsymbol{\mu}_n),
\end{equation}
where the random variable of the distribution is $\boldsymbol{\mu}_n$. The term $\Tilde{p}_{t_n}$ represents the noised stochastic process corresponding to the noising SDE, and approximates a Gaussian noise when $t_n \to T$. However, when the forward model $\mathcal{H} \circ \mathcal{A}$ has low uncertainty, then the likelihood $p(y|\bar{\boldsymbol{\mu}}_0(\boldsymbol{\mu}_n))$ plays a dominant role in the computation of the posterior. In this case, the resulting posterior distribution at the terminal time $t_n \to T$ may significantly differ from the Gaussian noise due to the strong likelihood, thereby violating the assumptions of the diffusion model. Furthermore, when we generate the samples from the posterior, the terminated distribution for us to start the denoising process for sampling via the standard conditional sampling algorithm (i.e., \cref{algo:GDM_prior} but with the score $s_\theta$ replaced by \eqref{score_post_fin}) is $\Tilde{p}_{T}$, which is the terminated distribution obtained by noising the prior. However, our objective is to generate the posterior sample, which requires us to utilize the noising posterior $p_{\rm p}(\cdot; T|y)$, i.e., \eqref{nondecay_post}. This means that we use the \textit{wrong} terminated distribution to initialize the denoising process. This could potentially be the reason why, with the constant $\rho = \frac{1}{\sigma_{\varepsilon}^2}$, the use of the standard conditional sampling algorithm leads to divergence issues, which, consequently, puts the framework in a setting that fails in generating posterior samples. In addition, simply replacing the terminated distribution $\Tilde{p}_{T}$ by $p_{\rm p}(\cdot; T|y)$ presents challenges since it is not a simple, tractable distribution for drawing samples. 

To address this issue, we propose to introduce a noising posterior process with exponential decay likelihood. This posterior stochastic process for the noising process is given by 
\begin{equation}\label{decay_post}
    p_{\rm p}(\boldsymbol{\mu}_n; t_n|y) \ \propto \ p^{\sigma_{\varepsilon}^2 \rho(t_n)}(y|\bar{\boldsymbol{\mu}}_0(\boldsymbol{\mu}_n)) \cdot \Tilde{p}_{t_n}(\boldsymbol{\mu}_n). 
\end{equation}
at time step $t_n$, with 
\begin{equation}\label{step_size}
    \rho(t) = \frac{\Delta t}{t} \frac{1}{\sigma_{\varepsilon}^2}.
\end{equation}
With the decay function, $\rho(t)$, and given that $\sigma_{\varepsilon}^2 \rho(t_n) \approx 0$ as $t_n \to T$, the posterior distribution at $T$ approximates $\Tilde{p}_{T}$ and essentially becomes approximated Gaussian noise. This approximated Gaussian distribution is exactly equivalent to the prior terminated distribution $\Tilde{p}_{T}$, thereby fulfilling the assumption of the diffusion model. Moreover, since $\rho(t_1) = \frac{1}{\sigma_{\varepsilon}^2}$, the initial distribution of this stochastic process essentially represents our target posterior distribution: $p_{\rm p}(\cdot; t_1|y) \ \propto \ p(y|\bar{\boldsymbol{\mu}}_1(\cdot)) \cdot \Tilde{p}_1(\cdot) \approx p(y|\bar{\boldsymbol{\mu}}_0(\cdot)) \cdot \Tilde{p}_0(\cdot)$ after the approximation of \eqref{llh_proxy}. The reverse of the stochastic process \eqref{decay_post} can not, in fact, be obtained via the corresponding reverse SDE iteration \eqref{SMLDR_EM_g}, \eqref{SMLDR_EM}, or \eqref{DDPMR_EM} in the denoising process. The additional Langevin Monte Carlo steps, which serve as corrector steps, are needed to adjust the intermediate samples in each denoising step to follow \eqref{decay_post}. Consequently, with Langevin Monte Carlo steps, the samples obtained via the sampling algorithm with the time-varying step size \eqref{step_size} converge to the samples following the true posterior distribution $p(\boldsymbol{\mu}_0|y)$ since the algorithm meets the requirements of the diffusion model \citep{song_score-based_2021}: the terminated distribution is indeed Gaussian, which is consistent with the initialization of the denoising process. 

The decay function in \eqref{decay_post} is equivalent to introducing a time-decreasing step size given in \eqref{step_size} into the posterior score governed by \eqref{score_post_fin}. Substituting \eqref{decay_post} and \eqref{step_size} into \eqref{score_post_fin} yields
\begin{equation}\label{score_post_tvss}
    \nabla_{\boldsymbol{\mu}} \ln{p(\boldsymbol{\mu}_n|y)} = \rho(t) \left(\left(\nabla_{\boldsymbol{\mu}} \mathcal{H} \circ \mathcal{A} (\bar{\boldsymbol{\mu}}_0(\boldsymbol{\mu}_n))\right)^T \cdot \left(y-\mathcal{H} \circ \mathcal{A} (\bar{\boldsymbol{\mu}}_0(\boldsymbol{\mu}_n))\right)\right) + s_\theta(\boldsymbol{\mu}_n, t_n). 
\end{equation}

There is one other possible cause of the divergence of the denoising process. Recall that the score-based generative diffusion model is a sampling tool that generates samples from noise (following $p_T$) and then uses the denoising process to reconstruct samples following $p_0$. Essentially, the samples at the beginning of the denoising process are primarily noise and do not provide much useful information. 

In our work, the forward model is approximated by a surrogate CNN which offers accurate predictions within the region on which the prior distribution $\Tilde{p}$ concentrates, but lacks information on the tail of the distribution. In other words, for those nearly noise samples, the surrogate CNN can barely provide useful predictions and gradients compared to the true model obtained from the PDE solver. Furthermore, in a PDE-constrained forward model, $\boldsymbol{\mu}$ has specific physical significance and constraints. For example, $\boldsymbol{\mu}(x)$ may be required to be smooth and positive for all $x \in \Omega$. The almost-noise samples may not satisfy these constraints, rendering the forward simulation based on the \textit{true} PDE solver unphysical since they violate physical laws. In contrast, the surrogate CNN can still provide \enquote{fake} predictions and gradients in this case, even though the input $\boldsymbol{\mu}$ is physically inconsistent. We argue that assigning a high weight to these \enquote{fake} predictions in \eqref{score_post_fin} likely contributes to the divergence of the denoising process.

\noindent \textbf{Time-decreasing step size for $\eta(t)$:}  The time-decreasing step size, $\rho(t)$, circumvents this divergence issue by using a very small step size at the beginning of the denoising process (corresponding to large $t$), assigning a tiny weight to the forward-model-based likelihood. Then, as the reversing process progresses, the step size increases and assigns the same weight as the constant $\frac{1}{\sigma_{\varepsilon}^2}$ to the likelihood when it arrives at the initial time step. 
 
Moreover, we note that, even with the time-decreasing step size \eqref{step_size}, the application of DDPM through \eqref{DDPMR_EM} and \eqref{DDPM_pos} in the denoising process may still carry the risk of divergence in our applications. To understand this effect, we draw a comparison between SMLD and DDPM.

On the one hand, SMLD utilizes a VE SDE, leading to the conditional solution given by \eqref{SMLD_c_g}. Since the diffusion process of this equation is a martingale \citep{klebaner_introduction_2012} (i.e.,  it maintains a constant mean during the process), the mean of the solution is not close to zero if the current state is not zero, i.e., $\mathbb{E}(\boldsymbol{\mu}(T)|\boldsymbol{\mu}(0)) = \boldsymbol{\mu}(0) \not\approx 0$. Furthermore, the variance increases as $t$ increases, leading to $(\sigma^2(t) - \sigma^2(0)) \gg 1$ for large $t$. Although the mean is highly dependent on the initial distribution $p_0$, considering that the ratio between the mean and the standard deviation is nearly $0$, we can treat this terminated distribution \eqref{SMLD_c_g} as akin to Gaussian noise. On the other hand, DDPM employs a VP SDE, resulting in a conditional solution given by \eqref{DDPM_c} with a bounded variance, i.e., $(1 - e^{- \int_0^t \beta(s) \, {\rm d}s}) \leq 1$. However, since the diffusion of the VP SDE is a supermartingale (i.e., a stochastic process in which the expectation conditional on the current state is less than the current state \citep{klebaner_introduction_2012}), it introduces shrinkage to the mean by a factor of $e^{-\frac{1}{2} \int_0^t \beta(s) \, {\rm d}s}$, which is approximately zero for large $t$. As a result, the mean $\boldsymbol{\mu}(0) e^{-\frac{1}{2} \int_0^t \beta(s) \, {\rm d}s}$ is approximately $0$. In this way, the distribution approximates Gaussian noise with zero mean.

While the different paths they take to reach Gaussian noise may not exhibit significant discrepancies in unconditional sampling, subtle differences emerge when considering conditional sampling, especially with the introduction of the approximation in \eqref{llh_proxy}. Upon comparing \eqref{SMLD_pos_g} with \eqref{DDPM_pos}, 
we observe that the mathematical formula corresponding to the DDPM introduces an additional factor $e^{\frac{1}{2} \int_0^t \beta(s) \, {\rm d}s } \gg 1$ multiplying $\boldsymbol{\mu}_n$, which significantly magnifies the current state $\boldsymbol{\mu}_n$. This observation is closely related to the effect discussed in the preceding paragraphs: the conditional distribution by DDPM contracts the initial state $\boldsymbol{\mu}_0$ to a small value. Conversely, the posterior formula \eqref{DDPM_pos} expands the current state $\boldsymbol{\mu}_n$ to a large value. It is crucial to note that applying \eqref{DDPM_pos} directly in posterior sampling warrants caution. At the start of the denoising process, the primary constituent of the state $\boldsymbol{\mu}_n$ is noise rather than information from the true initial state $\boldsymbol{\mu}_0$. Expanding noise to a large value via the factor $e^{\frac{1}{2} \int_0^t \beta(s) \, {\rm d}s } \gg 1$ can be precarious and lead to extreme instability in the denoising process.

\begin{algorithm}[!b]
    \caption{Score-based generative diffusion model for posterior distribution}
    \label{algo:GDM_post}
    \begin{algorithmic}[1] 
        \Input
        \State $\textbf{f}$ and $g$  \Comment{the drift and diffusion of the SDE in \eqref{SDE_kernel}}
        \State $s_\theta$  \Comment{the approximated score function for $\nabla_{\boldsymbol{\mu}} \ln{p_t(\boldsymbol{\mu})}$}
        \State $p_T$  \Comment{the tractable terminated distribution} 
        \State $\mathcal{H} \circ \mathcal{A}$  \Comment{the forward model} 
        \EndInput
        
        \State draw a sample $\{\boldsymbol{\mu}^i_{N_T}\}_{i=1}^{N_S} \sim p_T$ 
        \For {$n = N_T$ \texttt{:} $1$}
            \For {$i = 1$ \texttt{:} $N_S$}
                \State compute $\bar{\boldsymbol{\mu}}_0(\boldsymbol{\mu}_n^i)$  \Comment{via \eqref{SMLD_pos_g}, \eqref{SMLD_pos} or \eqref{DDPM_pos_s} at $t_n$}
                \State $\textbf{s} = \nabla_{\boldsymbol{\mu}} \ln{p(\boldsymbol{\mu}_{n}^i|y)}$  \Comment{compute for $t_n$ by \eqref{score_post_fin} with $\rho$ given in \eqref{step_size}}
                \State $\hat{\boldsymbol{\mu}}_{n-1}^i \gets \boldsymbol{\mu}_{n}^i  + \left(\textbf{f}(\boldsymbol{\mu}_{n}^i, t_{n}) - g^2(t_{n}) \textbf{s}\right) \Delta t$
                \State $\textbf{z} \sim \mathcal{N}(0, \textbf{I})$
                \State $\boldsymbol{\mu}_{n-1}^i \gets \hat{\boldsymbol{\mu}}_{n-1}^i + g(t_{n}) \sqrt{\Delta t} \, \textbf{z}$ \Comment{reverse SDE in \eqref{EM_RSDE}}
                \For {$j = 1$ \texttt{:} $K$}  \Comment{Langevin Monte Carlo in \eqref{LD}}
                    \State compute $\bar{\boldsymbol{\mu}}_0(\boldsymbol{\mu}_{n-1}^i)$  \Comment{via \eqref{SMLD_pos_g}, \eqref{SMLD_pos} or \eqref{DDPM_pos_s} at $t_n$}
                    \State $\textbf{s} = \nabla_{\boldsymbol{\mu}} \ln{p(\boldsymbol{\mu}_{n-1}^i|y)}$  \Comment{compute for $t_n$ by \eqref{score_post_fin} with $\rho$ given in \eqref{step_size}}
                    \State $\textbf{z} \sim \mathcal{N}(0, \textbf{I})$
                    \State $\epsilon \gets 2 (r \|\textbf{z}\|/\|\textbf{s}\|)^2$  \Comment{Langevin step size}
                    \State $\boldsymbol{\mu}_{n-1}^i \gets \boldsymbol{\mu}_{n-1}^i + \epsilon \, \textbf{s} + \sqrt{2\epsilon} \, \textbf{z}$
                \EndFor
            \EndFor
        \EndFor
        \Output
        \State $\{\boldsymbol{\mu}^i_0\}_{i=1}^{N_S}$  \Comment{the samples following the posterior $p({\boldsymbol{\mu}}|y)$}
        \EndOutput
    \end{algorithmic}
\end{algorithm}

To address the divergence caused by the effect discussed above, we adopt a strategy akin to the step size \eqref{step_size} and introduce a time-decreasing factor $\eta(t)$ to the posterior mean \eqref{DDPM_pos} for DDPM. The modified expression is given by
\begin{equation}\label{DDPM_pos_s}
    \bar{\boldsymbol{\mu}}_0(\boldsymbol{\mu}_n) = \frac{\eta(t_n)}{e^{- \frac{1}{2} \int_0^{t_n} \beta(s) \, {\rm d}s }} \left(\boldsymbol{\mu}_n + \left(1 - e^{- \int_0^{t_n} \beta(s) \, {\rm d}s}\right) \nabla_{\boldsymbol{\mu}} \ln{p_{t_n}(\boldsymbol{\mu}_n)}\right).
\end{equation}

The factor $\eta$ should adhere to the conditions $\eta(t \to T) \to 0$ and $\eta(t \to 0) \to 1$. This is inspired by the fact that as the denoising process progresses (i.e., $t$ approaches 0), the current state $\boldsymbol{\mu}_n$ contains more useful information about $\boldsymbol{\mu}_0$ and less noise. Consequently, we can mitigate the shrinkage effect introduced by the expansion factor. A natural choice for $\eta(t)$ is given by
\begin{equation}\label{step_size2}
    \eta(t) = e^{- \frac{1}{2} \int_0^{t} \beta(s) \, {\rm d}s }.
\end{equation}
Subsequently, the expression for the posterior mean of $\boldsymbol{\mu}_0$ for DDPM simplifies to
\begin{equation}
    \bar{\boldsymbol{\mu}}_0(\boldsymbol{\mu}_n) = \boldsymbol{\mu}_n + \left(1 - e^{- \int_0^{t_n} \beta(s) \, {\rm d}s}\right) \nabla_{\boldsymbol{\mu}} \ln{p_{t_n}(\boldsymbol{\mu}_n)}. 
\end{equation}

In conclusion, the posterior sampling algorithm is summarized in \cref{algo:GDM_post}.

\section{Forward model surrogate}\label{sec:PICNN}

In this section, we introduce a final key component of our proposed approach—the surrogate for the forward model $\mathcal{H}\circ\mathcal{A}$—using a convolutional neural network (CNN) \citep{goodfellow_generative_2020, lecun_deep_2015, schmidhuber_deep_2015}.

A convolutional neural network differs from a fully connected feedforward neural network in its sparse connections between layers. In CNN, only the nodes inside the convolutional kernel in the previous layer are connected to the nodes in the next layer, and all nodes share the same convolutional kernels. This sparse and shared nature significantly reduces the number of trainable parameters. Demonstrated to work very well in computer vision \citep{lecun_deep_2015, schmidhuber_deep_2015}, the CNN layer can extract features relevant to the neighboring nodes, such as edges and contrast, through the convolutional kernel, proving crucial in image processing. 

In this section, we detail the physical problems for the numerical experiments in this work, and address how to adapt the CNNs to these problems and to introduce a physics-informed strategy. The chosen problems include a hyper-elastic mechanics problem, and a more complex, multi-scale problem based on the hyper-elastic model.

\subsection{Model Problems}\label{subsec:CNN_PS} 

The two problems are presented in the following. 

\noindent \textbf{The hyper-elastic problem:} The hyper-elastic model is defined by the variational formulation:
\begin{align}
    u & = \arg \min_{u \in \hat{C}(\Omega)} \mathcal{E} \label{PDE_model} \\
    \mathcal{E} & = \int_{\Omega}\left(\frac{\mu}{2}({\rm Tr}(\textbf{F}^T\cdot\textbf{F}) - 2) - \mu \ln{(\det \textbf{F})} + \frac{\lambda}{2} \ln{(\det \textbf{F})}^2\right) \, {\rm d} V - \int_{\partial \Omega} F \cdot u \, {\rm d} S \label{vir_work}
\end{align}
where $\textbf{F} = \textbf{I} + \nabla_x \textit{\textbf{u}}$ represents the deformation gradient, and $\mathcal{E}$ denotes the virtual work of the system. The Lamé parameters $\lambda$ and $\mu$ are given by $\lambda = \frac{\nu E(x)}{(1 + \nu) (1 - 2 \nu)}$ and $\mu = \frac{E(x)}{2(1 + \nu)}$, where $\nu = 0.3$ is the fixed Poisson ratio. The spatially varying Young’s modulus $E(x)$, denoted as $\boldsymbol{\mu}$ in \cref{sec:Prob_set,sec:GDM}, is an unknown parameter to be estimated in the inverse problem. It serves as an input for the forward model $\mathcal{H} \circ \mathcal{A}$. The prescribed force $F$ imposes as the Neumann boundary condition at the boundary $\partial \Omega$, while the solution space, $\hat{C}(\Omega) := \{u \in C^1(\Omega): u(x) = 0 \text{ for } x \in \partial \Omega_0\}$, enforces a zero Dirichlet boundary at the fixed boundary $\partial \Omega_0 \subset \partial \Omega$.

\begin{figure}[!t]
\centering
\begin{subfigure}[t]{0.48\textwidth}
    \centering
    \captionsetup{width=0.95\linewidth}
    \includegraphics[width=0.6\linewidth]{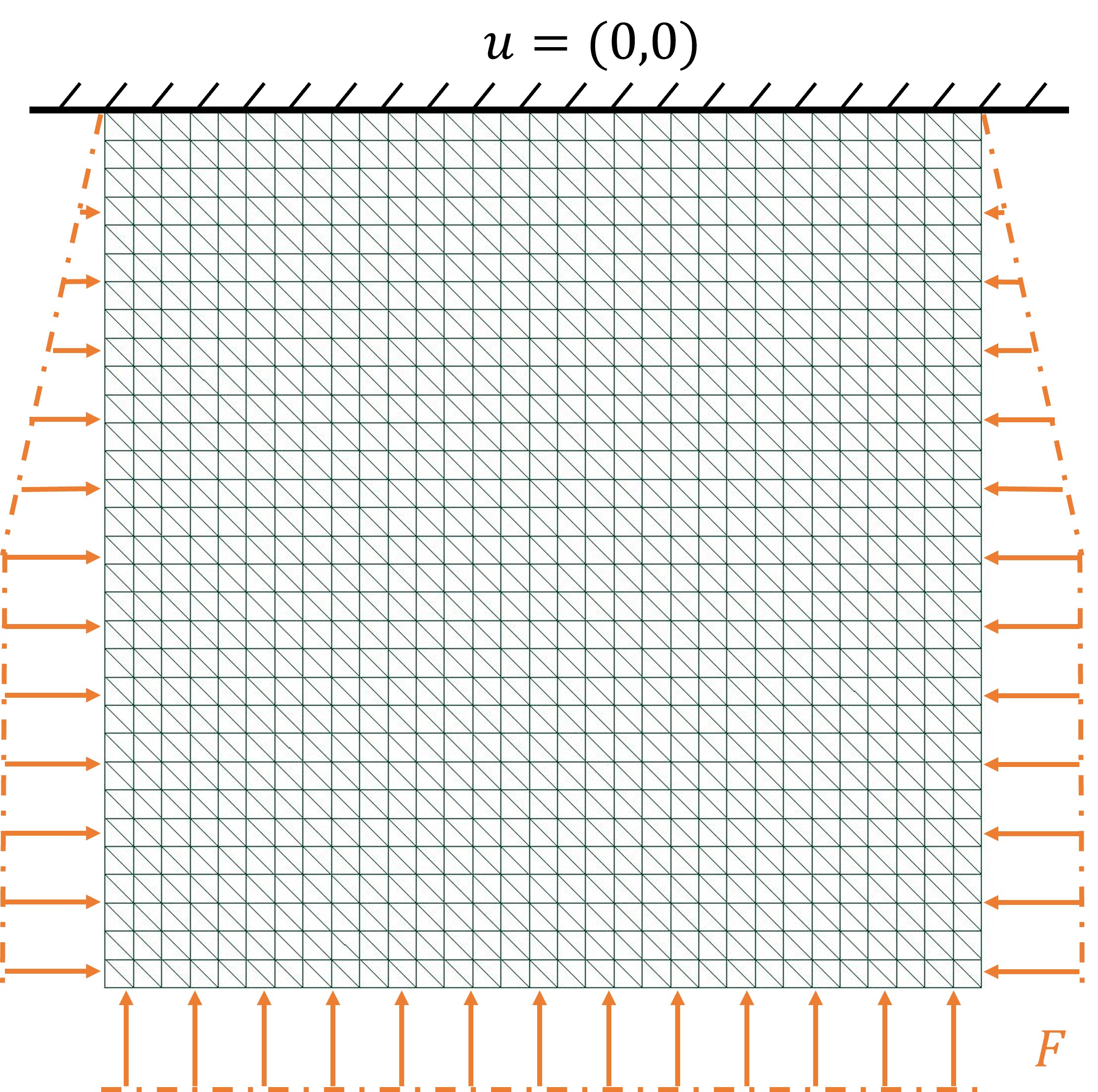}
    \caption{Computational domain $\Omega$. The fixed boundary $\partial \Omega_0$--the upper edge is shown in black, the load $F$ is shown in orange. }
    \label{fig:comp_domain}
\end{subfigure}
\begin{subfigure}[t]{0.48\textwidth}
    \centering
    \captionsetup{width=0.95\linewidth}
    \includegraphics[width=0.6\linewidth]{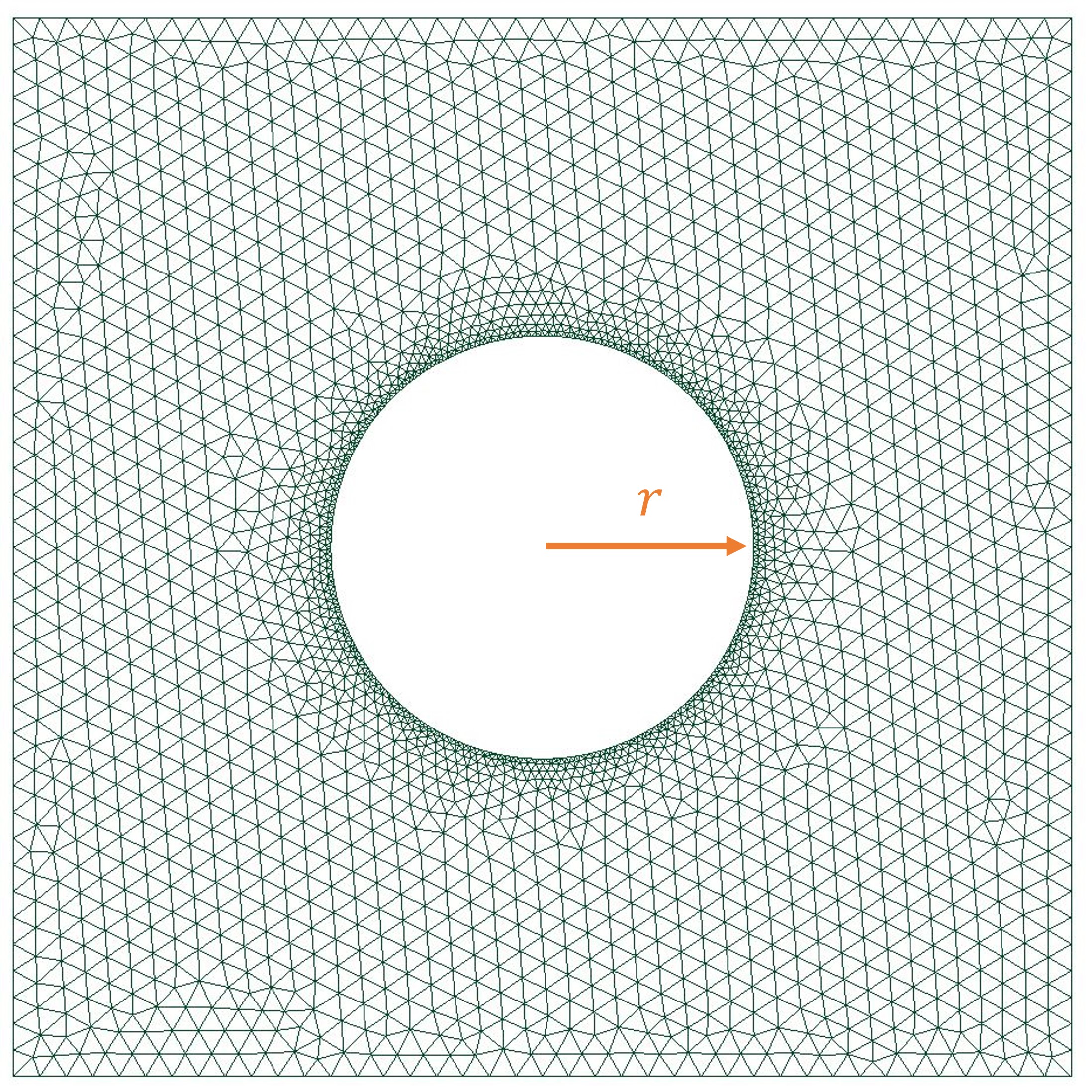}
    \caption{The RVE $\Omega_{\rm m}$. The parameter to be estimated, $r$, is the radius of the porosity as a function of the macro-scale location. }
    \label{fig:RVE}
\end{subfigure}

\caption{\small Computational domains and meshes.}
\label{fig:DigNN}
\end{figure}

The schematic graph of the computational domain $\Omega$ and its boundary are depicted in \cref{fig:comp_domain}. The domain is a square meshed with uniformly distributed node on a grid. The nodes can thus be treated akin to pixels in computer vision, and the various components of the unknown displacement $u = (u_x, u_y)$ are analogous to different channels of an image. 

In this experiment, given a boundary condition $F$ and Poisson ratio $\nu$, the displacement field $u$ is determined by the Young's modulus $E$. We thus aim to infer the spatial distribution of $E$, given observations of $u$. 

\noindent \textbf{The multi-scale problem:} A multi-scale model describes a system that incorporates small scale features in large scale computations, necessitating specialized techniques to avoid using excessively fine spatial discretizations. Computational homogenization methods (CHM) is one method that is commonly employed for such problems; they leverage, for example, the finite element squared method to decompose the solution of partial differential equations into micro-scale and macro-scale computations. Specifically, in solid mechanics problems, the micro-scale simulation computes the effective stress $\textbf{P} (\textbf{F}; \boldsymbol{\mu})$ within a representative volume element (RVE), denoted as $\Omega_{\rm m}$, for a given macro-scale deformation $\textbf{F}$ and parameter $\boldsymbol{\mu}$ at all macro-scale points. Subsequently, the macro-scale simulation employs the numerical effective constitutive law $(\textbf{F}, \boldsymbol{\mu}) \mapsto \textbf{P}$, thus accounting for the micro-structure's effects obtained from the micro-scale simulation. The micro-scale simulation is conducted within the RVE $\Omega_{\rm m}$, utilizing the \enquote{true} constitutive law, typically with periodic boundary conditions \citep{stein_homogenization_2017}. Comprehensive theorems on CHM can be found in, for example, \citep{stein_homogenization_2017, miehe_computational_2002}.

In this numerical experiment, the true constitutive law is assumed to be the hyper-elastic model, with its strain energy density represented by the first term in \eqref{vir_work}. For the micro-scale simulation, the computational domain $\Omega_{\rm m}$ is illustrated in \cref{fig:RVE}, featuring a porous micro-structure. Precisely, the micro-scale problem is: Find
\begin{align*}
    u_{\rm m} & = \arg \min_{u_{\rm m} \in \hat{C}(\Omega_{\rm m})} \mathcal{E}_{\rm m} \\
    \mathcal{E}_{\rm m} & = \int_{\Omega_{\rm m}}\left(\frac{\mu}{2}({\rm Tr}(\textbf{F}_{\rm m}^T\cdot\textbf{F}_{\rm m}) - 2) - \mu \ln{(\det \textbf{F}_{\rm m})} + \frac{\lambda}{2} \ln{(\det \textbf{F}_{\rm m})}^2\right) \, {\rm d} V,
\end{align*}
for $\hat{C}(\Omega_{\rm m}) := \{u_{\rm m} \in C(\Omega_{\rm m}): u_{\rm m}^+ - u_{\rm m}^- = \varepsilon_{\rm RVE} (\textbf{F} - \textbf{I}) \cdot (u^+ - u^-)\}$, where $\varepsilon_{\rm RVE}$ is the length of the edge of $\Omega_{\rm m}$, $u^+$ and $u^-$ stand for the right, top edges and left, bottom edges of $\Omega_{\rm m}$, respectively. We assume periodic boundary conditions \citep{stein_homogenization_2017}, i.e., $\textbf{P}_{\rm m}^+ \cdot \textit{\textbf{n}}^+ = - \textbf{P}_{\rm m}^- \cdot \textit{\textbf{n}}^-$, where $\textbf{P}_{\rm m} = \frac{\partial \mathcal{E}_{\rm m}}{\partial \textbf{F}_{\rm m}(u_{\rm m})}$. Then, $\textbf{P} = \int_{\Omega_{\rm m}} \textbf{P}_{\rm m} \, {\rm d}V$. In this micro-scale problem, $\textbf{F}$ and $\boldsymbol{\mu}$ serve as the parameters to determine $\textbf{P}$. 

In the context of a multi-scale model, our objective is to accelerate both the micro- and the macro-scale computations. First, in building a surrogate for the micro-scale simulation we address the challenges posed by nonlinearities while enhancing generalization accuracy under limited full-order training data. To achieve this, we employ a machine learning-based physics-informed Two-Tier Deep Network (TTDN), which approximates the effective constitutive law $\textbf{P}: (\textbf{F}, \boldsymbol{\mu}) \mapsto f_{\rm TTDN}(\textbf{F}, \boldsymbol{\mu})$ for the macro-scale simulation. TTDN is based on a two-tier strategy that introduces two networks for solving the PDE. In this method, a secondary network is trained to learn the physical information of the governing PDEs, and it is then used to train the primary network to predict the solution of the PDE. Numerical tests \citep{hong_physics-informed_2024} suggest that this method improves the generalization predictive accuracy while reducing the required training data generated by expensive finite element solver. As the focus of this work is not on TTDN details or accelerating micro-scale simulations, we refer the reader to \citep{hong_physics-informed_2024}. 

Second, in building a surrogate for the macro-scale solver, our objective is to find $u(x)$ such that
\begin{equation}\label{M_PDE}
    \nabla_x \textbf{P}(\textbf{F}(u), \boldsymbol{\mu}) = 0
\end{equation}
holds within the same computational domain $\Omega$ as illustrated in \cref{fig:comp_domain}, together with
\begin{equation*}
    u(x) = 0, \quad x \in \partial \Omega_0, \qquad \text{and} \qquad
    \textbf{P}(x) = F(x), \quad x \in \partial \Omega.
\end{equation*}

Since we use a neural network $f_{\rm TTDN}$ by TTDN as a surrogate for $\textbf{P}(\textbf{F}, \boldsymbol{\mu})$, the macro-scale problem in \eqref{M_PDE} is converted into a neural PDE. In this experiment, we aim to estimate the radius of the micro-scale porosity, $r(x)$, as a function of the macro-scale location. Hence, in this experiment, the input for the forward model $\mathcal{H} \circ \mathcal{A}$ and as the quantity for sampling, $\boldsymbol{\mu}$, is the porosity radius $r$. The solver for \eqref{M_PDE}, denoted as $\mathcal{A}$, will be approximated by the CNN surrogate.

\subsection{Convolutional neural network}\label{subsec:CNN}

In mechanics, it is common to assume that the impact of abnormalities in the properties, e.g., Young's modulus $E$ and the porosity radius $r$, diminishes as the distance increases. In other words, abnormalities in $E$ or $r$ significantly influence $u$ only in the neighboring area. Based on this assumption, a CNN is well-suited as the surrogate model for the forward PDE model, as it efficiently captures information relevant to the neighboring area. The convolutional kernel parameters in CNN allow for the control of the size of the neighborhood considered.

In terms of the architecture, we recommend employing a design with uniform width for all the layers. Unlike the U-net architecture \citep{ronneberger_u-net_2015} (which involves downscaling followed by upscaling) that was used as the surrogate for the score-based sampling, we maintain the same resolution for each channel as the resolution of the input $\boldsymbol{\mu}$ and the output $u$ at all layers. This approach ensures that the information from all the pixels (or nodes in finite element terminology) is retained across all the layers, thereby mitigating inaccuracies that may arise during upscaling.

Specifically, we introduce a bias layer as the first layer of the CNN, imposing independent biases for each pixel or node. This layer introduces specific trainable parameters to each pixel, i.e., the trainable parameters are not shared with other nodes. The bias layer captures the information specific to its corresponding node, and also assists in overcoming training difficulties when the data features many zero areas. Without this initial bias layer, identical zero areas in the training data would lead to identical values in the immediate layers, necessitating a very deep network to match the true output value in our mechanics problems. Additionally, the presence of zeros precisely at the non-differentiable points for activation functions like ReLU can pose challenges during the training stage. 

The CNN is trained through supervised learning on the generated training data $\Xi_{\rm sl} = \{(\boldsymbol{\mu}^i, \textbf{u}^i_{\rm PDE})\}_{i=1}^{N_{\rm sl}}$ of the size $N_{\rm sl}$, where $\textbf{u}^i_{\rm PDE}$ is the solution vector obtained by solving the PDE \eqref{PDE_model} using a finite element solver for a given $\boldsymbol{\mu}^i$. The training loss function is the Mean Squared Error (MSE) loss, and is defined as:
\begin{equation}\label{loss_sl}
    \mathcal{L}_{\rm sl}(f_{\rm CNN}, \Xi_{\rm sl}) = \sum_{(\boldsymbol{\mu}^i, \textbf{u}^i_{\rm PDE}) \in \Xi_{\rm sl}} \left\| f_{\rm CNN}(\boldsymbol{\mu}^i) - \textbf{u}^i_{\rm PDE}\right\|^2. 
\end{equation}

In this context, the observation operator $\mathcal{H}$ is represented as a matrix since the observation from the forward model is only the displacement $u$ at the boundary, excluding the known fixed homogeneous Dirichlet boundary. Achieving this involves adding a layer representing $\mathcal{H}$ to the CNN.

\subsection{Physics-informed strategy}\label{subsec:PICNN}

The input of the CNN is high-dimensional since it represents the parameters (here, the Young's modulus or the micro-scale pore radius) as a function of position (here, given at the finite element nodes). Given the high-dimensional nature of the input, the generalization ability of the network may be suboptimal. To enhance the generalization accuracy, we propose a physics-informed strategy.

The physics-informed CNN proposed in this work incorporates the PDE, i.e., \eqref{PDE_model} or \eqref{M_PDE} as an additional part of the loss function. We now discuss the physics-informed loss function for single-scale hyper-elastic and multi-scale problems.   

\noindent \textbf{The hyper-elastic problem:} In the case of the hyper-elastic problem, the PDE \eqref{PDE_model} is already in a variational formulation. In many scientific or engineering problems, the action to be minimized, denoted as $A$ (e.g., the energy density $\mathcal{E}$ in \eqref{vir_work} for mechanical problems) in Section \ref{sec:Prob_set}, is a function of the unknown $u$ and its gradient $\nabla_x u$. The gradient $\nabla_x u$ is linear in $u$, and if, as in the finite element framework, we denote the finite element vector of $u$ as $\textbf{u}$, we can determine the matrix $\textbf{G}$ for the gradient operator. This matrix ensures that $\textbf{v} := \textbf{G} \cdot \textbf{u}$ is the finite element vector corresponding to $\nabla_x u$. After reshaping $\textbf{v}$ into an image pixel array, $\textbf{v}$ becomes a $4$-channel array, $\textbf{v} \in \mathbb{R}^{4\times N_{\rm d}}$, since it is the gradient of a $2$-dimensional vector.

Next, the action $\mathcal{E}$ at every location (node or pixel) can be obtained by applying an activation function element-wise to its input. Taking \eqref{vir_work} as an example, the activation function is given by
\begin{equation*}
    f_{\rm a} (\textbf{v}, E) = \frac{E}{4(1 + \nu)}(\textbf{F}_h^T\cdot\textbf{F}_h) - 2) - \frac{E}{2(1 + \nu)} \ln{(\det \textbf{F}_h)} + \frac{\nu E}{2 (1 + \nu) (1 - 2 \nu)} \ln{(\det \textbf{F}_h)}^2,
\end{equation*}
where $\textbf{F}_h = \textbf{v} + (1,0,0,1)^T$ is the finite element discretization of the deformation gradient, $\textbf{F}_h := \nabla u + \textbf{I}$. Next, via the finite element method and using numerical integration, we can determine the integration vector $\textbf{i}$ such that for any integrable function $g$ and its corresponding finite element vector $\textbf{g}$, $\int_\Omega g \, {\rm d}V \approx \textbf{i}^T \cdot \textbf{g}$. We can also determine a vector $\textbf{f}$ such that $\textbf{f}^T \cdot \textbf{u} \approx  - \int_{\partial \Omega} F \cdot u \, {\rm d} S$ for any $u$.

Using the terms introduced above, we have 
\begin{equation}\label{vw_cnn}
    \mathcal{E}(\textbf{u}, E) \approx \textbf{i}^T \cdot f_{\rm a} (\textbf{G}\cdot \textbf{u}, E) + \textbf{f}^T \cdot \textbf{u}.     
\end{equation}
Note that $\mathcal{E}$ represents the discrete action functional of $u$, and is, in general problem-dependent.
Such a formulation is compatible with CNNs since it is actually a sparse connected linear layer by $\textbf{G}$ composed with an activation function $f_{\rm a}$ and another fully connected layer by $\textbf{i}^T$, as well as a sparse connected linear layer by $\textbf{f}^T$. 

Please note that, although in \eqref{vw_cnn} we use an approximately equal sign for the virtual work, it is assumed to be an accurate finite element approximation to $\mathcal{E}$ and so can be regarded as the true virtual work. 

\noindent \textbf{The multi-scale problem:} For the multi-scale problem, since the neural PDE \eqref{M_PDE} is not provided in a variational formulation, minimizing the action is not a convenient option. Instead, we aim to minimize the \textit{residual} of the PDE.

Analogous to the hyper-elastic problem, we express the deformation gradient tensor $\textbf{F} := \nabla_x u + \textbf{I}$ into vector form and write $\textbf{F}_h = \textbf{G} \cdot \textbf{u} + (1,0,0,1)^T$ and, subsequently, $\textbf{P} = f_{\rm TTDN}(\textbf{F}_h, r)$. To compute the residual, we also require a gradient matrix for $P$. Similar to the gradient matrix $\textbf{G}$ for $\textbf{u}$, the gradient matrix for $P$, denoted by $\textbf{G}_2$, can be obtained through the finite element method. Consequently, we define the loss tailored for this problem as follows:
\begin{equation}\label{res_cnn}
    \mathcal{E}(\textbf{u}, r) = \left\| \textbf{G}_2 \cdot f_{\rm TTDN}(\textbf{G} \cdot \textbf{u}, r) \right\|^2 + \left\| f_{\rm TTDN}(\textbf{G} \cdot \textbf{u}, r)|_{x \in \partial \Omega} - F \right\|^2. 
\end{equation}
Note that, for this example, $\mathcal{E}$ represents the residual of the weak formulation of the governing PDE. This formulation of the loss function is also compatible with CNN since it is a sparse connected linear layer by $\textbf{G}$, composed with an activation function $f_{\rm TTDN}$, followed by another sparse connected linear layer by $\textbf{G}_2$.

\begin{algorithm}[!b]
    \caption{Pre-training and training algorithm for the physics-informed CNN}
    \label{algo:PICNN}
    \begin{algorithmic}[1] 
        \Input
        \State $\Xi_{\rm sl}$  \Comment{the supervised training data}
        \State $\Xi_{\rm ul}$  \Comment{the unsupervised training data}
        \State $N_{\rm pt}$  \Comment{the number of epochs for pre-training}
        \State $N_{\rm st}$  \Comment{the number of epochs for semi-supervised training} 
        \State $\eta$  \Comment{learning rate schedule}
        \EndInput
         
        \For {$n = 1$ \texttt{:} $N_{\rm pt}$}  \Comment{pre-training}
            \For {each mini-batch $\Xi$ in $\Xi_{\rm sl}$}
                \State compute $\mathcal{L}_{\rm sl} (f_{\rm CNN}, \Xi)$
                \State apply Adam optimizer \citep{kingma_adam_2017} $\Delta \omega = {\rm Adam}(\frac{\partial \mathcal{L}_{\rm sl}}{\partial \omega})$  \Comment{$\omega$ is the trainable parameter of $f_{\rm CNN}$}
                \State update the parameters $\omega \gets \omega + \eta \Delta \omega$ 
            \EndFor
        \EndFor
        \For {$n = 1$ \texttt{:} $N_{\rm st}$}  \Comment{semi-supervised training}
            \For {each mini-batch $\Xi_{\rm s}$ in $\Xi_{\rm sl}$, $\Xi_{\rm u}$ in $\Xi_{\rm ul}$}
                \State compute $\mathcal{L}_{\rm PICNN} (f_{\rm CNN}, \Xi_{\rm s}, \Xi_{\rm u})$
                \State apply Adam optimizer $\Delta \omega = {\rm Adam}(\frac{\partial \mathcal{L}_{\rm PICNN}}{\partial \omega})$ 
                \State update the parameters $\omega \gets \omega + \eta \Delta \omega$ 
            \EndFor
        \EndFor
        
        \Output
        \State $f_{\rm CNN}$  \Comment{the trained CNN based on $\omega$}
        \EndOutput
    \end{algorithmic}
\end{algorithm}

\noindent \textbf{Physics-informed strategy:} The loss function for \eqref{vw_cnn} or \eqref{res_cnn}, with the training data $\Xi_{\rm ul} = \{\boldsymbol{\mu}^i\}_{i=1}^{N_{\rm ul}}$, is given by 
\begin{equation}\label{loss_ul}
    \mathcal{L}_{\rm ul}(f_{\rm CNN}, \Xi_{\rm ul}) = \sum_{\boldsymbol{\mu}^i \in \Xi_{\rm ul}} \mathcal{E}(f_{\rm CNN}(\boldsymbol{\mu}^i), \boldsymbol{\mu}^i). 
\end{equation}
This corresponds to unsupervised learning since it involves only the input feature $\boldsymbol{\mu}$ and does not require the labeled data $\textbf{u}_{\rm PDE}$. To generate the unsupervised learning training data $\Xi_{\rm ul}$, we can make use of the given prior sample $S$ and also generate samples via the unconditional generative diffusion model introduced in Section \ref{subsec:GDMprior}. Different from the supervised learning training data $\Xi_{\rm sl}$, generating $\Xi_{\rm ul}$ only requires us to pay the cost of sampling, rather than the cost of solving the PDE using the expensive finite element solver. Hence, we can generate much more training data in $\Xi_{\rm ul}$ to improve the generalization ability of the CNN. 

The proposed physics-informed CNN makes use of the two loss functions defined in \eqref{loss_sl} and \eqref{loss_ul}, thereby combining supervised learning and unsupervised learning, leading to a semi-supervised learning method. The loss function is then given by
\begin{equation}
\begin{split}
    \mathcal{L}_{\rm PICNN} (f_{\rm CNN}, \Xi_{\rm sl}, \Xi_{\rm ul}) & = \mathcal{L}_{\rm sl}(f_{\rm CNN}, \Xi_{\rm sl}) + \mathcal{L}_{\rm ul}(f_{\rm CNN}, \Xi_{\rm ul}) \\
    & = \sum_{(\boldsymbol{\mu}^i, \textbf{u}^i_{\rm PDE}) \in \Xi_{\rm sl}} \left\| f_{\rm CNN}(\boldsymbol{\mu}^i) - \textbf{u}^i_{\rm PDE}\right\|^2 + \sum_{\boldsymbol{\mu}^i \in \Xi_{\rm ul}} \mathcal{E}(f_{\rm CNN}(\boldsymbol{\mu}^i), \boldsymbol{\mu}^i). 
\end{split}
\end{equation}

The physics-informed CNN is trained by a pre-training and training algorithm, shown in \cref{algo:PICNN}.

\section{Results and discussion}\label{sec:Res}

The description of the numerical experiments—a hyper-elastic mechanics problem and a multi-scale mechanics problem—is presented in Section \ref{subsec:CNN}. Our implementation is carried out using the Python programming language. The entire code is executed within Ubuntu 22.04 LTS, hosted on a Windows Subsystem for Linux (WSL2) within a Windows 11 platform. The platform consists of an AMD Ryzen 9 7950X 16-Core CPU with \SI{64}{\giga\byte} of RAM and an NVIDIA GeForce RTX 4080 GPU with \SI{16}{\giga\byte} of VRAM. The conventional finite element solver is based on the open-source package FEniCS \citep{alnaes_fenics_2015}, and the artificial neural network related codes are based on PyTorch \citep{paszke_pytorch_2019} using the CUDA accelerator. 

The Jupyter Notebooks of the implementations can be found in \href{https://github.com/Yankun-Hong/Diffusion-model-for-inverse-problems}{the Github page}.

\subsection{Hyper-elastic mechanics problem}\label{subsec:hemp}

We now discuss the results of the CNN surrogate model for the forward model, $\mathcal{H} \circ \mathcal{A}$, in the hyper-elastic mechanics problem. In this experiment, we implement the CNN regression with and without the physics-informed strategy, as discussed in \cref{subsec:CNN,subsec:PICNN}. Due to the sparsity of CNN, our surrogate network has only \num{63304} trainable network parameters, making the prediction efficient, i.e., we need less than \SI{1}{\ms} for a mini-batch with a size of \num{64} and less than \SI{2}{\ms} for both prediction and back-propagation for the derivative. 

\begin{figure}[!b]
\centering
    \captionsetup{width=0.95\linewidth}
    \includegraphics[width=0.4\linewidth]{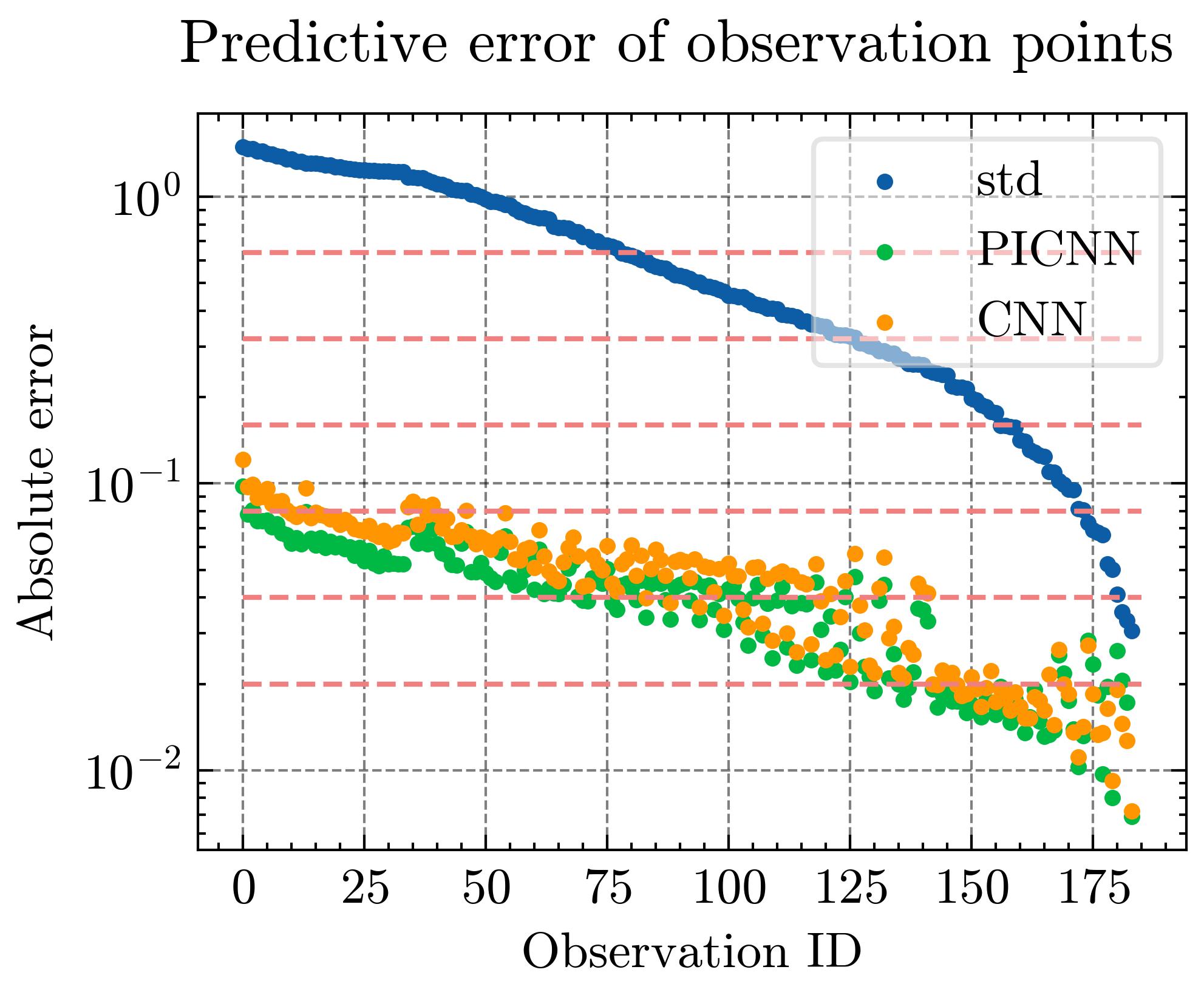}
    \caption{The absolute errors of the CNN and physics-informed CNN surrogates (orange and green, respectively), and the standard deviation within the prior sample $S_0$ (blue) at each boundary observation point. The observation points along the $x$-axis are sorted in decreasing order of the standard deviation. The red lines represent different observation error settings. }
    \label{fig:FWestd}
\end{figure}

The accuracy of this surrogate model is presented in \cref{fig:FWestd}, measured at each boundary observation point; the observation identification numbers are shown in the $x$-axis. On average, compared to standard (i.e., non-physics-based) CNN surrogates, the physics-informed strategy reduces the prediction error by a factor of $20\%$. The blue points in \cref{fig:FWestd} represent the standard deviation of boundary observations obtained using the given prior samples. Intuitively, the ratio of this standard deviation to the summation of the predictive error and the observation noise represents a signal-to-noise ratio of the forward model. The larger this signal-to-noise ratio, the more accurate we can expect the inverse estimation to be. In this work, considering different observation noise levels, shown as red lines in \cref{fig:FWestd}, and the standard deviation, both the (standard) CNN and physics-informed CNN surrogates have satisfactory accuracy for the inverse estimation.

\begin{remark}
    Note that there is a difference in the meaning of the signal-to-noise ratio here to that of the signal-to-noise ratio hyper-parameter $r$ appearing in Algorithms \ref{algo:GDM_prior} and \ref{algo:GDM_post}. The latter is only a tunable hyper-parameter for the performance of the diffusion model, whereas the former, a commonly used notion in inverse problems, represents the ratio of the information to the noise in the observations. 
\end{remark}

\begin{figure}[!t]
\centering
    \begin{subfigure}[t]{0.24\textwidth}
    \centering
    \captionsetup{width=0.99\linewidth}
    \includegraphics[width=0.9\linewidth]{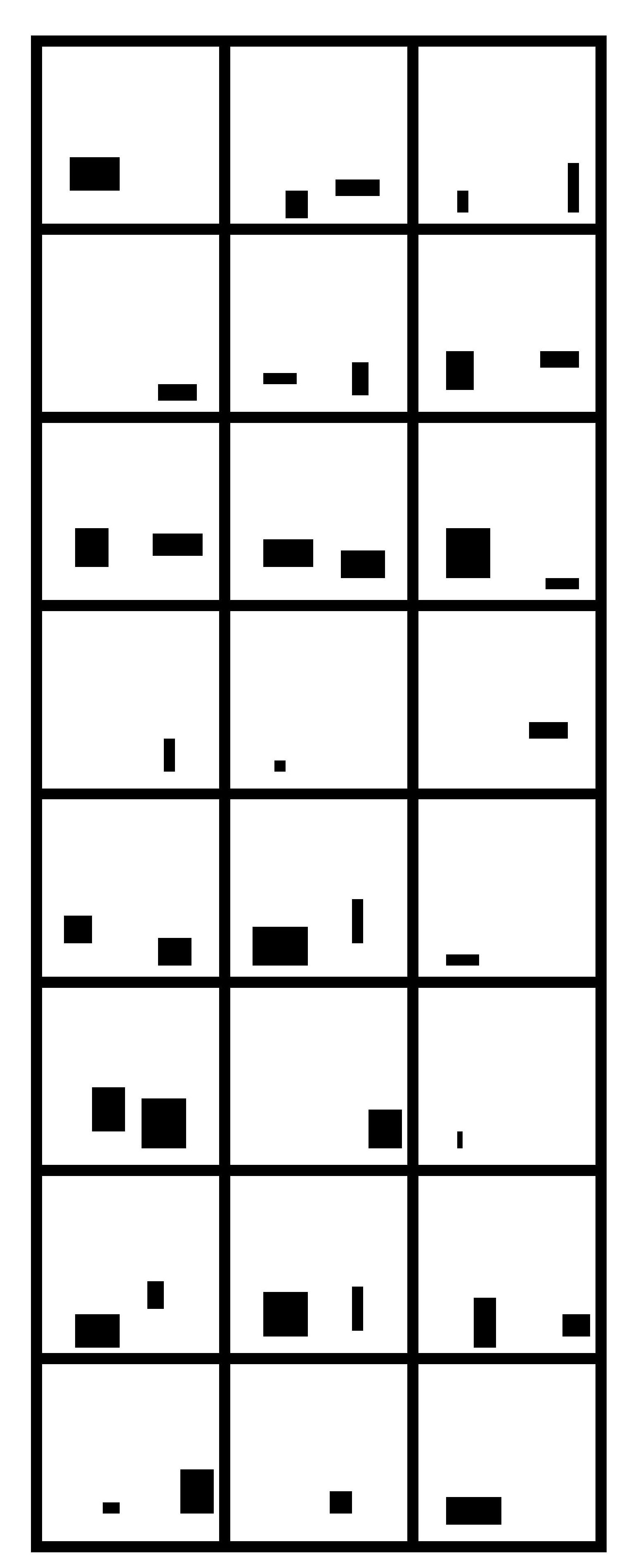}
    \caption{Given samples $S_0$; }
    \label{fig:refpri}
    \end{subfigure}
    \hspace{0.005\textwidth}
    \begin{subfigure}[t]{0.36\textwidth}
    \centering
    \captionsetup{width=0.99\linewidth}
    \includegraphics[width=0.6\linewidth]{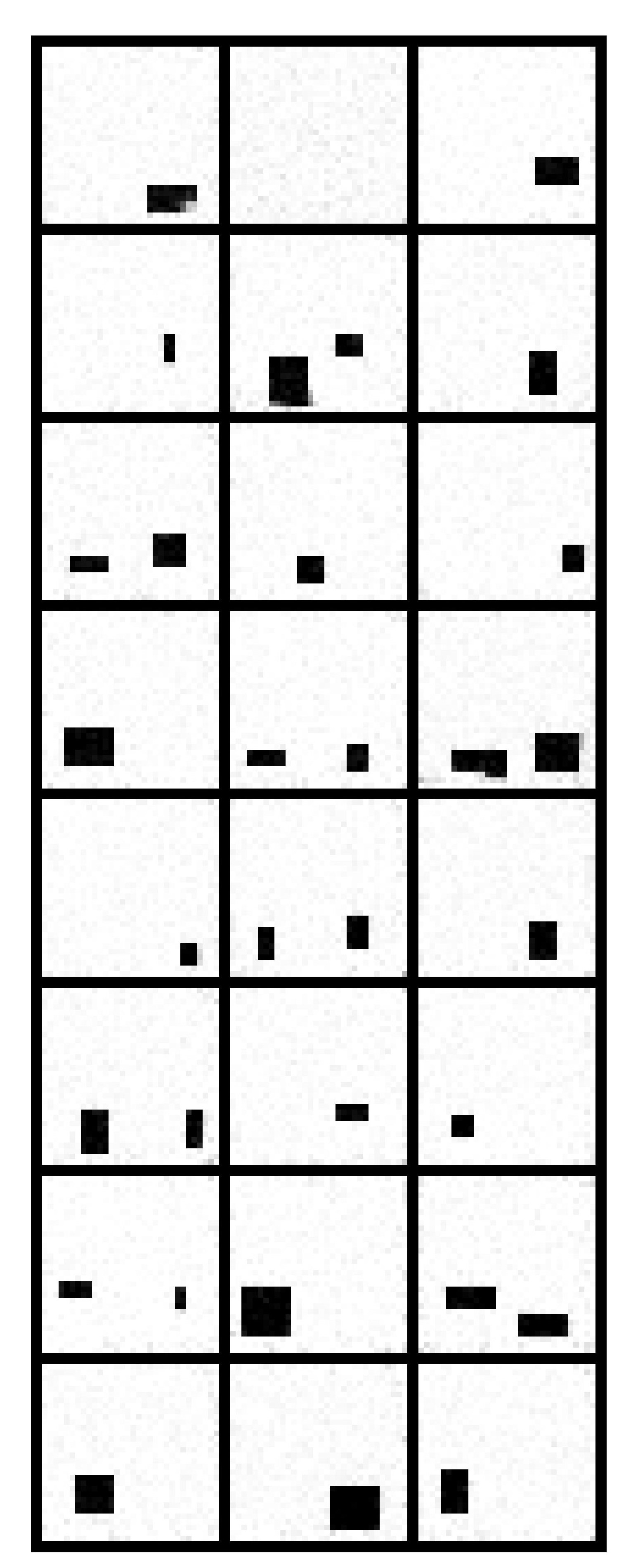}
    \caption{Samples generated using SMLD; }
    \label{fig:compprive}
    \end{subfigure}
    \hspace{0.005\textwidth}
    \begin{subfigure}[t]{0.36\textwidth}
    \centering
    \captionsetup{width=0.99\linewidth}
    \includegraphics[width=0.6\linewidth]{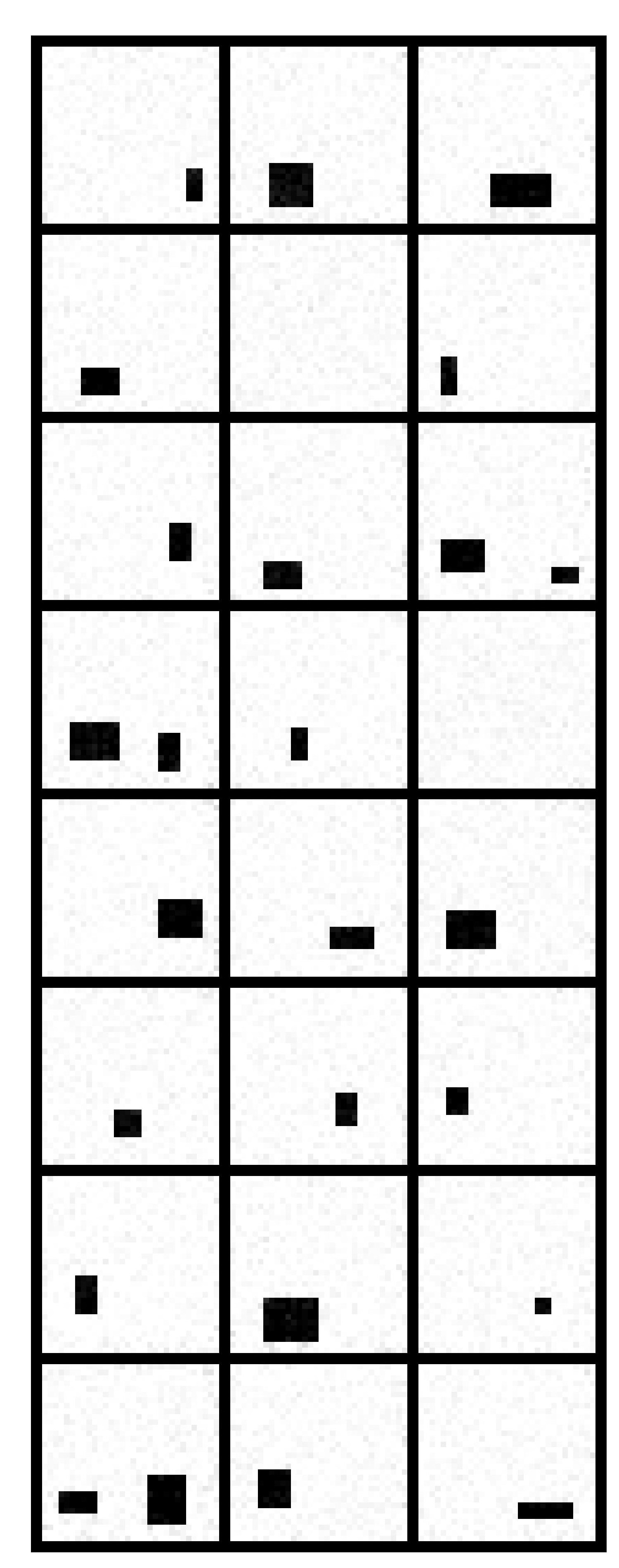}
    \caption{Samples generated using DDPM. }
    \label{fig:compprivp}
    \end{subfigure}
\caption{24 samples of prior}
\label{fig:pri}
\end{figure}

Next, we present the sampling quality of the prior samples via the two distinct diffusion model-based sampling methods, SMLD (using VE SDE) and DDPM (using VP SDE), using the unconditional sampling method presented in \cref{algo:GDM_prior}. In this numerical experiment, the true underlying prior distribution for the Young's modulus $E$ has $50\%$ probability to yield \num{2} blocks at the bottom of the square domain, $25\%$ probability to yield a single block at the left bottom, and the remaining $25\%$ to yield a block at the right bottom. Here, the block represents an area that has a different Young's modulus from the rest of the physical domain. The size and position of the blocks follow an independent uniform distribution. 
The given prior sample $S_0$ is drawn from this distribution. \Cref{fig:refpri} presents a sample from $S_0$, serving as a reference for the sampling techniques. \Cref{fig:compprive,fig:compprivp} illustrate the samples generated via SMLD, proposed in \citep{song_generative_2020}, and DDPM, proposed in \citep{ho_denoising_2020}, respectively. Clearly, both the score-based generative diffusion models via different SDEs can capture the shape of the blocks and generate a sharp interface in the sample. However, both of these methods underestimate the number of blocks in the sample—the sample with \num{2} blocks occurs with less than \num{50} percent chance, and there are even samples that have no blocks inside the physical domain. Despite this effect, we will see that it does not affect the quality of the results of the inverse estimation.

\begin{figure}[!t]
\centering
    \captionsetup{width=0.95\linewidth}
    \includegraphics[width=0.8\linewidth]{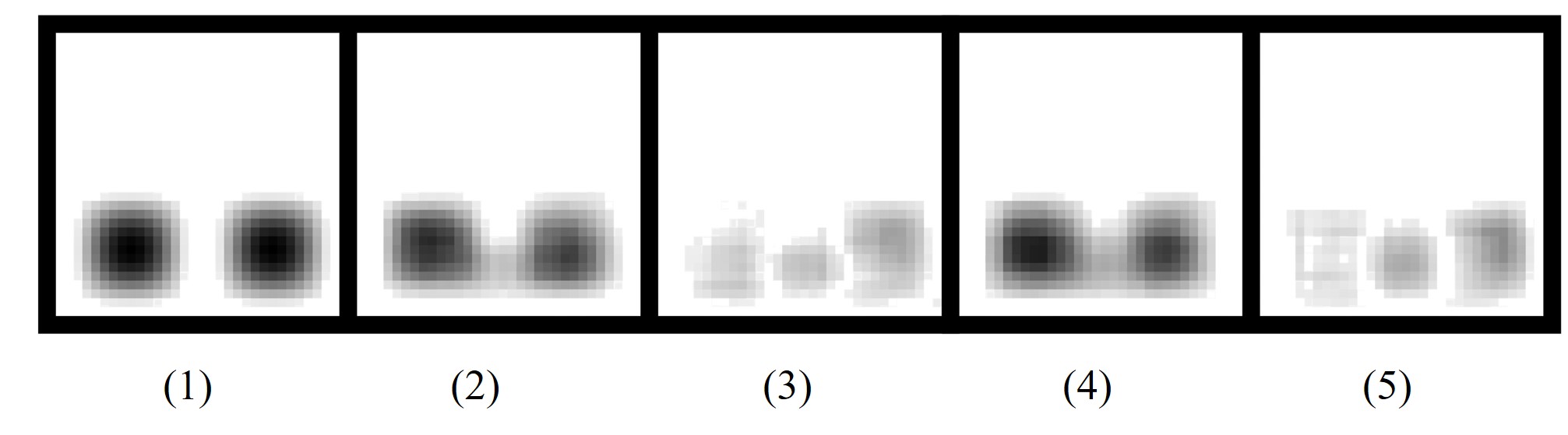}
    \caption{(1) the \enquote{true} mean of the given prior $S$; (2) the generated mean via SMLD; (3) the difference between (1) and (2); (4) the generated mean via DDPM; (5) the difference between (1) and (4). }
    \label{fig:primean}
\end{figure}

\begin{figure}[hbt!]
\centering
    \captionsetup{width=0.95\linewidth}
    \includegraphics[width=0.8\linewidth]{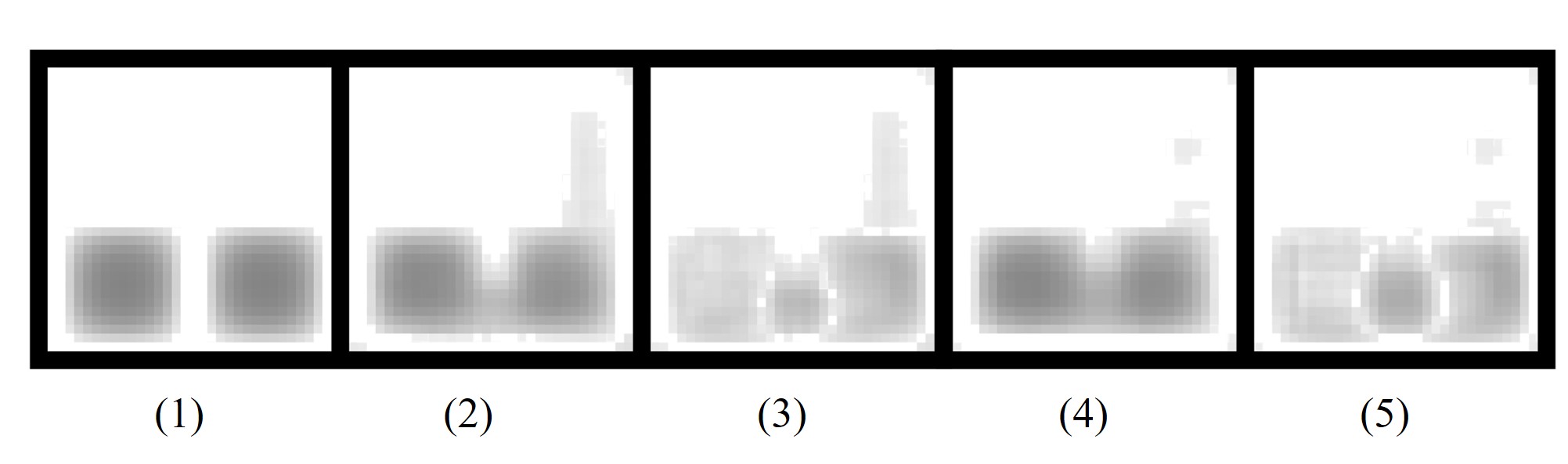}
    \caption{(1) the \enquote{true} standard deviation of the given prior $S$; (2) the generated standard deviation via SMLD; (3) the difference between (1) and (2); (4) the generated standard deviation via DDPM; (5) the difference between (1) and (5). }
    \label{fig:pricov}
\end{figure}

\Cref{fig:primean,fig:pricov} present the mean and standard deviation obtained from the prior. It is important to note that the terms \enquote{true} mean and \enquote{true} standard deviation refer to statistically estimated quantities derived from the given sample $S$, rather than analytical values. When considering these statistical estimates, it becomes evident that a simplistic Gaussian prior is an oversimplification, as it fails to capture the intricate geometrical features. For instance, relying solely on the Gaussian prior with the mean and standard deviation shown in these figures results in the loss of information regarding the square-shaped block inside the domain and the presence of at most 2 blocks. The mean and standard deviation obtained through SMLD and DDPM exhibit shapes similar to the reference true ones, thereby demonstrating that the diffusion model is able to learn the prior from the given sample. 

\begin{figure}[!b]
\centering
    \begin{subfigure}[t]{0.48\textwidth}
    \centering
    \captionsetup{width=0.95\linewidth}
    \includegraphics[width=0.7\linewidth]{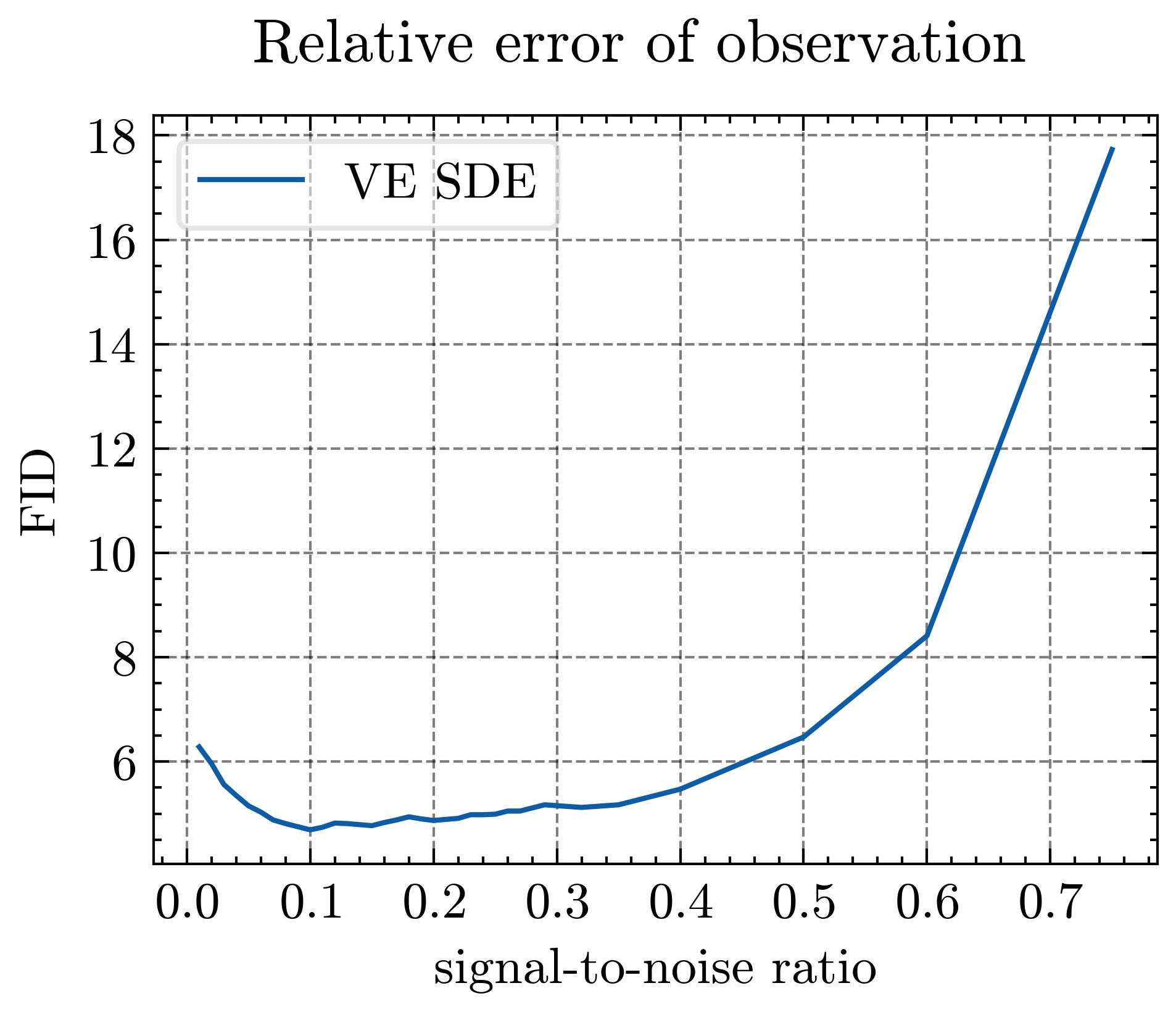}
    \caption{The FID using SMLD; }
    \label{fig:snrFIDve}
    \end{subfigure}
    \hspace{0.02\textwidth}
    \begin{subfigure}[t]{0.48\textwidth}
    \centering
    \captionsetup{width=0.95\linewidth}
    \includegraphics[width=0.7\linewidth]{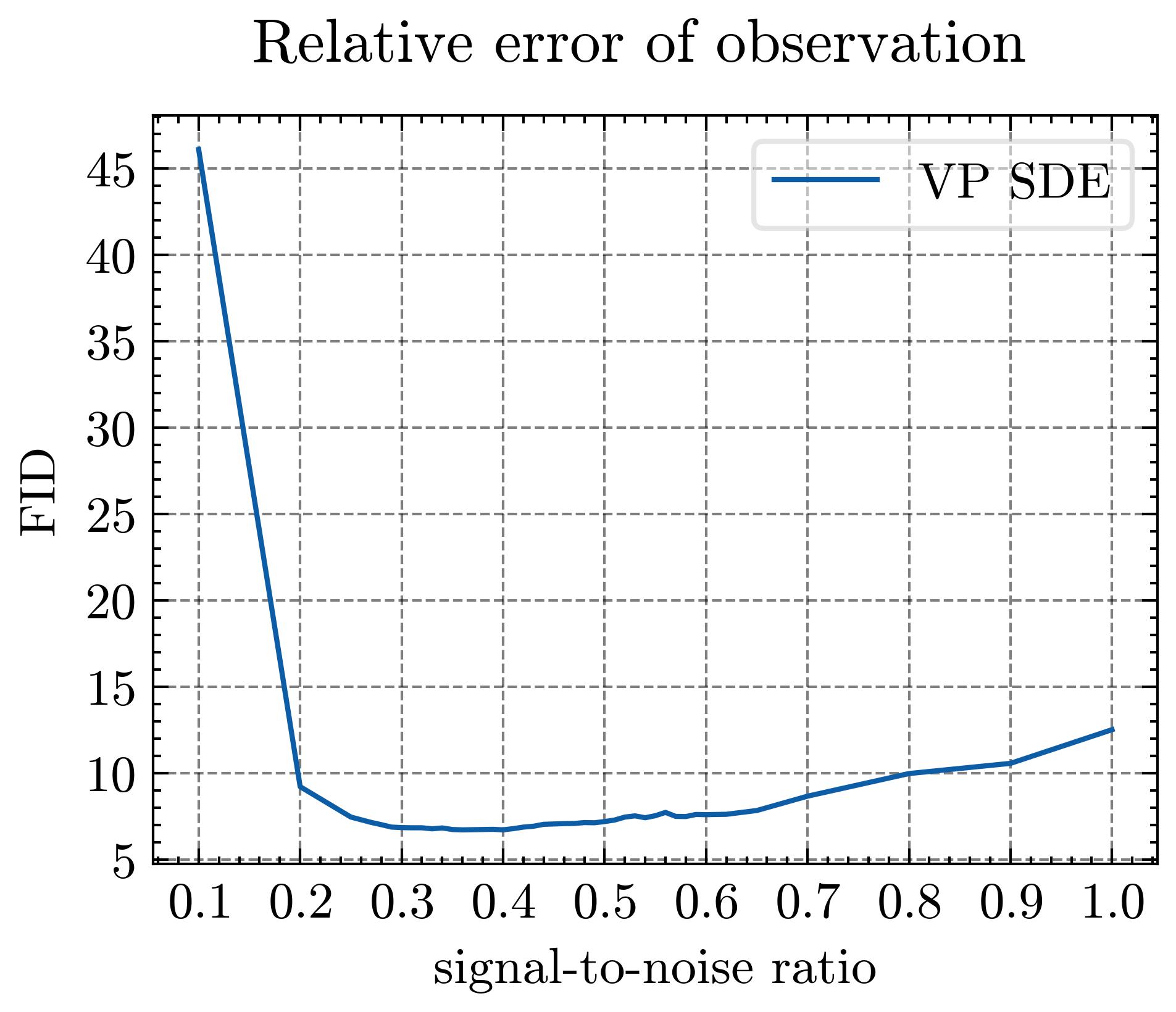}
    \caption{The FID using DDPM.}
    \label{fig:snrFIDvp}
    \end{subfigure}
\caption{The FID of the prior sampling with respect to the signal-to-noise ratio $r$.}
\label{fig:snrFID}
\end{figure}

In computer vision, the similarity between two sets of samples, denoted as $A$ and $B$, is typically quantified using the Fréchet Inception Distance (FID) \citep{heusel_gans_2017}, defined as follows:
\begin{equation*}
    d^2((\textbf{m}_A, \textbf{C}_A), (\textbf{m}_B, \textbf{C}_B)) = \|\textbf{m}_A - \textbf{m}_B\|^2 + {\rm Tr}(\textbf{C}_A + \textbf{C}_B - 2(\textbf{C}_A \cdot \textbf{C}_B)^{\frac{1}{2}}).
\end{equation*}
Here, $\textbf{m}_A$ and $\textbf{C}_A$ represent the mean and covariance matrix of the set of samples $A$, while $\textbf{m}_B$ and $\textbf{C}_B$ denote the mean and covariance matrix of $B$. Essentially, FID measures the Fréchet distance between two Gaussian distributions $\mathcal{N}(\textbf{m}_A, \textbf{C}_A)$ and $\mathcal{N}(\textbf{m}_B, \textbf{C}_B)$. A lower FID indicates a higher similarity between the underlying distributions of the two samples. \Cref{fig:snrFID} illustrate the FID curves between the given prior sample set and the generated sample set using the SMLD and DDPM methods. The represented FID curves are functions of the signal-to-noise ratio $r$, which is a tunable hyperparameter in Line 13 of \cref{algo:GDM_prior}. In our numerical experiments, including the generation of \Cref{fig:compprive,fig:compprivp,fig:primean,fig:pricov}, we set the signal-to-noise ratio to $r = 0.1$ for SMLD and $r = 0.36$ for DDPM, which correspond to the optimal points that achieve the best sampling quality, resulting in FID values of $4.69$ and $6.71$ for SMLD and DDPM, respectively; see \cref{fig:snrFID}. These FID levels are considered satisfactory in the field of computer vision and we use this measure for image-like parameters in the scope of the inverse problem at hand.

Finally, we shift our attention to the results of the inverse estimation, specifically the outcomes of conditional sampling. For comparison, we not only employ SMLD and DDPM for posterior sampling shown in \cref{algo:GDM_post}, but also implement the Ensemble Kalman Inversion (EnKI) method \citep{iglesias_ensemble_2013, schillings_analysis_2017}, which is an inverse solver inspired by the ensemble Kalman filter. EnKI, essentially a Newton-like method, utilizes a Monte Carlo-estimated covariance matrix to approximate the Hessian matrix, which is essential in a standard Newton's method. It mimics the temporal dynamics of the ensemble Kalman filter through iteration and applies the theory and methods of the ensemble Kalman filter for dynamic systems to find the maximum likelihood point. Tikhonov regularization is also introduced in EnKI, along with other improvement techniques to enhance convergence speed and stability \citep{chada_convergence_2021, chada_tikhonov_2020}. It should be noted that EnKI requires the size of the ensemble, i.e., the size of the sample for Monte Carlo estimation, to be larger than the dimension of the quantity to be estimated. To this end, in this work, given a grid resolution of $32 \times 32$, the ensemble size is set to be $4096$, which is greater than $1024$. For more details on EnKI, please refer to relevant works such as \citep{chada_convergence_2021, chada_tikhonov_2020, iglesias_ensemble_2013, schillings_analysis_2017}.

\begin{figure}[bthp]
\begin{subfigure}[t]{0.48\textwidth}
    \centering
    \captionsetup{width=0.95\linewidth}
    \includegraphics[width=0.6\linewidth]{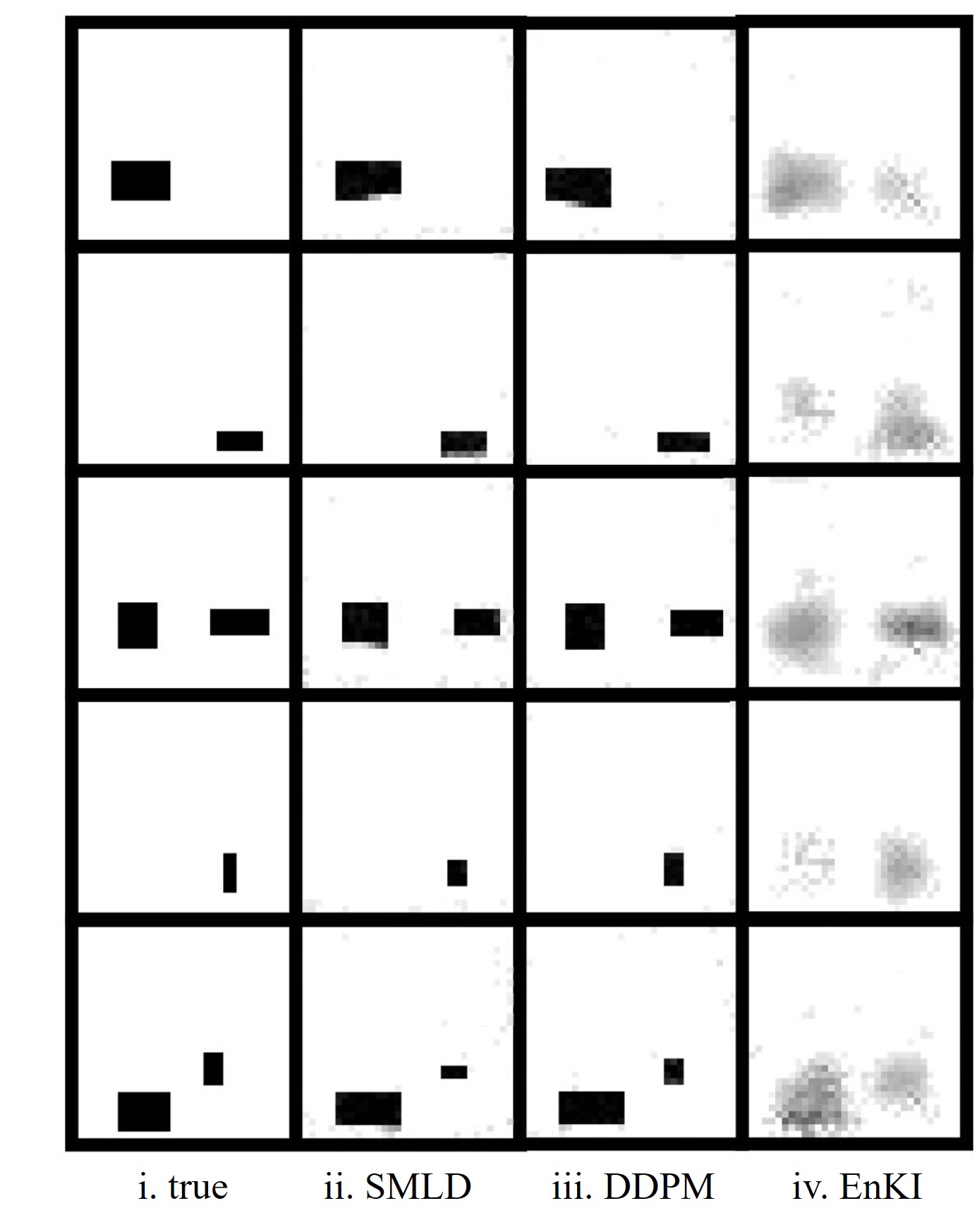}
    \caption{Noise level 1}
    \label{fig:postnl1}
    \vspace*{3mm}    
\end{subfigure}
\hspace{0.02\textwidth}
\begin{subfigure}[t]{0.48\textwidth}
    \centering
    \captionsetup{width=0.95\linewidth}
    \includegraphics[width=0.6\linewidth]{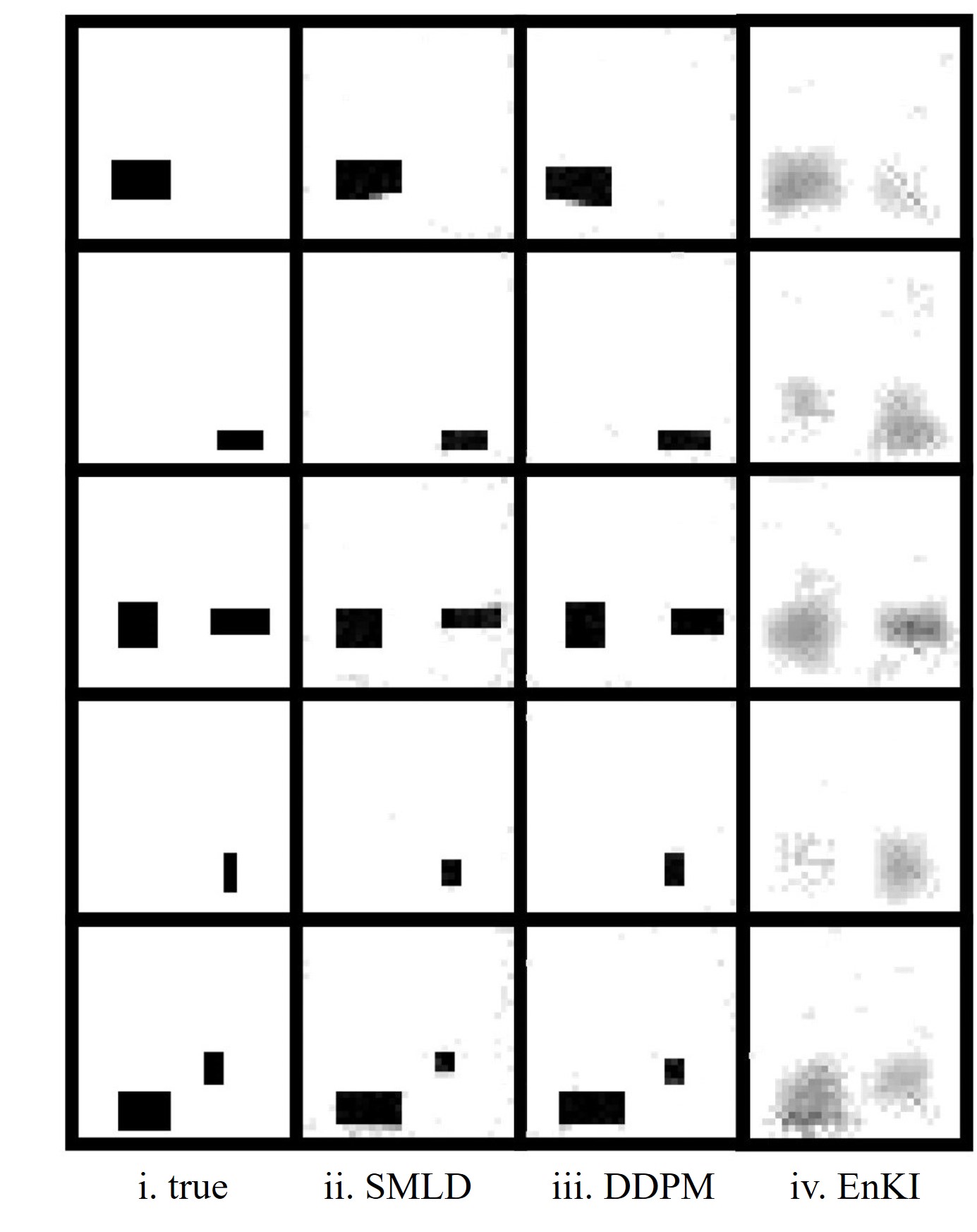}
    \caption{Noise level 2}
    \label{fig:postnl2}
    \vspace*{3mm}    
\end{subfigure}

\begin{subfigure}[t]{0.48\textwidth}
    \centering
    \captionsetup{width=0.95\linewidth}
    \includegraphics[width=0.6\linewidth]{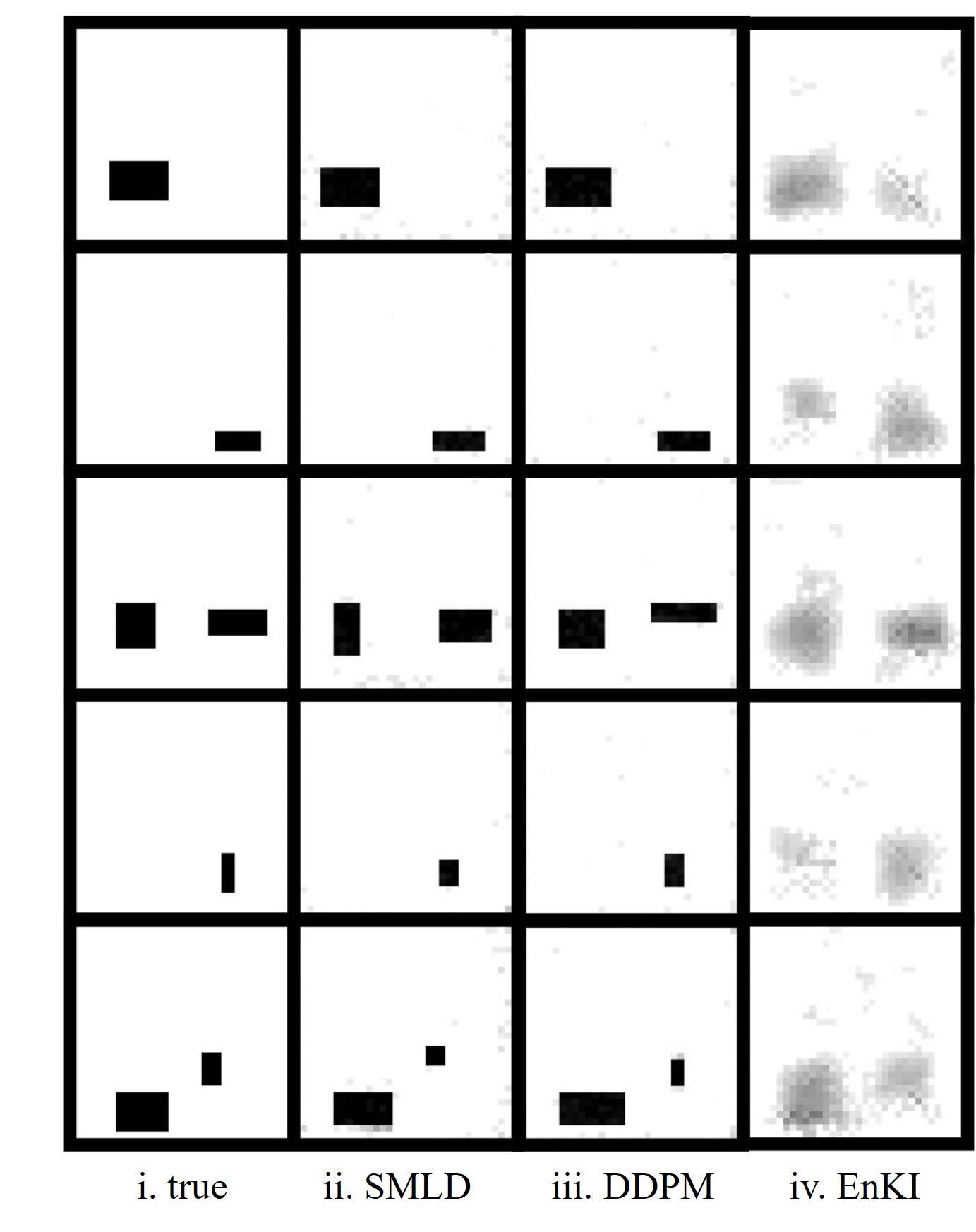}
    \caption{Noise level 3}
    \label{fig:postnl3}
    \vspace*{3mm}    
\end{subfigure}
\hspace{0.02\textwidth}
\begin{subfigure}[t]{0.48\textwidth}
    \centering
    \captionsetup{width=0.95\linewidth}
    \includegraphics[width=0.6\linewidth]{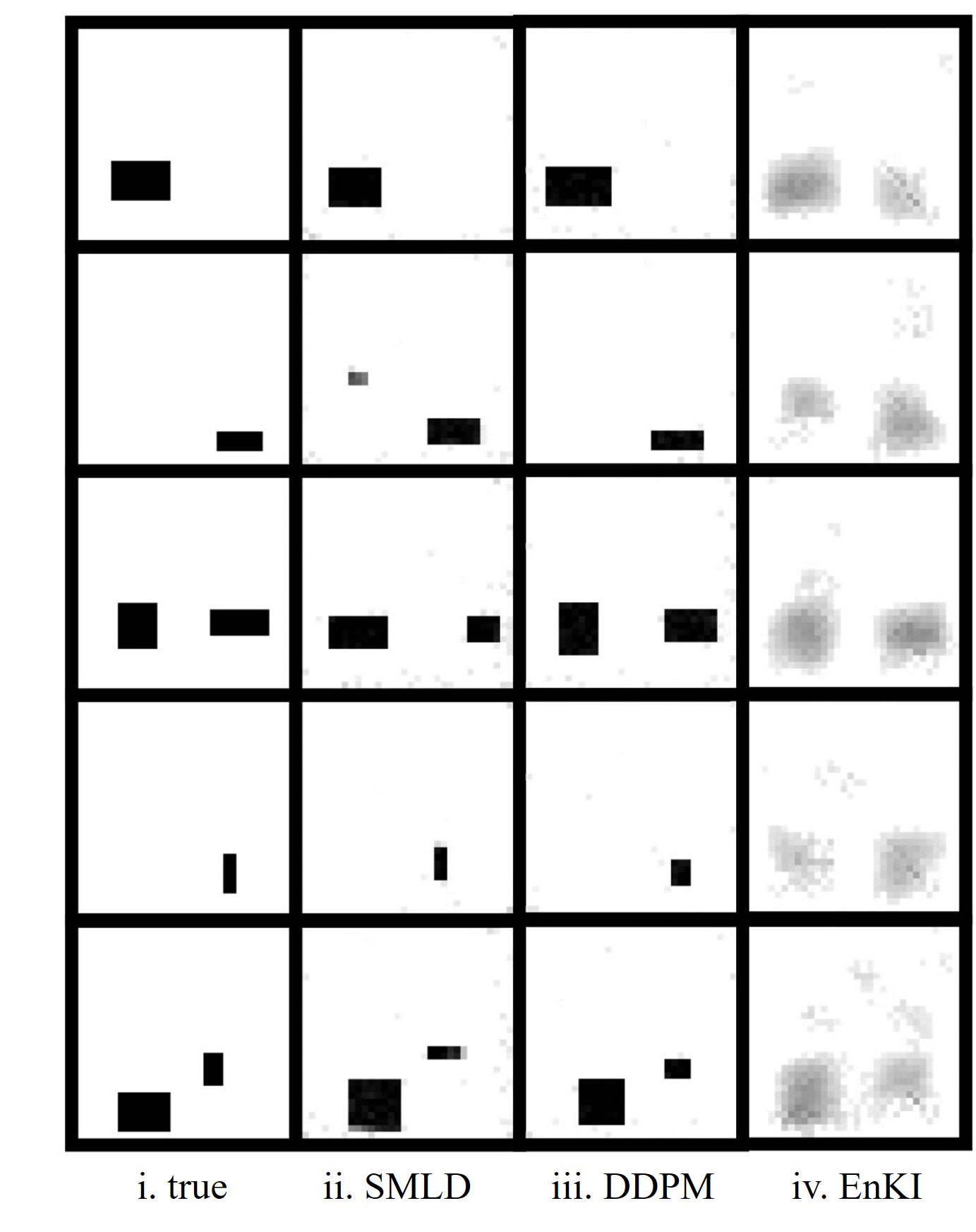}
    \caption{Noise level 4}
    \label{fig:postnl4}
    \vspace*{3mm}    
\end{subfigure}

\begin{subfigure}[t]{0.48\textwidth}
    \centering
    \captionsetup{width=0.95\linewidth}
    \includegraphics[width=0.6\linewidth]{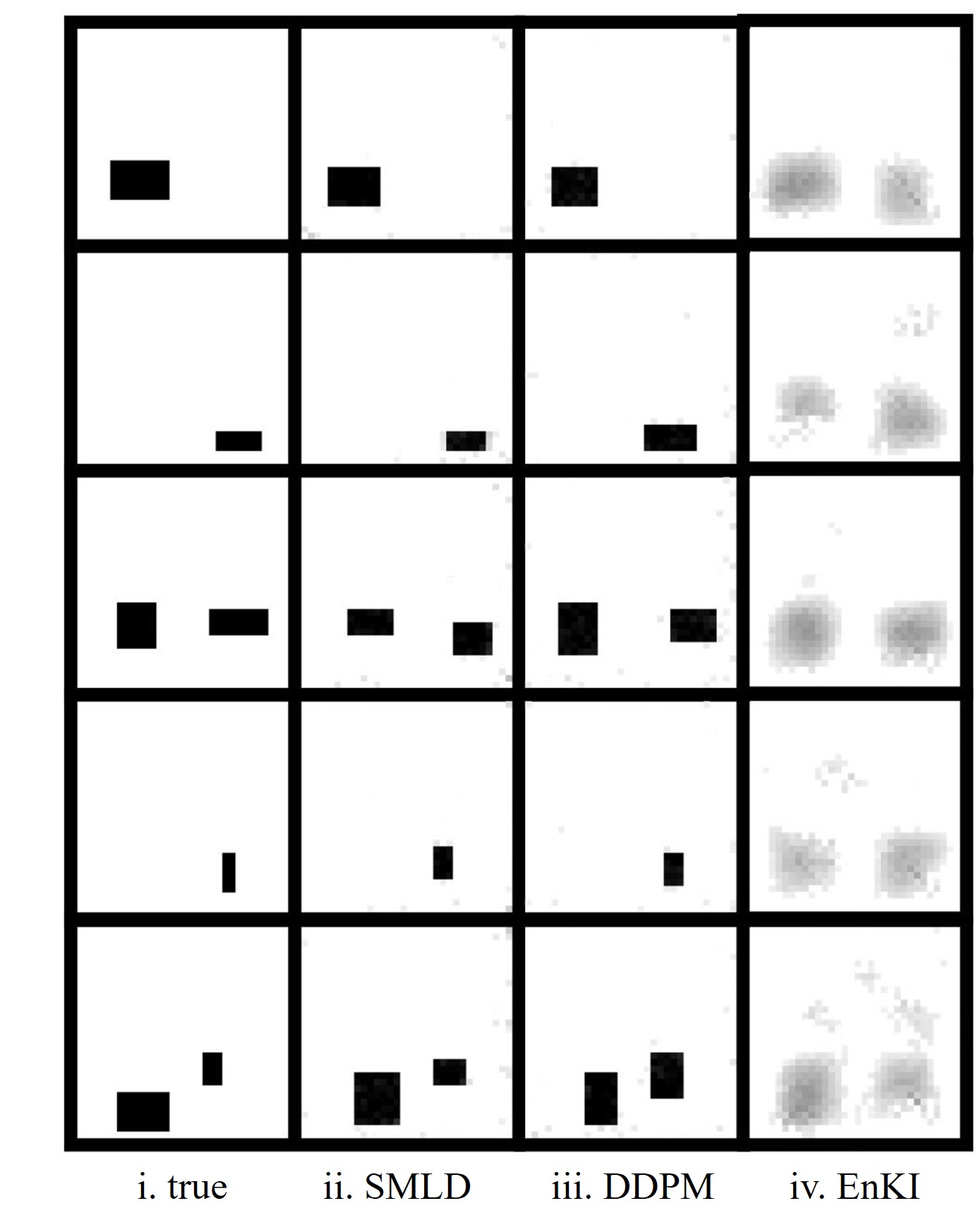}
    \caption{Noise level 5}
    \label{fig:postnl5}
\end{subfigure}
\hspace{0.02\textwidth}
\begin{subfigure}[t]{0.48\textwidth}
    \centering
    \captionsetup{width=0.95\linewidth}
    \includegraphics[width=0.6\linewidth]{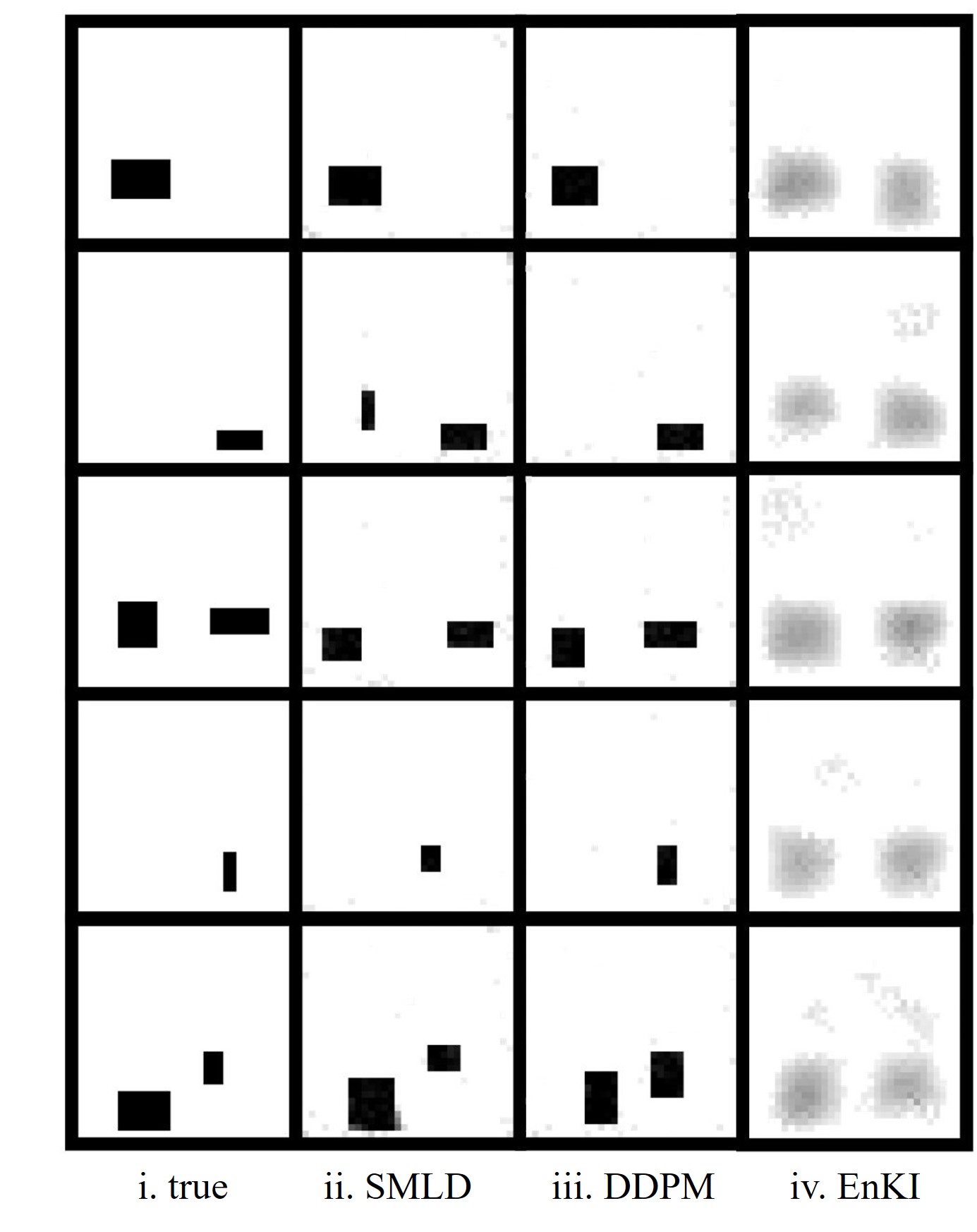}
    \caption{Noise level 6}
    \label{fig:postnl6}
\end{subfigure}
\caption{MLAPS points of the hyper-elastic inverse problems for each (increasing) observation noise level. For each subfigure, the i.~column presents the true solutions, while the ii., iii., and iv.~columns are the solutions via SMLD, DDPM and EnKI, respectively. The rows represent \num{5} different settings of parameters to be estimated. }
\label{fig:postsamp}
\end{figure}

We conduct the inverse estimation for various observational noise levels, with the Gaussian noise standard deviation $\sigma_\varepsilon$ illustrated by grey lines in \cref{fig:FWestd}, to assess the robustness of the algorithm with respect to the observation noise. From level 1 to level 6, the noise $\sigma_\varepsilon$ with increased strength is imposed to pollute the simulated true observation and then obtain the observation data for the inverse estimation. Considering the error due to the CNN surrogate, the uncertainty of the model's observation $y$ is set to be the square of the sum of the observation noise variance and the CNN's square error. For each noise level, we randomly select \num{5} out-of-sample parameter settings to be estimated and implement the inverse estimation using SMLD, DDPM, and EnKI. The results are presented in \cref{fig:postsamp}.

Recall that the inverse solution via the score-based generative diffusion model, i.e., SMLD and DDPM, is a posterior sample rather than a point in the parameter space. The mean of the posterior sample does not serve as a suitable inverse solution because averaging would compromise the geometry obtained at each sample point, thereby leading to the smoothening of the block interface. For instance, the given prior sample $S_0$ has a sharp interface between the inside block and the matrix for all its sample points (see \cref{fig:refpri}), but its mean, illustrated in \cref{fig:primean}, is as smooth as a quadratic function, and thus fails to preserve the geometrical features. In this work, as shown in the second and third columns of each subfigure in \cref{fig:postsamp}, we consider the maximum likelihood point amongst the posterior samples (MLAPS point) as the solution. \Cref{app:results} illustrates more detailed results for this problem, including the definition of the MLAPS point and the uncertainty of the inverse results. 

As shown in \cref{fig:postsamp}, when the observation noise is small, both SMLD and DDPM yield relatively good inverse results in the sense that they successfully capture the underlying geometrical features—up to 2 square blocks at the bottom of the domain. Moreover, the accuracy of the size and shape of the blocks is satisfactory, considering that only noisy boundary measurements are available. As the noise level increases, we experience a slight loss of accuracy. However, the algorithms still preserve the geometrical features obtained by the posterior sampling. The ability to capture these geometrical features for the diffusion model seems relatively independent of the noise level. 

However, EnKI fails to provide satisfactory inverse results, consistent with our expectations. This is attributed to its use of a Gaussian distribution as the prior, which oversimplifies the given prior sample $S$ by discarding crucial information about the underlying geometrical features. Essentially, the inverse problem is highly ill-posed, as the measurement is only \num{184}-dimensional while the parameter to be estimated has \num{1024} dimensions. In such cases, a high-quality prior is crucial for a successful inverse estimation. Unfortunately, a Gaussian distribution falls short in this application. In contrast, the score-based diffusion model excels in learning a high-quality prior from the given sample and utilizing it in posterior sampling, highlighting its advantage in handling complex, high-dimensional inverse problems.

It is worth noting that the standard conditional sampling in \eqref{score_post_fin} with a constant step size $\rho = \frac{1}{\sigma_\varepsilon^2}$ for posterior sampling resulted in divergence in our experiments for both SMLD and DDPM. The reason of this effect is discussed in \cref{subsubsec:TVSS}. By employing the time-varying step size in \eqref{step_size} for both SMLD and DDPM, and also \eqref{step_size2} for DDPM, the iteration of these algorithms becomes stable, leading to the aforementioned results.

\subsection{Multi-scale mechanics problem}\label{subsec:msmp}

In the multi-scale problem, we utilize the same prior sample $S_0$ as in the hyper-elastic problem, yielding the same prior sampling performance as in \cref{subsec:hemp}. The only difference in the prior is that, in this experiment, the parameter to be estimated is the radius $r$ of the pore in the micro-structure and, hence, the block represents an area that has a different pore radius from the rest of the physical domain. 

\begin{figure}[!t]
\centering
    \captionsetup{width=0.95\linewidth}
    \includegraphics[width=0.4\linewidth]{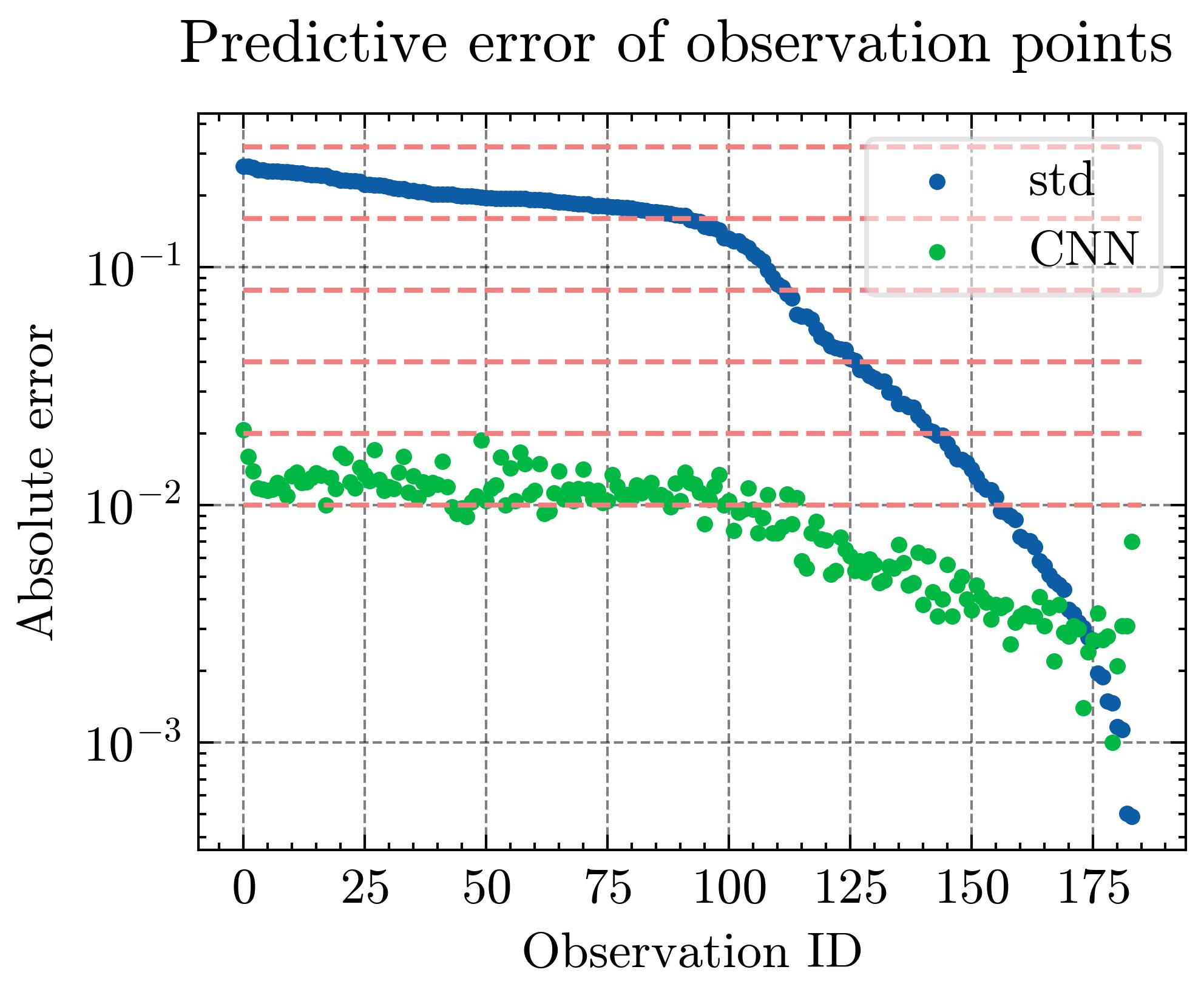}
    \caption{The absolute error of the CNN surrogate (green), and the standard deviation within the prior sample $S_0$ (blue) at each boundary observation point. The observation points along the $x$-axis are sorted in decreasing order of the standard deviation. The red lines represent different observation error settings. }
    \label{fig:msFWestd}
\end{figure}

The surrogate model, as detailed in \cref{subsec:CNN}, is also implemented with an identical CNN architecture to the one used for the hyper-elastic mechanics problem in \cref{subsec:hemp}. As a consequence, the number of trainable parameters and the predictive time remain the same as in the previous subsection. \Cref{fig:msFWestd} displays the observation standard deviation for the prior sample $S_0$ and the predictive error of the CNN without using the physics-informed strategy. The results illustrate that the (standard) CNN already exhibits satisfactory predictive accuracy compared to the standard deviation, with a significantly lower absolute predictive error than the hyper-elastic surrogate model. Given the challenge of further reducing the predictive error from an already low level, we chose not to implement the physics-informed strategy for this problem. In fact, the relative error of this surrogate model is at the level of \num{e-3} and, hence, the accuracy of this neural network regression already reaches $99.9\%$. 

\begin{figure}[bthp]
\begin{subfigure}[t]{0.48\textwidth}
    \centering
    \captionsetup{width=0.95\linewidth}
    \includegraphics[width=0.6\linewidth]{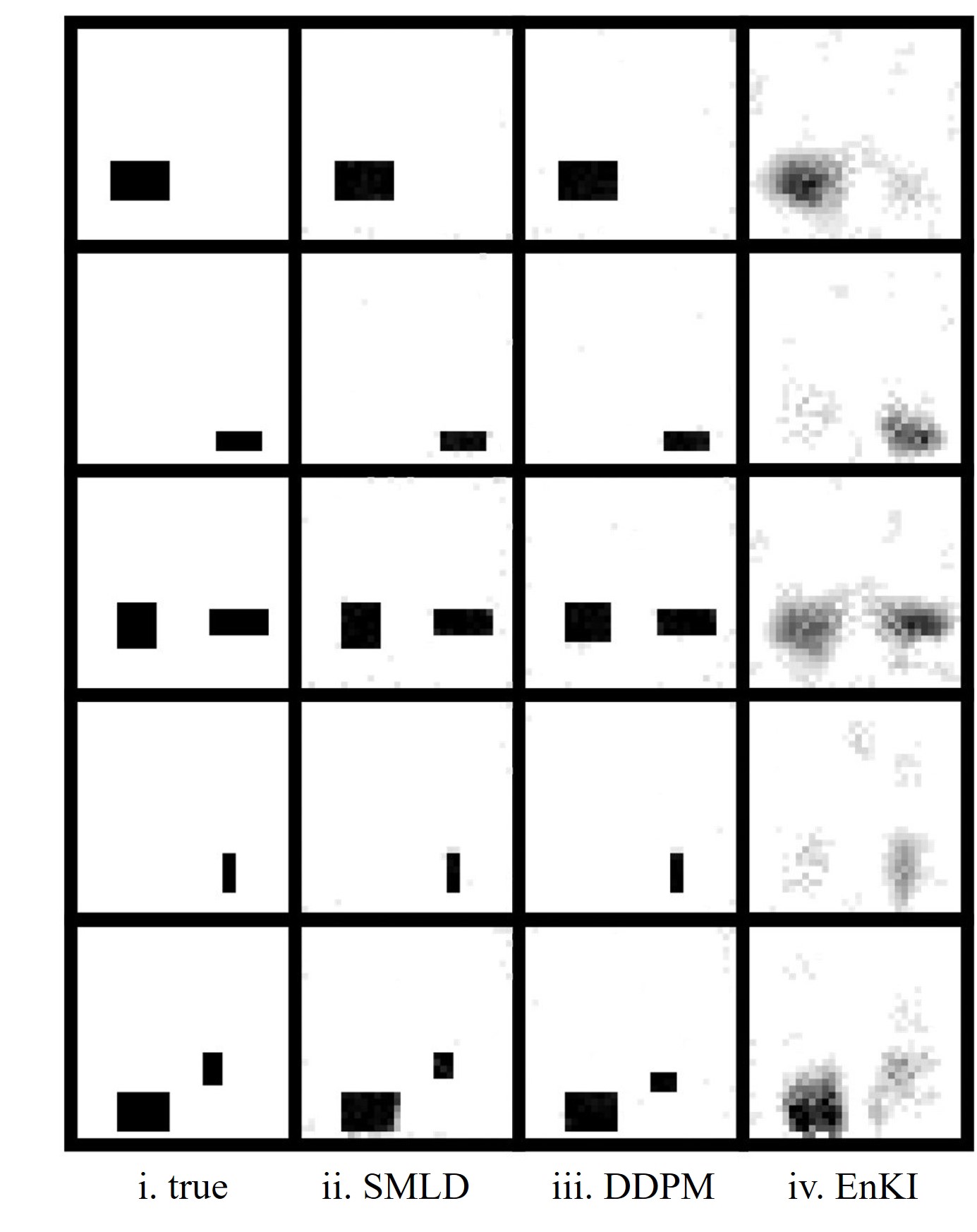}
    \caption{Noise level 1}
    \label{fig:ms_postnl1}
    \vspace*{3mm}    
\end{subfigure}
\hspace{0.02\textwidth}
\begin{subfigure}[t]{0.48\textwidth}
    \centering
    \captionsetup{width=0.95\linewidth}
    \includegraphics[width=0.6\linewidth]{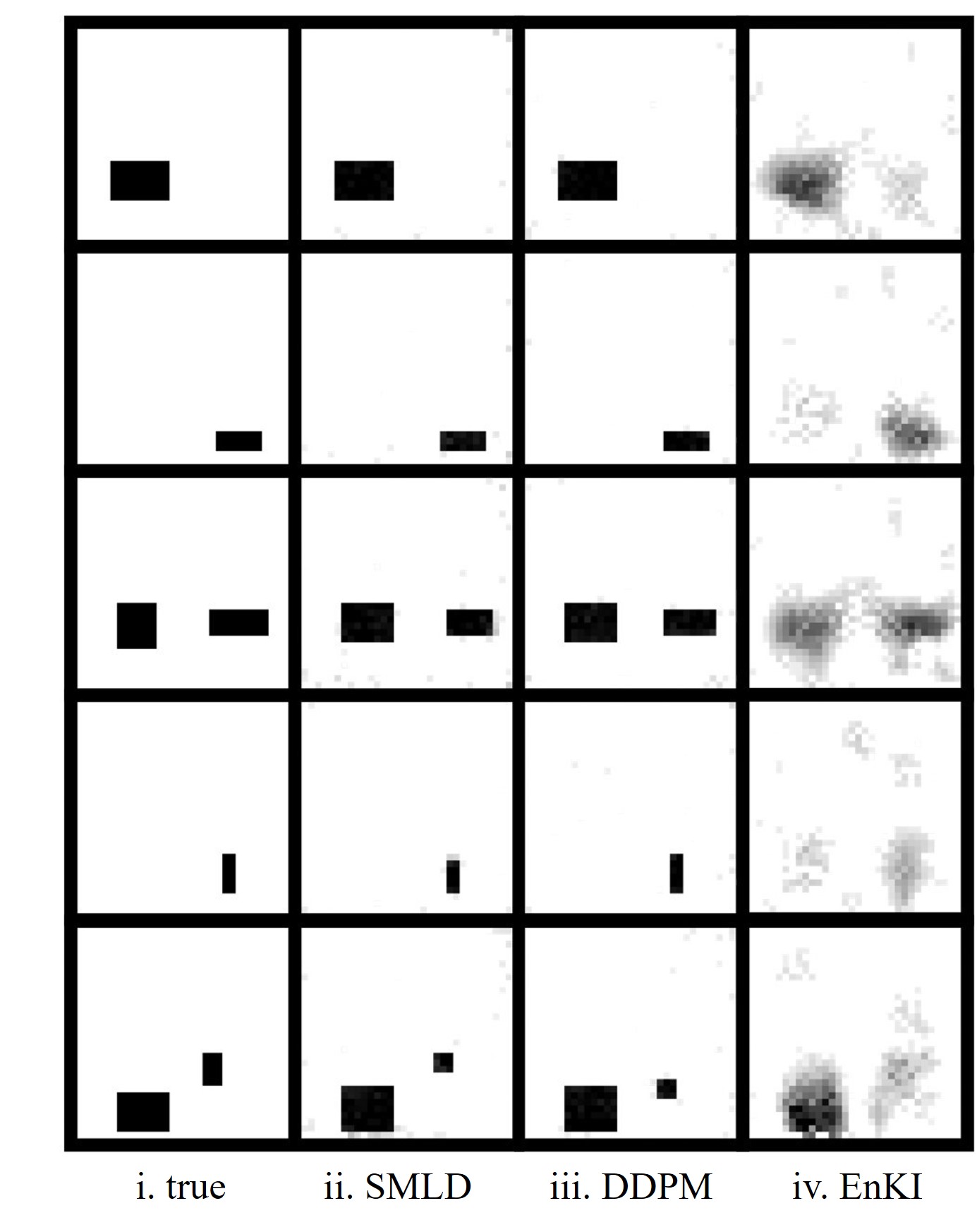}
    \caption{Noise level 2}
    \label{fig:ms_postnl2}
    \vspace*{3mm}    
\end{subfigure}

\begin{subfigure}[t]{0.48\textwidth}
    \centering
    \captionsetup{width=0.95\linewidth}
    \includegraphics[width=0.6\linewidth]{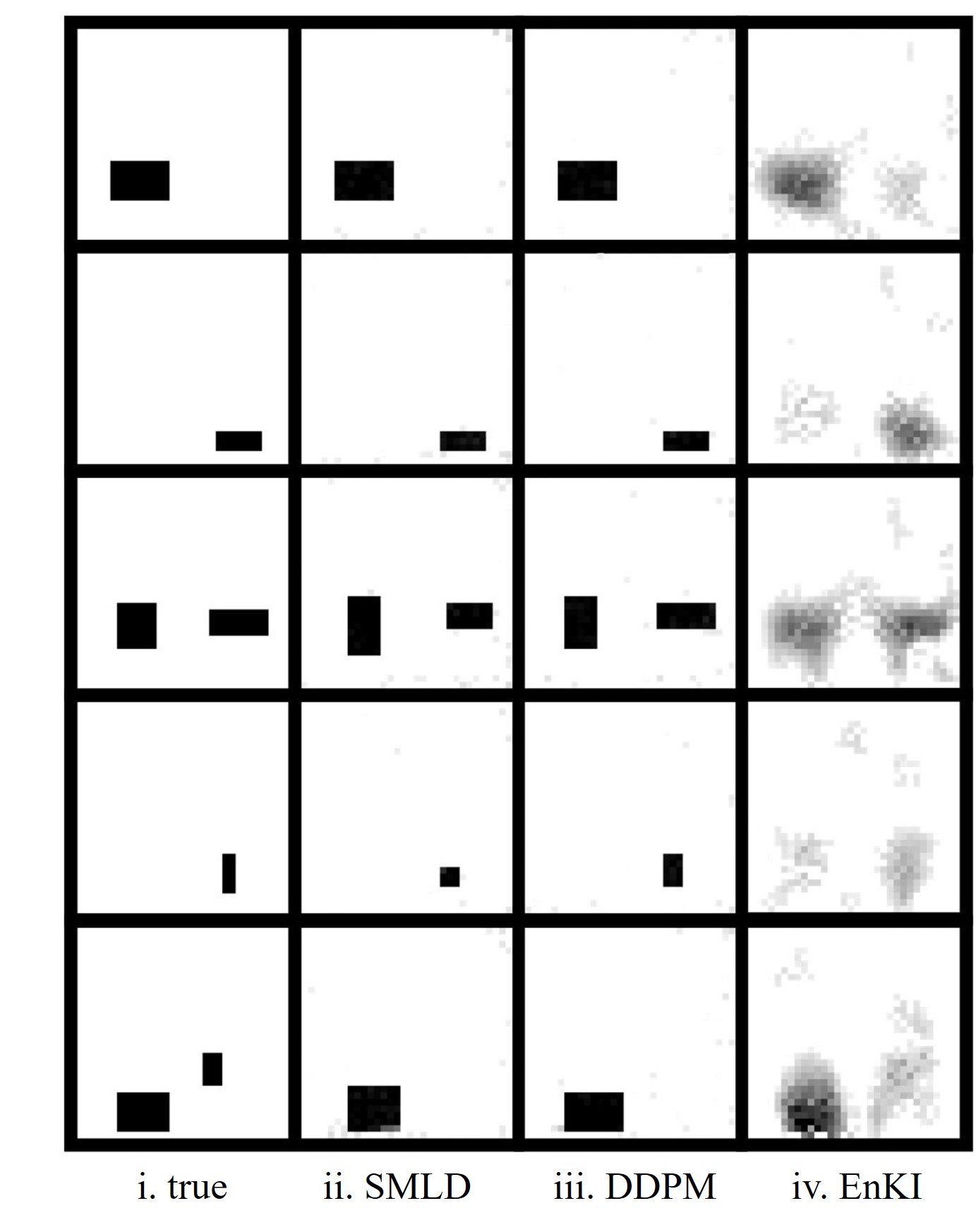}
    \caption{Noise level 3}
    \label{fig:ms_postnl3}
    \vspace*{3mm}    
\end{subfigure}
\hspace{0.02\textwidth}
\begin{subfigure}[t]{0.48\textwidth}
    \centering
    \captionsetup{width=0.95\linewidth}
    \includegraphics[width=0.6\linewidth]{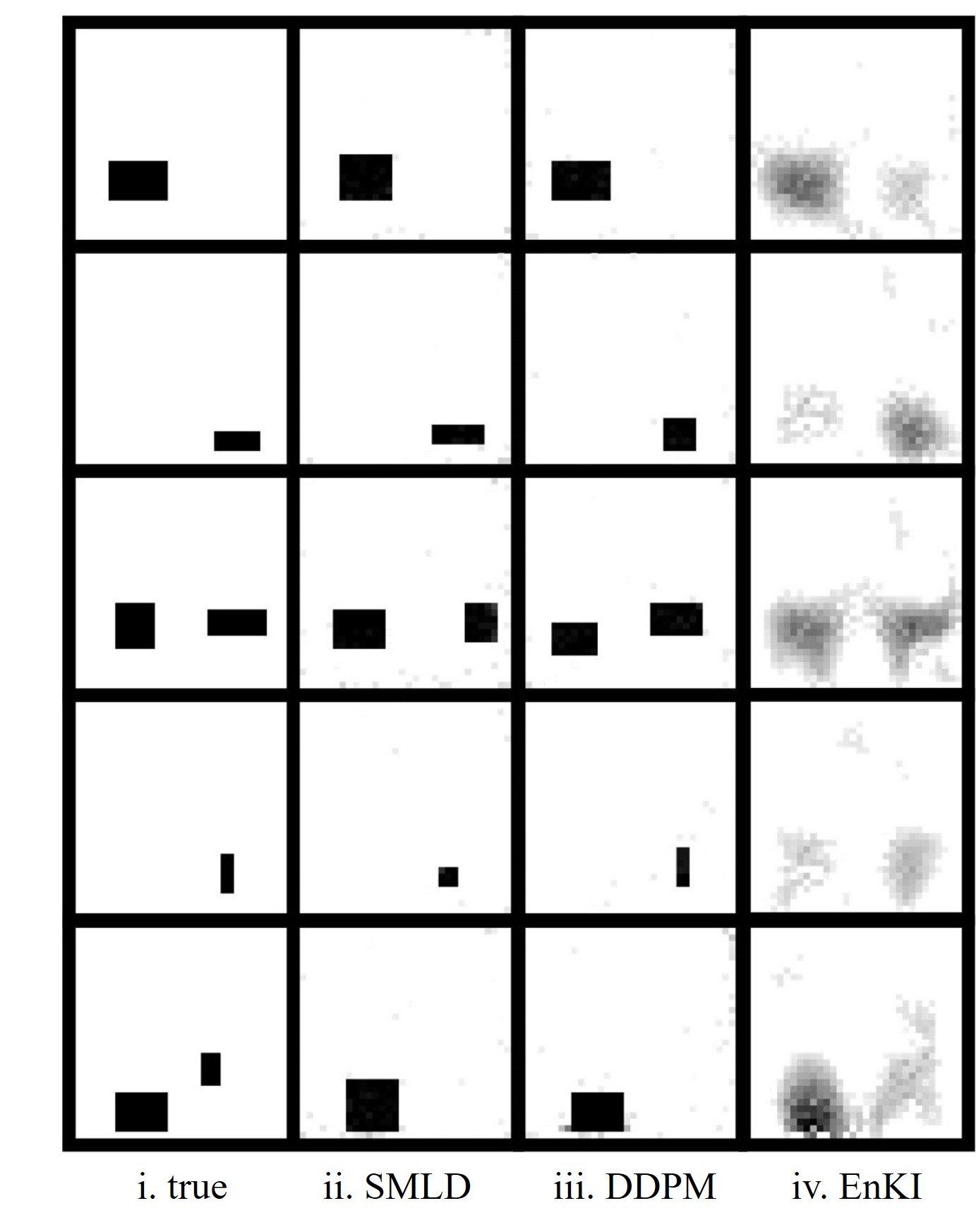}
    \caption{Noise level 4}
    \label{fig:ms_postnl4}
    \vspace*{3mm}    
\end{subfigure}

\begin{subfigure}[t]{0.48\textwidth}
    \centering
    \captionsetup{width=0.95\linewidth}
    \includegraphics[width=0.6\linewidth]{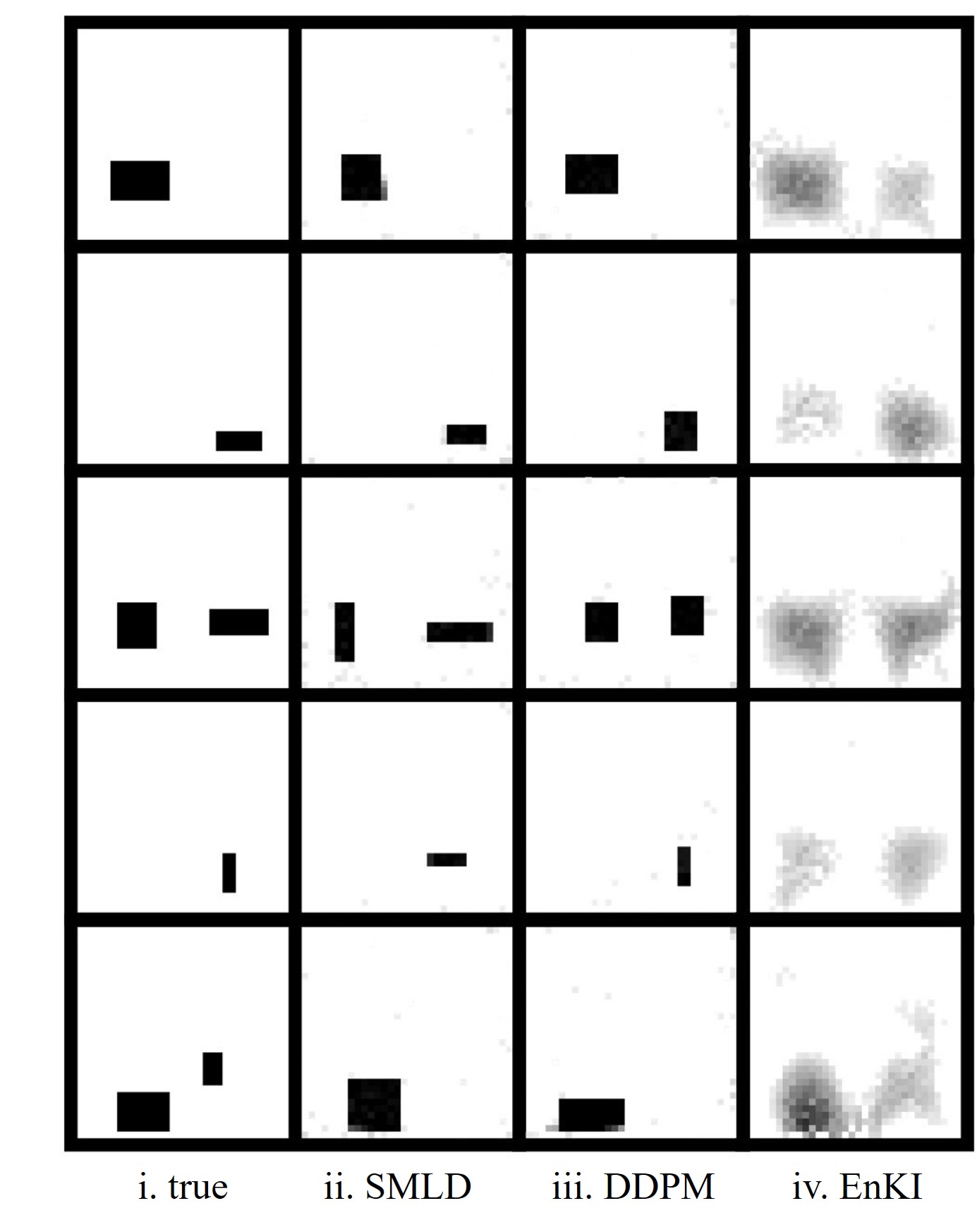}
    \caption{Noise level 5}
    \label{fig:ms_postnl5}
\end{subfigure}
\hspace{0.02\textwidth}
\begin{subfigure}[t]{0.48\textwidth}
    \centering
    \captionsetup{width=0.95\linewidth}
    \includegraphics[width=0.6\linewidth]{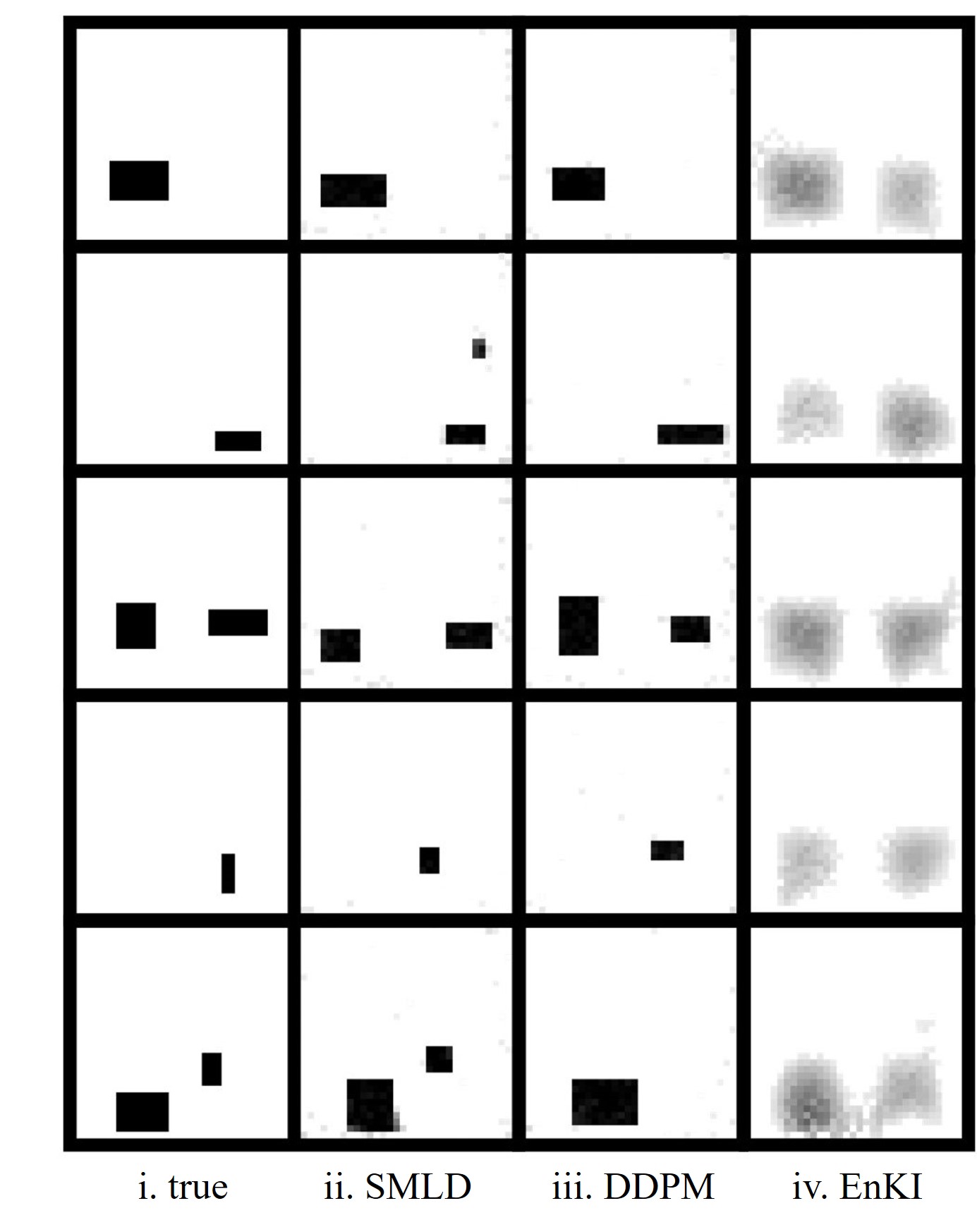}
    \caption{Noise level 6}
    \label{fig:ms_postnl6}
\end{subfigure}
\caption{MLAPS points of the multi-scale inverse problems for each (increasing) observation noise level. For each subfigure, the i.~column presents the true solutions, while the ii., iii., and iv.~columns are the solutions via SMLD, DDPM and EnKI, respectively. The rows represent \num{5} different settings of parameters to be estimated. }
\label{fig:ms_postsamp}
\end{figure}

For the inverse results via posterior sampling, we once again employ SMLD, DDPM, and EnKI for comparison. The observational noise levels are also depicted by the red lines in \cref{fig:msFWestd}. The uncertainty of the observation of the model is again set to be the square of the sum of the observation noise variance and the CNN's square error. The results, presented in a similar manner as in \cref{fig:postsamp}, are displayed in \cref{fig:ms_postsamp}. As illustrated in \cref{fig:ms_postsamp}, we observe outcomes similar to the hyper-elastic problem discussed in Section \ref{subsec:hemp}. The diffusion models, i.e., both SMLD and DDPM, successfully capture the geometrical features from the prior sample, providing better inverse estimation when the observation noise is relatively. However, the impact of increasing observation noise appears more significant for this multi-scale mechanics problem than for the single-scale hyper-elastic problem. If we scrutinize the noise levels depicted in \cref{fig:msFWestd}, the largest observation noise setting has a noise standard deviation even larger than the boundary observation standard deviation. This implies that the signal-to-noise ratio of this inverse problem is low, making the inverse estimation challenging. Despite this challenging setting, both SMLD and DDPM are capable of providing reasonably accurate inverse estimations, thereby showcasing the power of the score-based diffusion model. EnKI encounters similar challenges as observed in the hyper-elastic problem. The estimated results exhibit even poorer quality, given that the observation standard deviation, and consequently, the signal-to-noise ratio of the problem, is lower than that of the hyper-elastic problem. In fact, we see a similar outcome with EnKI-based inverse solver in the scope of the thermal multi-scale problem in \citep{abdulle_ensemble_2020}. \Cref{app:results} illustrates more detailed results for this problem.

\subsection{Computational costs}\label{subsec:compcost}

In terms of computational time for both the hyper-elastic problem and multi-scale problem, the entire offline training stage takes less than 4 hours in total. This includes: the running time of a single algorithm (SMLD or DDPM), including the time for training the score surrogate with the noising process, for training the CNN surrogate for the forward model, and for the generation of its training dataset, $\Xi_{\rm sl} = \{(\boldsymbol{\mu}^i, \textbf{u}^i_{\rm PDE})\}_{i=1}^{N_{\rm sl}}$. The primary computational bottleneck arises during the execution of the finite element solver for generating the training dataset, $\Xi_{\rm sl} = \{(\boldsymbol{\mu}^i, \textbf{u}^i_{\rm PDE})\}_{i=1}^{N_{\rm sl}}$, for the forward model since this step requires over 3 hours. The online time for inverse estimation with the diffusion model (i.e., \cref{algo:GDM_post} which involves generating 256 sample points through posterior sampling) takes around 2.5 minutes for SMLD and 3.5 minutes for DDPM. Here, the number of iteration steps for the noising-denoising process is set to be $N_T = 2000$ and each step involves $K = 1$ Langevin step. The running time for EnKI inverse estimation is approximately 4 minutes, configured with 4096 particles in the ensemble and 100 iteration steps.

\section{Conclusions}\label{sec:Concl}

In this study, we evaluate the efficacy of the score-based generative diffusion model as an inverse solver for PDE-constrained inverse problems, showcasing its adaptability from computer vision to PDE-constrained inverse problem scenarios. We summarize our findings below.

First, in highly ill-posed inverse problems, a high-quality prior is essential for obtaining accurate results. We propose the use of a diffusion model to exploit its ability to capture the underlying features and patterns of the given prior sample, ultimately enhancing estimated performance. Our numerical experiments demonstrate that, compared to conventional methods, e.g., EnKI method, the diffusion model learns and leverages complex, statistical priors, improving estimation quality, even for complex PDE-constrained forward models, such as multi-scale models. In addition, the score-based generative diffusion model exhibits good performance in handling high-dimensional problems, extending its applicability. Also, the method shows robustness under different levels of observation noise.

Second, in the denoising algorithm for posterior sampling, our proposed denoising stochastic process and the time-varying step schedule strategy proves crucial for addressing potential instabilities in the context of PDE-constrained inverse problems. This improvement mitigates the divergence issue for both SMLD and DDPM, thereby enhancing robustness for conditional sampling. 

Finally, the CNN architecture serves as an efficient surrogate for the forward model, offering high predictive speed and accuracy. The application of our proposed physics-informed strategy, to some extent, further enhances generalization accuracy. With this efficient surrogate forward model, inverse estimation via the score-based diffusion model, including SMLD and DDPM, is rapid. 

Looking ahead to future prospects in this field, it is important to acknowledge that, in this work, the numerical experiments are based on a square computational domain, common in computer vision but not necessarily realistic for PDE-constrained problems. Hence, there is a need to extend the method to irregular domains. In such cases, adjustments to the U-net architecture for the score function and the CNN for the forward surrogate may be necessary.

\appendix
\section{Additional numerical results}\label{app:results}

\begin{figure}[bthp]
\begin{subfigure}[t]{1\textwidth}
    \centering
    \captionsetup{width=0.95\linewidth}
    \includegraphics[width=0.6\linewidth]{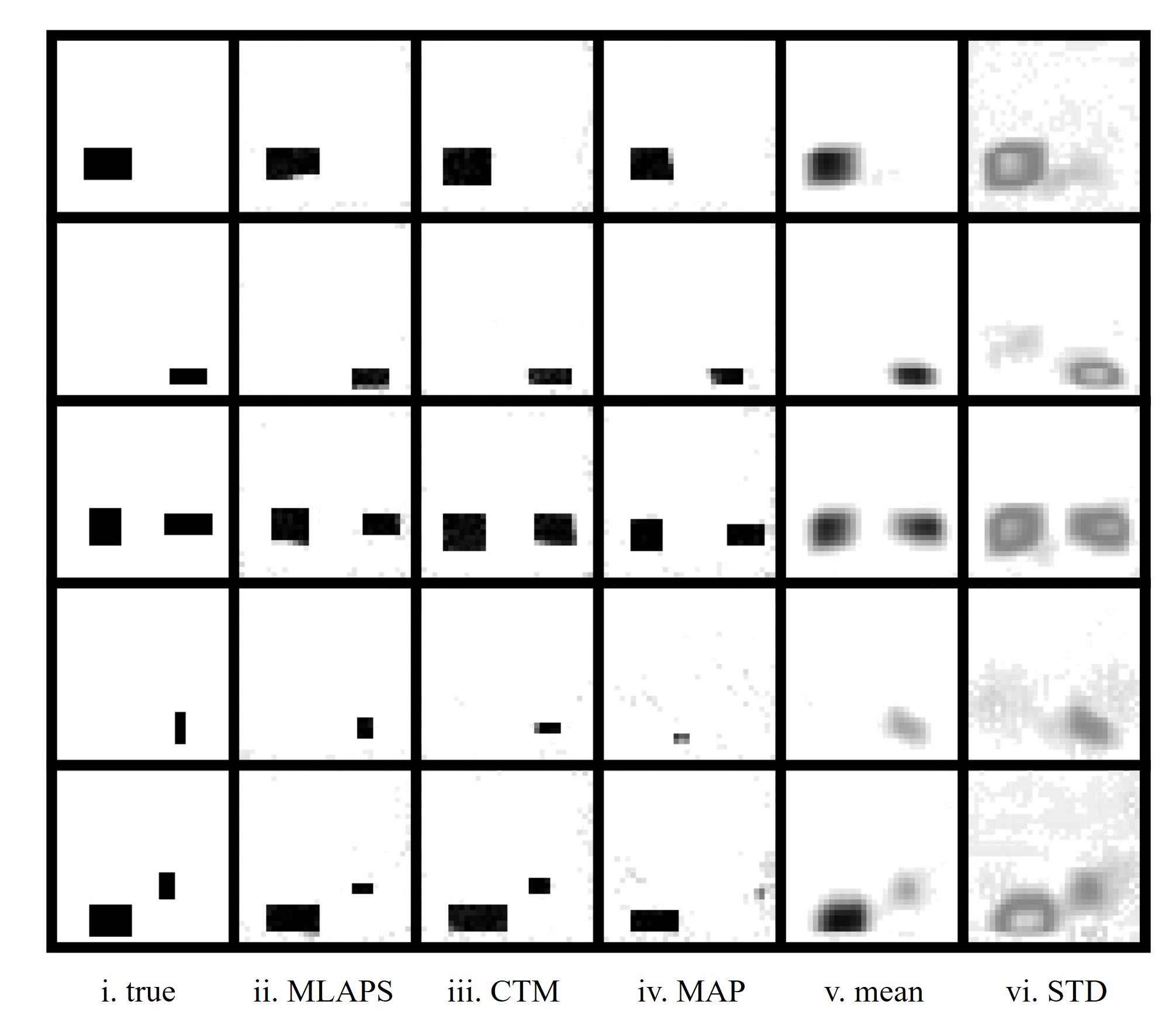}
    \caption{i.~the true solution, ii.~the MLAPS point, $\boldsymbol{\mu}_{\rm MLAPS}$, iii.~the CTM point, $\boldsymbol{\mu}_{\rm CTM}$, iv.~the MAP point, $\boldsymbol{\mu}_{\rm MAP}$, v.~the mean of the posterior samples, $\boldsymbol{\mu}_{\rm mean}$, and vi.~the STD of the posterior samples. The rows represent \num{5} different settings of parameters to be estimated. }
    \label{fig:otpostve}
    \vspace*{3mm}    
\end{subfigure}

\begin{subfigure}[t]{1\textwidth}
    \centering
    \captionsetup{width=0.95\linewidth}
    \includegraphics[width=0.8\linewidth]{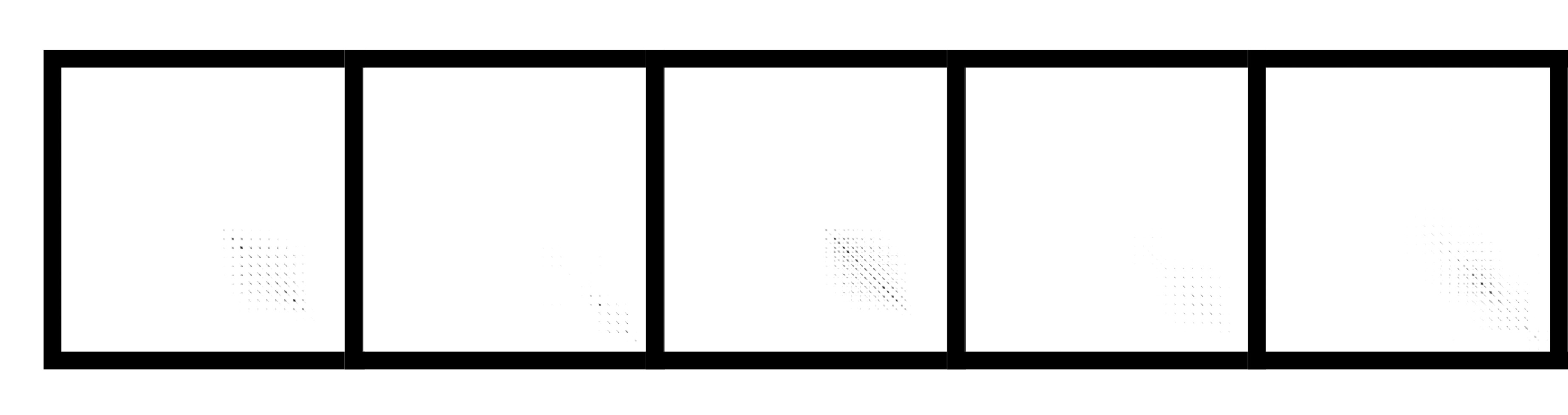}
    \caption{Covariance of the posterior samples for the 5 different settings of parameters to be estimated. The values are scaled by \num{4} for visibility. }
    \label{fig:otpostcovve}
    \vspace*{3mm}    
\end{subfigure}

\caption{Additional results for the hyper-elastic mechanics problem, noise level 1, shown in \cref{subsec:hemp} via SMLD. Note that the last column in \cref{fig:otpostve} is, in fact, the square root of the diagonal in \cref{fig:otpostcovve}. }
\label{fig:otpostvet}
\end{figure}

\begin{figure}[bthp]
\begin{subfigure}[t]{1\textwidth}
    \centering
    \captionsetup{width=0.95\linewidth}
    \includegraphics[width=0.6\linewidth]{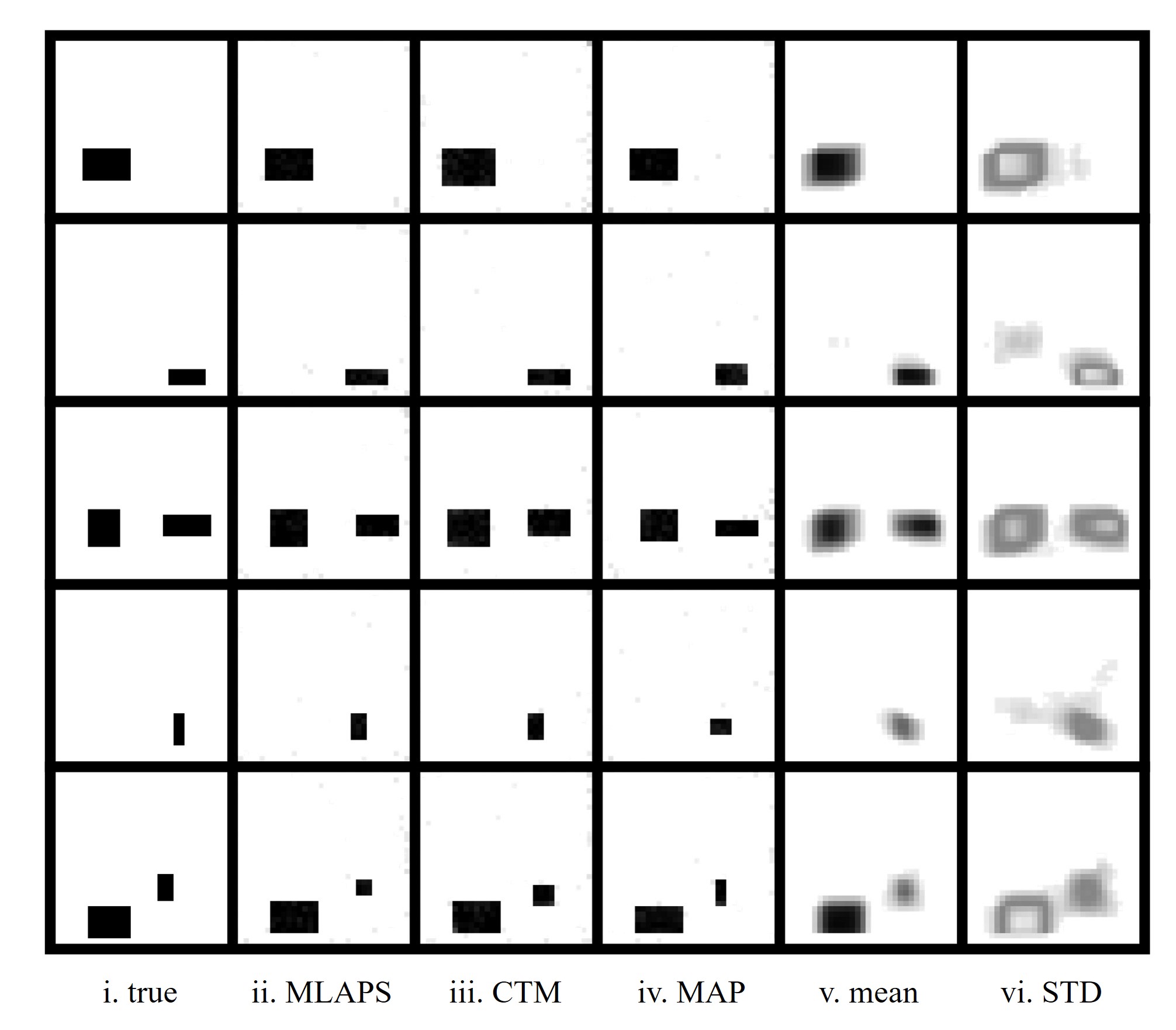}
    \caption{i.~the true solution, ii.~the MLAPS point, $\boldsymbol{\mu}_{\rm MLAPS}$, iii.~the CTM point, $\boldsymbol{\mu}_{\rm CTM}$, iv.~the MAP point, $\boldsymbol{\mu}_{\rm MAP}$, v.~the mean of the posterior samples, $\boldsymbol{\mu}_{\rm mean}$, and vi.~the STD of the posterior samples. The rows represent \num{5} different settings of parameters to be estimated. }
    \label{fig:otpostvp}
    \vspace*{3mm}    
\end{subfigure}

\begin{subfigure}[t]{1\textwidth}
    \centering
    \captionsetup{width=0.95\linewidth}
    \includegraphics[width=0.8\linewidth]{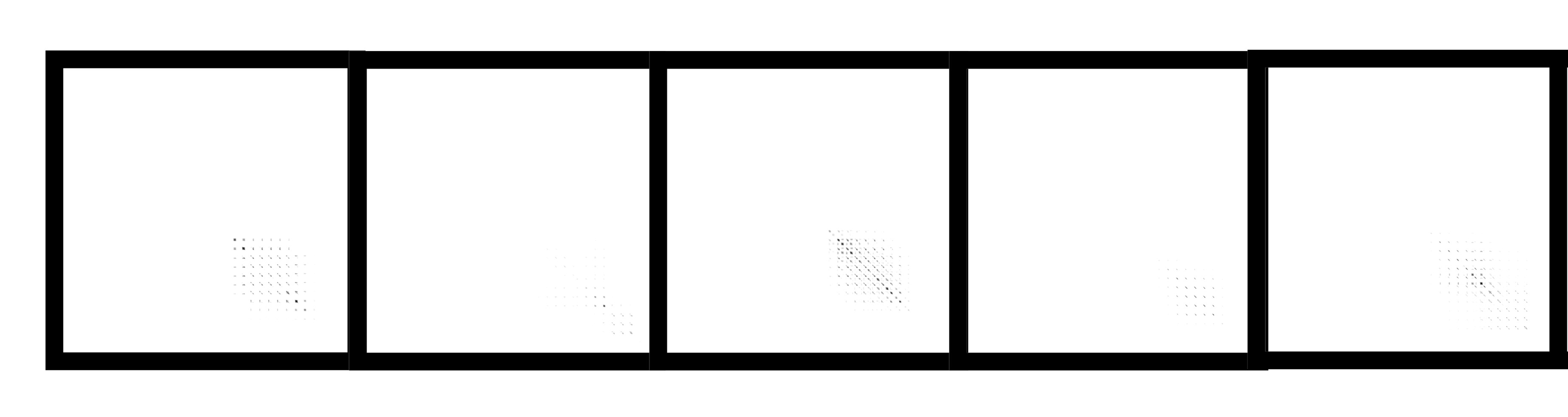}
    \caption{Covariance of the posterior samples for the 5 different settings of parameters to be estimated. The values are scaled by \num{4} for visibility. }
    \label{fig:otpostcovvp}
    \vspace*{3mm}    
\end{subfigure}

\caption{Additional results for the hyper-elastics mechanics problem, noise level 1, shown in \cref{subsec:hemp} via DDPM. }
\label{fig:otpostvpt}
\end{figure}

\begin{figure}[bthp]
\begin{subfigure}[t]{1\textwidth}
    \centering
    \captionsetup{width=0.95\linewidth}
    \includegraphics[width=0.6\linewidth]{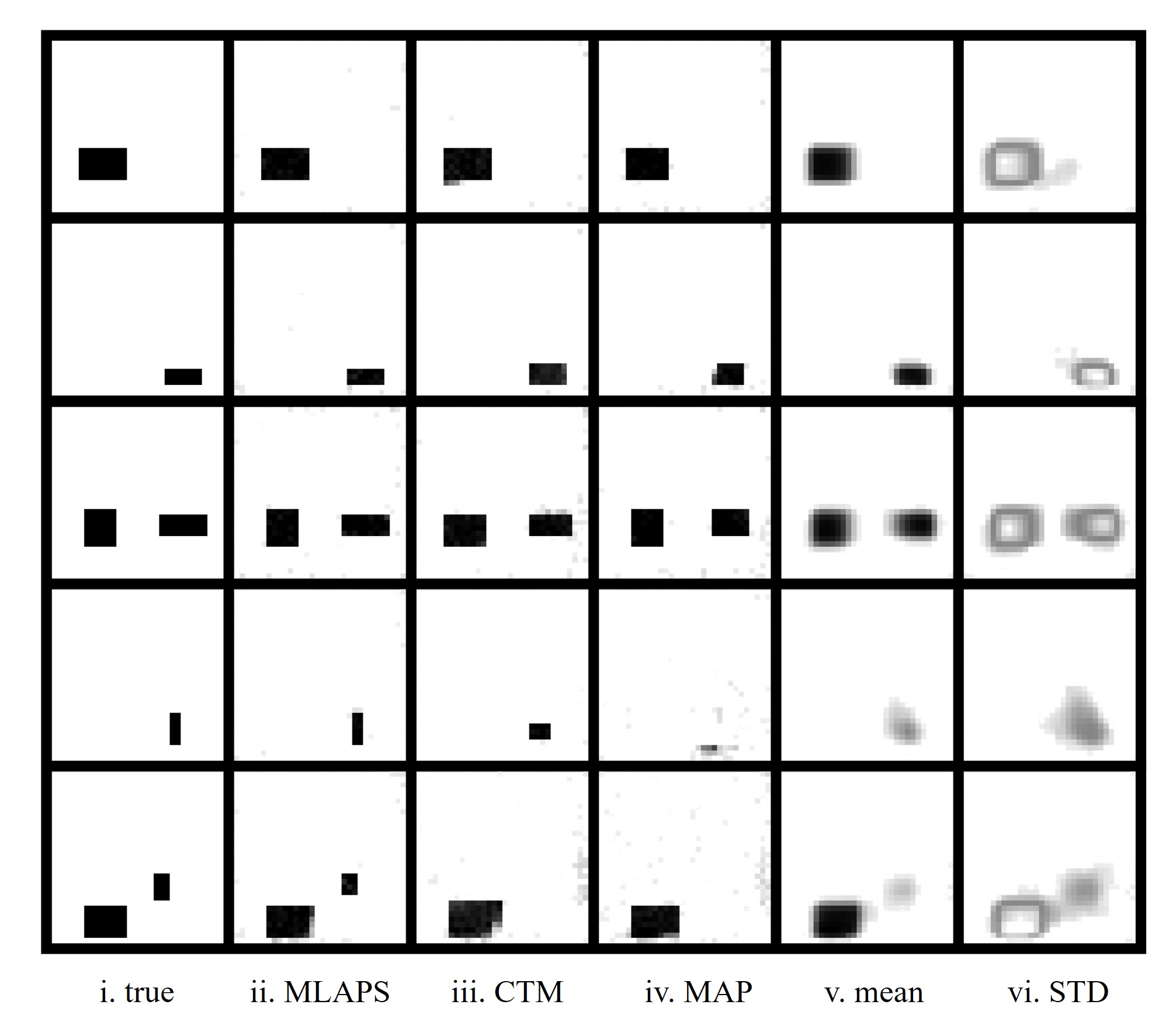}
    \caption{i.~the true solution, ii.~the MLAPS point, $\boldsymbol{\mu}_{\rm MLAPS}$, iii.~the CTM point, $\boldsymbol{\mu}_{\rm CTM}$, iv.~the MAP point, $\boldsymbol{\mu}_{\rm MAP}$, v.~the mean of the posterior samples, $\boldsymbol{\mu}_{\rm mean}$, and vi.~the STD of the posterior samples. The rows represent \num{5} different settings of parameters to be estimated. }
    \label{fig:msotpostve}
    \vspace*{3mm}    
\end{subfigure}

\begin{subfigure}[t]{1\textwidth}
    \centering
    \captionsetup{width=0.95\linewidth}
    \includegraphics[width=0.8\linewidth]{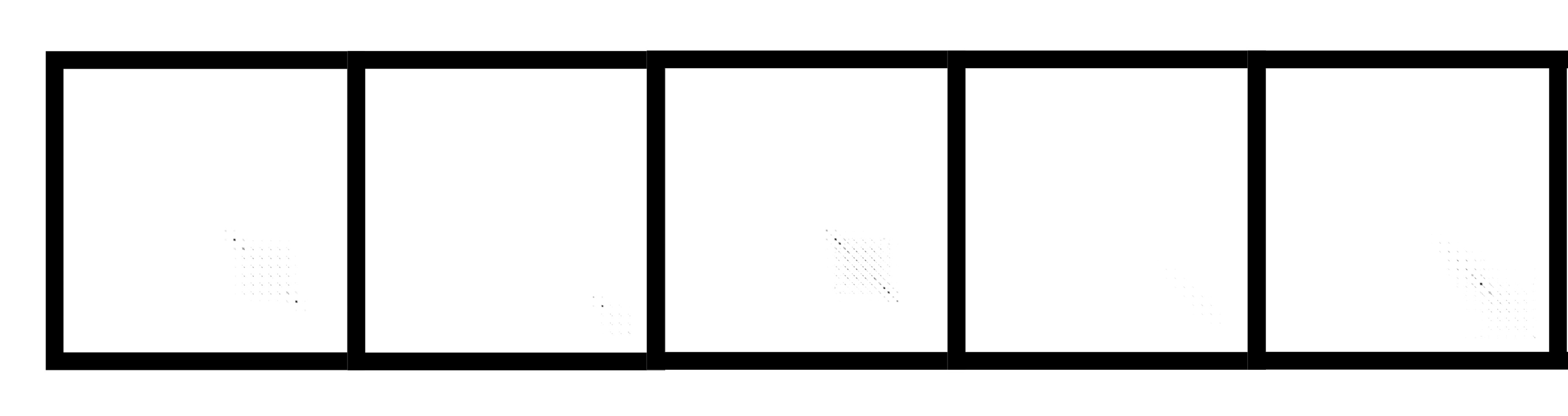}
    \caption{Covariance of the posterior samples for the 5 different settings of parameters to be estimated. The values are scaled by \num{4} for visibility. }
    \label{fig:msotpostcovve}
    \vspace*{3mm}    
\end{subfigure}

\caption{Additional results for the multi-scale mechanics problem, noise level 1, shown in \cref{subsec:msmp} via SMLD. }
\label{fig:msotpostvet}
\end{figure}

\begin{figure}[bthp]
\begin{subfigure}[t]{1\textwidth}
    \centering
    \captionsetup{width=0.95\linewidth}
    \includegraphics[width=0.6\linewidth]{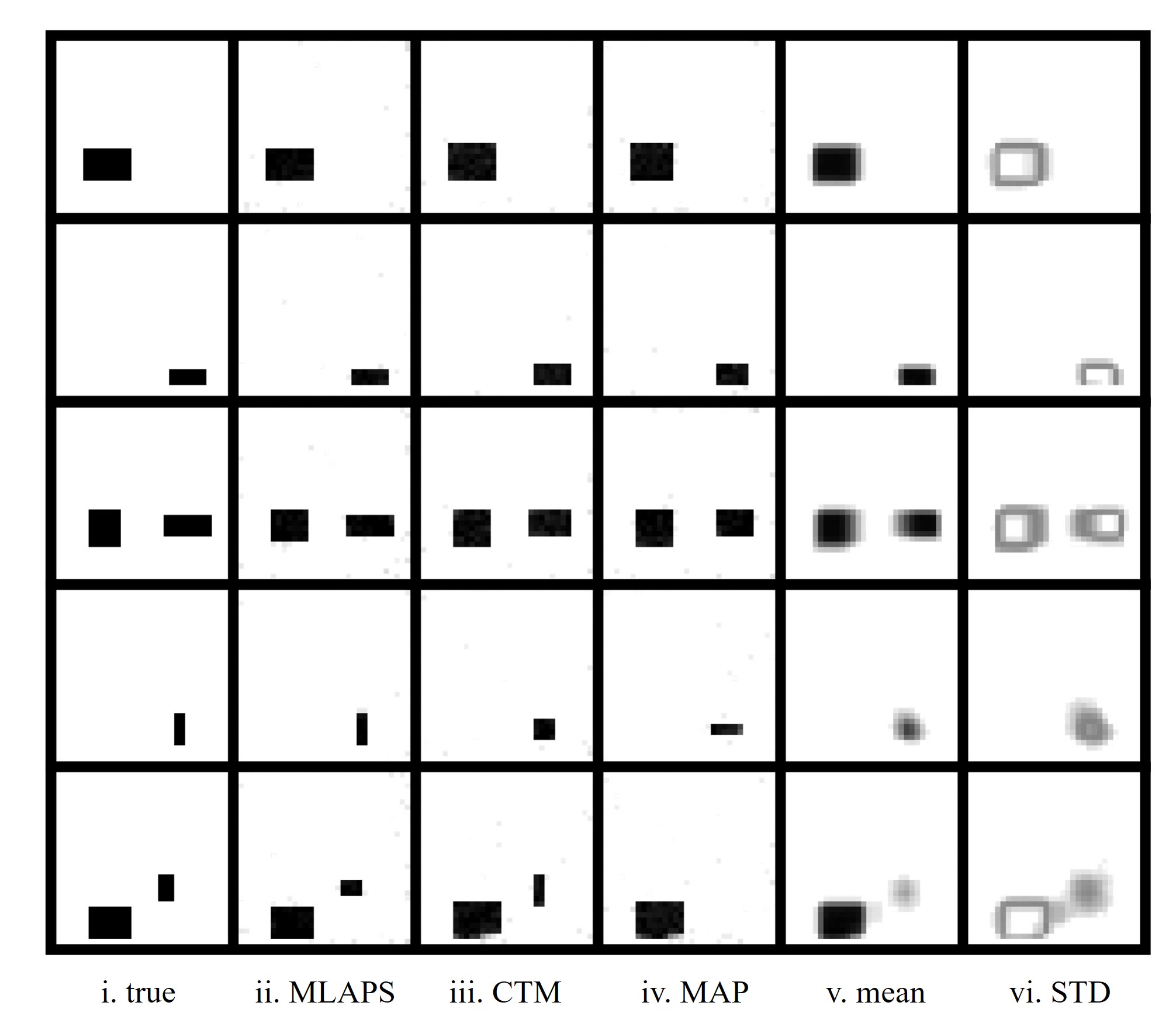}
    \caption{i.~the true solution, ii.~the MLAPS point, $\boldsymbol{\mu}_{\rm MLAPS}$, iii.~the CTM point, $\boldsymbol{\mu}_{\rm CTM}$, iv.~the MAP point, $\boldsymbol{\mu}_{\rm MAP}$, v.~the mean of the posterior samples, $\boldsymbol{\mu}_{\rm mean}$, and vi.~the STD of the posterior samples. The rows represent \num{5} different settings of parameters to be estimated. }
    \label{fig:msotpostvp}
    \vspace*{3mm}    
\end{subfigure}

\begin{subfigure}[t]{1\textwidth}
    \centering
    \captionsetup{width=0.95\linewidth}
    \includegraphics[width=0.8\linewidth]{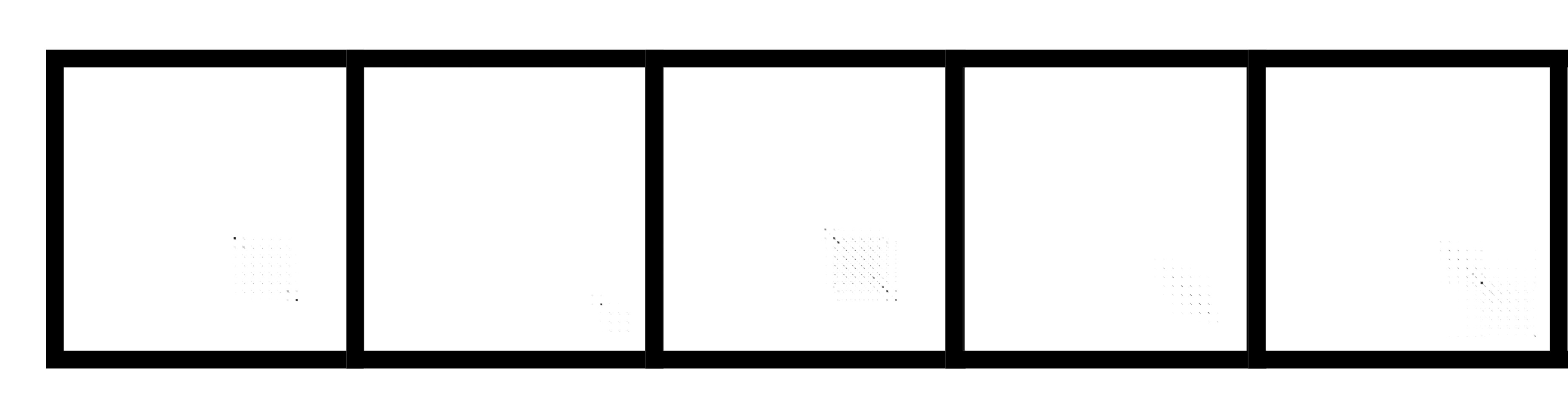}
    \caption{Covariance of the posterior samples for the 5 different settings of parameters to be estimated. The values are scaled by \num{4} for visibility. }
    \label{fig:msotpostcovvp}
    \vspace*{3mm}    
\end{subfigure}

\caption{Additional results for the multi-scale mechanics problem, noise level 1, shown in \cref{subsec:msmp} via DDPM. }
\label{fig:msotpostvpt}
\end{figure}

We provide a detailed exposition of the results obtained for the hyper-elastic and multi-scale problems, utilizing the score-based generative diffusion model as a sampling-based inverse solver. The outcomes extend beyond a singular solution point, offering a set of posterior samples that enable thorough uncertainty quantification. \Cref{fig:otpostvet,fig:otpostvpt} showcase the results through SMLD (using VE SDE) and DDPM (using VP SDE), respectively, for the hyper-elastic problem. Similarly, \cref{fig:msotpostvet,fig:msotpostvpt} present results via SMLD (using VE SDE) and DDPM (using VP SDE), respectively, for the multi-scale problem. The observation noise is set to be the lowest level (level 1 in \cref{sec:Res}). The noteworthy aspects of these results include:

\emph{(1)} The maximum likelihood point among the posterior samples, denoted as the MLAPS point, $\boldsymbol{\mu}_{\rm MLAPS}$, is defined as
\begin{equation*}
    \boldsymbol{\mu}_{\rm MLAPS} := \arg \max_{\boldsymbol{\mu} \in \hat{S}} - \left(y-\mathcal{H} \circ \mathcal{A} ({\boldsymbol{\mu}})\right)^T \cdot \left(y-\mathcal{H} \circ \mathcal{A} ({\boldsymbol{\mu}})\right), 
\end{equation*}
where $\hat{S}$ represents the set of posterior samples. 

\emph{(2)} The point that is closest to the mean $\boldsymbol{\mu}_{\rm mean}$, denoted as CTM point, is given by
\begin{equation*}
    \boldsymbol{\mu}_{\rm CTM} := \arg \min_{\boldsymbol{\mu} \in \hat{S}} \left\| \boldsymbol{\mu} - \boldsymbol{\mu}_{\rm mean} \right\|^2.
\end{equation*}

\emph{(3)} The maximum a posterior point (MAP point), $\boldsymbol{\mu}_{\rm MAP}$, is defined within the discretized samples $\hat{S}$ as 
\begin{equation*}
    \boldsymbol{\mu}_{\rm MAP} := \arg \max_{\boldsymbol{\mu} \in \hat{S}} \sum_{\tilde{\boldsymbol{\mu}} \in \hat{S}} \textbf{1}(\left\| \boldsymbol{\mu} - \tilde{\boldsymbol{\mu}} \right\| < d_{\rm neighbor}), 
\end{equation*}
where $\sum_{\tilde{\boldsymbol{\mu}} \in \hat{S}} \textbf{1}(\left\| \boldsymbol{\mu} - \tilde{\boldsymbol{\mu}} \right\| < d_{\rm neighbor})$ denotes the number of neighbors for $\boldsymbol{\mu}$, and $d_{\rm neighbor}$ is a hyper-parameter.

\emph{(4)} The mean of the posterior samples is denoted as $\boldsymbol{\mu}_{\rm mean}$.

\emph{(5)} The standard deviation (STD) of the posterior samples represents the uncertainty of the inverse results.

Additionally, we provide the complete covariance matrix of the posterior samples. The standard deviation array, which can be reshaped into the image formulation, is the square root of the diagonal of the covariance.

\section*{Acknowledgments}
The research leading to these results received funding from the European Research Council (ERC) under the European Union’s Horizon 2020 Research and Innovation Programme (Grant Agreement No. 818473).

\bibliography{SBDM}

\begin{thebibliography}{67}
\providecommand{\natexlab}[1]{#1}
\providecommand{\url}[1]{\texttt{#1}}
\expandafter\ifx\csname urlstyle\endcsname\relax
  \providecommand{\doi}[1]{doi: #1}\else
  \providecommand{\doi}{doi: \begingroup \urlstyle{rm}\Url}\fi

\bibitem[Abdulle et~al.(2020)Abdulle, Garegnani, and Zanoni]{abdulle_ensemble_2020}
A.~Abdulle, G.~Garegnani, and A.~Zanoni.
\newblock Ensemble {Kalman} filter for multiscale inverse problems.
\newblock \emph{Multiscale Modeling \& Simulation}, 18\penalty0 (4):\penalty0 1565--1594, Jan. 2020.
\newblock \doi{10.1137/20M1348431}.

\bibitem[Alnæs et~al.(2015)Alnæs, Blechta, Hake, Johansson, Kehlet, Logg, Richardson, Ring, Rognes, and Wells]{alnaes_fenics_2015}
M.~Alnæs, J.~Blechta, J.~Hake, A.~Johansson, B.~Kehlet, A.~Logg, C.~Richardson, J.~Ring, M.~E. Rognes, and G.~N. Wells.
\newblock The {FEniCS} project version 1.5.
\newblock \emph{Archive of Numerical Software}, Vol 3, 2015.
\newblock \doi{10.11588/ANS.2015.100.20553}.

\bibitem[Anderson(1982)]{anderson_reverse-time_1982}
B.~D. Anderson.
\newblock Reverse-time diffusion equation models.
\newblock \emph{Stochastic Processes and their Applications}, 12\penalty0 (3):\penalty0 313--326, May 1982.
\newblock \doi{10.1016/0304-4149(82)90051-5}.

\bibitem[Arridge et~al.(2019)Arridge, Maass, Öktem, and Schönlieb]{arridge_solving_2019}
S.~Arridge, P.~Maass, O.~Öktem, and C.-B. Schönlieb.
\newblock Solving inverse problems using data-driven models.
\newblock \emph{Acta Numerica}, 28:\penalty0 1--174, May 2019.
\newblock \doi{10.1017/S0962492919000059}.

\bibitem[Baldassari et~al.(2023)Baldassari, Siahkoohi, Garnier, Solna, and de~Hoop]{baldassari_conditional_2023}
L.~Baldassari, A.~Siahkoohi, J.~Garnier, K.~Solna, and M.~V. de~Hoop.
\newblock Conditional score-based diffusion models for {Bayesian} inference in infinite dimensions, Oct. 2023.

\bibitem[Benning and Burger(2018)]{benning_modern_2018}
M.~Benning and M.~Burger.
\newblock Modern regularization methods for inverse problems, Jan. 2018.

\bibitem[Bohra et~al.(2023)Bohra, Pham, Dong, and Unser]{bohra_bayesian_2023}
P.~Bohra, T.-a. Pham, J.~Dong, and M.~Unser.
\newblock Bayesian inversion for nonlinear imaging models using deep generative priors, May 2023.

\bibitem[Chada and Tong(2021)]{chada_convergence_2021}
N.~Chada and X.~Tong.
\newblock Convergence acceleration of ensemble {Kalman} inversion in nonlinear settings.
\newblock \emph{Mathematics of Computation}, Dec. 2021.
\newblock \doi{10.1090/mcom/3709}.

\bibitem[Chada et~al.(2020)Chada, Stuart, and Tong]{chada_tikhonov_2020}
N.~K. Chada, A.~M. Stuart, and X.~T. Tong.
\newblock Tikhonov regularization within ensemble {Kalman} inversion.
\newblock \emph{SIAM Journal on Numerical Analysis}, 58\penalty0 (2):\penalty0 1263--1294, Jan. 2020.
\newblock \doi{10.1137/19M1242331}.

\bibitem[Chung et~al.(2023)Chung, Kim, Mccann, Klasky, and Ye]{chung_diffusion_2023}
H.~Chung, J.~Kim, M.~T. Mccann, M.~L. Klasky, and J.~C. Ye.
\newblock Diffusion posterior sampling for general noisy inverse problems, Feb. 2023.

\bibitem[Dasgupta et~al.(2023)Dasgupta, Murgoitio-Esandi, Ray, and Oberai]{dasgupta_conditional_2023}
A.~Dasgupta, J.~Murgoitio-Esandi, D.~Ray, and A.~Oberai.
\newblock Conditional score-based generative models for solving physics-based inverse problems.
\newblock In \emph{{NeurIPS} 2023 {Workshop} on {Deep} {Learning} and {Inverse} {Problems}}, Nov. 2023.
\newblock URL \url{https://openreview.net/forum?id=ZL5wlFMg0Y}.

\bibitem[De~Bortoli et~al.(2022)De~Bortoli, Mathieu, Hutchinson, Thornton, Teh, and Doucet]{de_bortoli_riemannian_2022}
V.~De~Bortoli, E.~Mathieu, M.~Hutchinson, J.~Thornton, Y.~W. Teh, and A.~Doucet.
\newblock Riemannian score-based generative modelling, Nov. 2022.

\bibitem[Efron(2011)]{efron_tweedies_2011}
B.~Efron.
\newblock Tweedie’s formula and selection bias.
\newblock \emph{Journal of the American Statistical Association}, 106\penalty0 (496):\penalty0 1602--1614, Dec. 2011.
\newblock URL \url{http://www.tandfonline.com/doi/abs/10.1198/jasa.2011.tm11181}.

\bibitem[Engl et~al.(2000)Engl, Hanke-Bourgeois, and Neubauer]{engl_regularization_2000}
H.~W. Engl, M.~Hanke-Bourgeois, and A.~Neubauer.
\newblock \emph{Regularization of {Inverse} {Problems}}.
\newblock Number 375 in Mathematics and {Its} {Applications}. Kluwer Acad. Publ, Dordrecht, 2000.
\newblock ISBN 978-0-7923-6140-4 978-0-7923-4157-4.

\bibitem[Gao et~al.(2021)Gao, Sun, and Wang]{gao_phygeonet_2021}
H.~Gao, L.~Sun, and J.-X. Wang.
\newblock {PhyGeoNet}: physics-informed geometry-adaptive convolutional neural networks for solving parameterized steady-state {PDEs} on irregular domain.
\newblock \emph{Journal of Computational Physics}, 428:\penalty0 110079, Mar. 2021.
\newblock \doi{10.1016/j.jcp.2020.110079}.

\bibitem[Gao et~al.(2020)Gao, Sitharam, and Roitberg]{gao_bounds_2020}
X.~Gao, M.~Sitharam, and A.~E. Roitberg.
\newblock Bounds on the {Jensen} gap, and implications for mean-concentrated distributions, Aug. 2020.

\bibitem[Geers et~al.(2017)Geers, Kouznetsova, Matouš, and Yvonnet]{stein_homogenization_2017}
M.~G.~D. Geers, V.~G. Kouznetsova, K.~Matouš, and J.~Yvonnet.
\newblock Homogenization methods and multiscale modeling: nonlinear problems.
\newblock In \emph{Encyclopedia of {Computational} {Mechanics} {Second} {Edition}}, pages 1--34. Wiley, United States, 1 edition, Dec. 2017.
\newblock ISBN 978-1-119-00379-3 978-1-119-17681-7.
\newblock \doi{10.1002/9781119176817.ecm2107}.

\bibitem[Goodfellow et~al.(2016)Goodfellow, Bengio, and Courville]{goodfellow_deep_2016}
I.~Goodfellow, Y.~Bengio, and A.~Courville.
\newblock \emph{Deep {Learning}}.
\newblock Adaptive {Computation} and {Machine} {Learning}. The MIT Press, Cambridge, Massachusetts, 2016.
\newblock ISBN 978-0-262-03561-3.

\bibitem[Goodfellow et~al.(2020)Goodfellow, Pouget-Abadie, Mirza, Xu, Warde-Farley, Ozair, Courville, and Bengio]{goodfellow_generative_2020}
I.~Goodfellow, J.~Pouget-Abadie, M.~Mirza, B.~Xu, D.~Warde-Farley, S.~Ozair, A.~Courville, and Y.~Bengio.
\newblock Generative adversarial networks.
\newblock \emph{Communications of the ACM}, 63\penalty0 (11):\penalty0 139--144, Oct. 2020.
\newblock \doi{10.1145/3422622}.

\bibitem[Grenander and Miller(1994)]{grenander_representations_1994}
U.~Grenander and M.~I. Miller.
\newblock Representations of knowledge in complex systems.
\newblock \emph{Journal of the Royal Statistical Society. Series B (Methodological)}, 56\penalty0 (4):\penalty0 549--603, 1994.
\newblock URL \url{https://www.jstor.org/stable/2346184}.

\bibitem[Gruber et~al.(2022)Gruber, Gunzburger, Ju, and Wang]{gruber_comparison_2022}
A.~Gruber, M.~Gunzburger, L.~Ju, and Z.~Wang.
\newblock A comparison of neural network architectures for data-driven reduced-order modeling.
\newblock \emph{Computer Methods in Applied Mechanics and Engineering}, 393:\penalty0 114764, Apr. 2022.
\newblock \doi{10.1016/j.cma.2022.114764}.

\bibitem[Guo and Hesthaven(2018)]{guo_reduced_2018}
M.~Guo and J.~S. Hesthaven.
\newblock Reduced order modeling for nonlinear structural analysis using {Gaussian} process regression.
\newblock \emph{Computer Methods in Applied Mechanics and Engineering}, 341:\penalty0 807--826, Nov. 2018.
\newblock \doi{10.1016/j.cma.2018.07.017}.

\bibitem[He et~al.(2015)He, Zhang, Ren, and Sun]{he_deep_2015}
K.~He, X.~Zhang, S.~Ren, and J.~Sun.
\newblock Deep residual learning for image recognition, Dec. 2015.

\bibitem[Hesthaven and Ubbiali(2018)]{hesthaven_non-intrusive_2018}
J.~S. Hesthaven and S.~Ubbiali.
\newblock Non-intrusive reduced order modeling of nonlinear problems using neural networks.
\newblock \emph{Journal of Computational Physics}, 363:\penalty0 55--78, June 2018.
\newblock \doi{10.1016/j.jcp.2018.02.037}.

\bibitem[Hesthaven et~al.(2016)Hesthaven, Rozza, and Stamm]{hesthaven_certified_2016}
J.~S. Hesthaven, G.~Rozza, and B.~Stamm.
\newblock \emph{Certified {Reduced} {Basis} {Methods} for {Parametrized} {Partial} {Differential} {Equations}}.
\newblock Springer {Briefs} in {Mathematics}. Springer, Cham Heidelberg New York Dordrecht London, 2016.
\newblock ISBN 978-3-319-22469-5.
\newblock \doi{10.1007/978-3-319-22470-1}.

\bibitem[Heusel et~al.(2017)Heusel, Ramsauer, Unterthiner, Nessler, and Hochreiter]{heusel_gans_2017}
M.~Heusel, H.~Ramsauer, T.~Unterthiner, B.~Nessler, and S.~Hochreiter.
\newblock {GANs} trained by a two time-scale update rule converge to a local {Nash} equilibrium.
\newblock In \emph{Advances in {Neural} {Information} {Processing} {Systems}}, volume~30. Curran Associates, Inc., 2017.
\newblock URL \url{https://proceedings.neurips.cc/paper/2017/hash/8a1d694707eb0fefe65871369074926d-Abstract.html}.

\bibitem[Ho et~al.(2020)Ho, Jain, and Abbeel]{ho_denoising_2020}
J.~Ho, A.~Jain, and P.~Abbeel.
\newblock Denoising diffusion probabilistic models, Dec. 2020.

\bibitem[Holzschuh et~al.(2023)Holzschuh, Vegetti, and Thuerey]{holzschuh_solving_2023}
B.~J. Holzschuh, S.~Vegetti, and N.~Thuerey.
\newblock Solving inverse physics problems with score matching, Dec. 2023.

\bibitem[Hong et~al.(2024)Hong, Bansal, and Veroy]{hong_physics-informed_2024}
Y.~Hong, H.~Bansal, and K.~Veroy.
\newblock Physics-informed two-tier neural network for non-linear model order reduction.
\newblock preprint, In Review, Jan. 2024.

\bibitem[Hyvarinen and Dayan(2005)]{hyvarinen_estimation_2005}
A.~Hyvarinen and P.~Dayan.
\newblock Estimation of non-normalized statistical models by score matching.
\newblock \emph{Journal of Machine Learning Research}, 6, 2005.

\bibitem[Iglesias et~al.(2013)Iglesias, Law, and Stuart]{iglesias_ensemble_2013}
M.~A. Iglesias, K.~J.~H. Law, and A.~M. Stuart.
\newblock Ensemble {Kalman} methods for inverse problems.
\newblock \emph{Inverse Problems}, 29\penalty0 (4):\penalty0 045001, Apr. 2013.
\newblock \doi{10.1088/0266-5611/29/4/045001}.

\bibitem[Jaiswal et~al.(2023)Jaiswal, Zhang, Chan, and Wang]{jaiswal_physics-driven_2023}
A.~Jaiswal, X.~Zhang, S.~H. Chan, and Z.~Wang.
\newblock Physics-driven turbulence image restoration with stochastic refinement, July 2023.

\bibitem[Jalal et~al.(2021)Jalal, Arvinte, Daras, Price, Dimakis, and Tamir]{jalal_robust_2021}
A.~Jalal, M.~Arvinte, G.~Daras, E.~Price, A.~G. Dimakis, and J.~I. Tamir.
\newblock Robust compressed sensing {MRI} with deep generative priors, Dec. 2021.

\bibitem[Karniadakis et~al.(2021)Karniadakis, Kevrekidis, Lu, Perdikaris, Wang, and Yang]{karniadakis_physics-informed_2021}
G.~E. Karniadakis, I.~G. Kevrekidis, L.~Lu, P.~Perdikaris, S.~Wang, and L.~Yang.
\newblock Physics-informed machine learning.
\newblock \emph{Nature Reviews Physics}, 3\penalty0 (6):\penalty0 422--440, May 2021.
\newblock \doi{10.1038/s42254-021-00314-5}.

\bibitem[Kashefi et~al.(2023)Kashefi, Guibas, and Mukerji]{kashefi_physics-informed_2023}
A.~Kashefi, L.~J. Guibas, and T.~Mukerji.
\newblock Physics-informed {PointNet}: {On} how many irregular geometries can it solve an inverse problem simultaneously? application to linear elasticity, Sept. 2023.

\bibitem[Kharazmi et~al.(2019)Kharazmi, Zhang, and Karniadakis]{kharazmi_variational_2019}
E.~Kharazmi, Z.~Zhang, and G.~E. Karniadakis.
\newblock Variational physics-informed neural networks for solving partial differential equations, Nov. 2019.

\bibitem[Kim and Ye(2021)]{kim_noise2score_2021}
K.~Kim and J.~C. Ye.
\newblock {Noise2Score}: {Tweedie}'s approach to self-supervised image denoising without clean images, Oct. 2021.

\bibitem[Kingma and Ba(2017)]{kingma_adam_2017}
D.~P. Kingma and J.~Ba.
\newblock Adam: a method for stochastic optimization, Jan. 2017.

\bibitem[Klebaner(2012)]{klebaner_introduction_2012}
F.~C. Klebaner.
\newblock \emph{Introduction to {Stochastic} {Calculus} with {Applications}}.
\newblock ICP, Imperial College Press, London, third edition edition, 2012.
\newblock ISBN 978-1-84816-831-2 978-1-84816-832-9.

\bibitem[Kovachki et~al.(2023)Kovachki, Li, Liu, Azizzadenesheli, Bhattacharya, and Stuart]{kovachki_neural_2023}
N.~Kovachki, Z.~Li, B.~Liu, K.~Azizzadenesheli, K.~Bhattacharya, and A.~Stuart.
\newblock Neural {Operator}: learning maps between function spaces with applications to {PDEs}, Apr. 2023.

\bibitem[Lagaris et~al.(1998)Lagaris, Likas, and Fotiadis]{lagaris_artificial_1998}
I.~Lagaris, A.~Likas, and D.~Fotiadis.
\newblock Artificial neural networks for solving ordinary and partial differential equations.
\newblock \emph{IEEE Transactions on Neural Networks}, 9\penalty0 (5):\penalty0 987–1000, 1998.
\newblock \doi{10.1109/72.712178}.

\bibitem[LeCun et~al.(2015)LeCun, Bengio, and Hinton]{lecun_deep_2015}
Y.~LeCun, Y.~Bengio, and G.~Hinton.
\newblock Deep learning.
\newblock \emph{Nature}, 521\penalty0 (7553):\penalty0 436--444, May 2015.
\newblock \doi{10.1038/nature14539}.

\bibitem[Legin et~al.(2023)Legin, Ho, Lemos, Perreault-Levasseur, Ho, Hezaveh, and Wandelt]{legin_posterior_2023}
R.~Legin, M.~Ho, P.~Lemos, L.~Perreault-Levasseur, S.~Ho, Y.~Hezaveh, and B.~Wandelt.
\newblock Posterior sampling of the initial conditions of the universe from non-linear large scale structures using score-based generative models, Apr. 2023.

\bibitem[Lu et~al.(2021)Lu, Jin, Pang, Zhang, and Karniadakis]{lu_learning_2021}
L.~Lu, P.~Jin, G.~Pang, Z.~Zhang, and G.~E. Karniadakis.
\newblock Learning nonlinear operators via {DeepONet} based on the universal approximation theorem of operators.
\newblock \emph{Nature Machine Intelligence}, 3\penalty0 (3):\penalty0 218--229, Mar. 2021.
\newblock ISSN 2522-5839.
\newblock \doi{10.1038/s42256-021-00302-5}.

\bibitem[Ma et~al.(2023)Ma, Zhang, Zhu, and Feng]{ma_preconditioned_2023}
H.~Ma, L.~Zhang, X.~Zhu, and J.~Feng.
\newblock Preconditioned score-based generative models, Dec. 2023.

\bibitem[Mao et~al.(2020)Mao, Jagtap, and Karniadakis]{mao_physics-informed_2020}
Z.~Mao, A.~D. Jagtap, and G.~E. Karniadakis.
\newblock Physics-informed neural networks for high-speed flows.
\newblock \emph{Computer Methods in Applied Mechanics and Engineering}, 360:\penalty0 112789, Mar. 2020.
\newblock \doi{10.1016/j.cma.2019.112789}.

\bibitem[Miehe and Koch(2002)]{miehe_computational_2002}
C.~Miehe and A.~Koch.
\newblock Computational micro-to-macro transitions of discretized microstructures undergoing small strains.
\newblock \emph{Archive of Applied Mechanics}, 72\penalty0 (4):\penalty0 300--317, July 2002.
\newblock \doi{10.1007/s00419-002-0212-2}.

\bibitem[Moliner et~al.(2023)Moliner, Lehtinen, and Välimäki]{moliner_solving_2023}
E.~Moliner, J.~Lehtinen, and V.~Välimäki.
\newblock Solving audio inverse problems with a diffusion model, Mar. 2023.

\bibitem[Paszke et~al.(2019)Paszke, Gross, Massa, Lerer, Bradbury, Chanan, Killeen, Lin, Gimelshein, Antiga, Desmaison, Kopf, Yang, DeVito, Raison, Tejani, Chilamkurthy, Steiner, Fang, Bai, and Chintala]{paszke_pytorch_2019}
A.~Paszke, S.~Gross, F.~Massa, A.~Lerer, J.~Bradbury, G.~Chanan, T.~Killeen, Z.~Lin, N.~Gimelshein, L.~Antiga, A.~Desmaison, A.~Kopf, E.~Yang, Z.~DeVito, M.~Raison, A.~Tejani, S.~Chilamkurthy, B.~Steiner, L.~Fang, J.~Bai, and S.~Chintala.
\newblock {PyTorch}: an imperative style, high-performance deep learning library.
\newblock In \emph{Advances in {Neural} {Information} {Processing} {Systems}}, volume~32, Vancouver, 2019. Curran Associates, Inc.
\newblock URL \url{https://proceedings.neurips.cc/paper_files/paper/2019/hash/bdbca288fee7f92f2bfa9f7012727740-Abstract.html}.

\bibitem[Pichi et~al.(2023)Pichi, Moya, and Hesthaven]{pichi_graph_2023}
F.~Pichi, B.~Moya, and J.~S. Hesthaven.
\newblock A graph convolutional autoencoder approach to model order reduction for parametrized {PDEs}, Nov. 2023.

\bibitem[Quarteroni et~al.(2016)Quarteroni, Manzoni, and Negri]{quarteroni_reduced_2016}
A.~Quarteroni, A.~Manzoni, and F.~Negri.
\newblock \emph{Reduced {Basis} {Methods} for {Partial} {Differential} {Equations}: an {Introduction}}.
\newblock Number~92 in {UNITEXT} {La} {Matematica} per il 3+2. Springer, Cham Heidelberg, 2016.
\newblock ISBN 978-3-319-15431-2 978-3-319-15430-5.
\newblock \doi{10.1007/978-3-319-15431-2}.

\bibitem[Raissi et~al.(2019)Raissi, Perdikaris, and Karniadakis]{raissi_physics-informed_2019}
M.~Raissi, P.~Perdikaris, and G.~E. Karniadakis.
\newblock Physics-informed neural networks: a deep learning framework for solving forward and inverse problems involving nonlinear partial differential equations.
\newblock \emph{Journal of Computational Physics}, 378:\penalty0 686--707, Feb. 2019.
\newblock \doi{10.1016/j.jcp.2018.10.045}.

\bibitem[Ronneberger et~al.(2015)Ronneberger, Fischer, and Brox]{ronneberger_u-net_2015}
O.~Ronneberger, P.~Fischer, and T.~Brox.
\newblock U-{Net}: convolutional networks for biomedical image segmentation, May 2015.

\bibitem[Rozet and Louppe(2023)]{rozet_score-based_2023}
F.~Rozet and G.~Louppe.
\newblock Score-based data assimilation, Oct. 2023.

\bibitem[Runkel et~al.(2023)Runkel, Moeller, Schönlieb, and Etmann]{runkel_learning_2023}
C.~Runkel, M.~Moeller, C.-B. Schönlieb, and C.~Etmann.
\newblock Learning posterior distributions in underdetermined inverse problems.
\newblock In \emph{Scale {Space} and {Variational} {Methods} in {Computer} {Vision}}, pages 187--209, Cham, 2023. Springer International Publishing.
\newblock ISBN 978-3-031-31975-4.
\newblock \doi{10.1007/978-3-031-31975-4_15}.

\bibitem[Scarlett et~al.(2023)Scarlett, Heckel, Rodrigues, Hand, and Eldar]{scarlett_theoretical_2023}
J.~Scarlett, R.~Heckel, M.~R.~D. Rodrigues, P.~Hand, and Y.~C. Eldar.
\newblock Theoretical perspectives on deep learning methods in inverse problems, Jan. 2023.

\bibitem[Schillings and Stuart(2017)]{schillings_analysis_2017}
C.~Schillings and A.~M. Stuart.
\newblock Analysis of the ensemble {Kalman} filter for inverse problems.
\newblock \emph{SIAM Journal on Numerical Analysis}, 55\penalty0 (3):\penalty0 1264--1290, Jan. 2017.
\newblock \doi{10.1137/16M105959X}.

\bibitem[Schmidhuber(2015)]{schmidhuber_deep_2015}
J.~Schmidhuber.
\newblock Deep learning in neural networks: an overview.
\newblock \emph{Neural Networks}, 61:\penalty0 85--117, Jan. 2015.
\newblock URL \url{https://linkinghub.elsevier.com/retrieve/pii/S0893608014002135}.

\bibitem[Sharma et~al.(2018)Sharma, Farimani, Gomes, Eastman, and Pande]{sharma_weakly-supervised_2018}
R.~Sharma, A.~B. Farimani, J.~Gomes, P.~Eastman, and V.~Pande.
\newblock Weakly-supervised deep learning of heat transport via physics informed loss, Aug. 2018.

\bibitem[Sohl-Dickstein et~al.(2015)Sohl-Dickstein, Weiss, Maheswaranathan, and Ganguli]{sohl-dickstein_deep_2015}
J.~Sohl-Dickstein, E.~A. Weiss, N.~Maheswaranathan, and S.~Ganguli.
\newblock Deep unsupervised learning using nonequilibrium thermodynamics, Nov. 2015.

\bibitem[Song and Ermon(2020)]{song_generative_2020}
Y.~Song and S.~Ermon.
\newblock Generative modeling by estimating gradients of the data distribution, Oct. 2020.

\bibitem[Song et~al.(2021)Song, Sohl-Dickstein, Kingma, Kumar, Ermon, and Poole]{song_score-based_2021}
Y.~Song, J.~Sohl-Dickstein, D.~P. Kingma, A.~Kumar, S.~Ermon, and B.~Poole.
\newblock Score-based generative modeling through stochastic differential equations, Feb. 2021.

\bibitem[Song et~al.(2022)Song, Shen, Xing, and Ermon]{song_solving_2022}
Y.~Song, L.~Shen, L.~Xing, and S.~Ermon.
\newblock Solving inverse problems in medical imaging with score-based generative models, June 2022.

\bibitem[Tancik et~al.(2020)Tancik, Srinivasan, Mildenhall, Fridovich-Keil, Raghavan, Singhal, Ramamoorthi, Barron, and Ng]{tancik_fourier_2020}
M.~Tancik, P.~P. Srinivasan, B.~Mildenhall, S.~Fridovich-Keil, N.~Raghavan, U.~Singhal, R.~Ramamoorthi, J.~T. Barron, and R.~Ng.
\newblock Fourier features let networks learn high frequency functions in low dimensional domains, June 2020.

\bibitem[Ulyanov et~al.(2018)Ulyanov, Vedaldi, and Lempitsky]{ulyanov_deep_2018}
D.~Ulyanov, A.~Vedaldi, and V.~Lempitsky.
\newblock Deep image prior.
\newblock In \emph{Proceedings of the {IEEE} {Conference} on {Computer} {Vision} and {Pattern} {Recognition}}, pages 9446--9454, 2018.
\newblock URL \url{https://openaccess.thecvf.com/content_cvpr_2018/html/Ulyanov_Deep_Image_Prior_CVPR_2018_paper.html}.

\bibitem[Vincent(2011)]{vincent_connection_2011}
P.~Vincent.
\newblock A connection between score matching and denoising autoencoders.
\newblock \emph{Neural Computation}, 23\penalty0 (7):\penalty0 1661--1674, July 2011.
\newblock URL \url{https://ieeexplore.ieee.org/abstract/document/6795935}.

\bibitem[Vlassis and Sun(2023)]{vlassis_denoising_2023}
N.~N. Vlassis and W.~Sun.
\newblock Denoising diffusion algorithm for inverse design of microstructures with fine-tuned nonlinear material properties.
\newblock \emph{Computer Methods in Applied Mechanics and Engineering}, 413:\penalty0 116126, Aug. 2023.
\newblock \doi{10.1016/j.cma.2023.116126}.

\end{thebibliography}

\end{document}